\pdfoutput=1
\documentclass[aps,prd,amsmath,floats,floatfix, twocolumn,
superscriptaddress,nofootinbib,showpacs,longbibliography]{revtex4-1}
\usepackage{aas_macros}
\usepackage[T1]{fontenc}
\usepackage[utf8]{inputenc}
\usepackage{txfonts}
\usepackage{verbatim}
\usepackage[dvipsnames, usenames]{xcolor}
\definecolor{linkcolor}{rgb}{0.0,0.3,0.5}
\usepackage[hypertexnames=false, unicode, colorlinks=true, linkcolor=linkcolor,
citecolor=linkcolor, filecolor=linkcolor,urlcolor=linkcolor,
pdfusetitle]{hyperref}
\usepackage[all]{hypcap}
\usepackage{graphicx}
\usepackage{xspace}
\usepackage{amssymb}
\usepackage{microtype}

\usepackage{url}

\usepackage[english]{babel}
\usepackage{blindtext}

\usepackage[normalem]{ulem} %for \sout
\usepackage{bm} % boldmath
\usepackage{mathrsfs}
\usepackage{xspace} % for \xspace
\usepackage{fontawesome}
\usepackage[caption=false]{subfig}

\DeclareMathAlphabet{\mathpzc}{OT1}{pzc}{m}{it}
\usepackage[nomessages]{fp}

\newcommand{\sk}[1]{}
\defcitealias{Islam:2025drw}{IW25}

\newcommand{\rapster}{\textcolor{linkcolor}{\texttt{rapster}}}

\newcommand{\gwModel}{\textcolor{linkcolor}{\texttt{gwModelRemP\_flow}}}
\newcommand{\HLZ}{\textcolor{linkcolor}{\texttt{HLZ}}}
\newcommand{\gwModelS}{\textcolor{linkcolor}{\texttt{gwModelRemS}}}
\newcommand{\gwModelP}{\textcolor{linkcolor}{\texttt{gwModelRemP}}}
\newcommand{\gwModelSE}{\textcolor{linkcolor}{\texttt{gwModelRemSE}}}
\newcommand{\gwModelPE}{\textcolor{linkcolor}{\texttt{gwModelRemPE}}}
\newcommand{\gwModelEMRI}{\textcolor{linkcolor}{\texttt{gwModelRemEMRI}}}
\newcommand{\HBR}{\textcolor{linkcolor}{\texttt{HBR}}}
\newcommand{\UIBa}{\textcolor{linkcolor}{\texttt{UIB2016}}}
\newcommand{\UIBb}{\textcolor{linkcolor}{\texttt{UIB2024}}}
\newcommand{\NRSur}{\textcolor{linkcolor}{\texttt{NRSur7dq4Remnant}}}
\newcommand{\NRSurEmri}{\textcolor{linkcolor}{\texttt{NRSur7dq4EmriRemnant}}}
\newcommand{\chihat}{\hat \chi}
\newcommand{\chia}{\chi_a}
\newcommand{\dm}{\delta_m}

\newcommand{\Sperp}{S_\perp}

\begin{document}
\title{Unified remnant models for aligned-spin, precessing, and eccentric binary black hole mergers}

\author{Tousif Islam}
\email{tousifislam@ucsb.edu}
\affiliation{Kavli Institute for Theoretical Physics, University of California Santa Barbara, Kohn Hall, Lagoon Rd, Santa Barbara, CA 93106}

\author{Digvijay Wadekar}
\affiliation{\mbox{Weinberg Institute, University of Texas at Austin, Austin, TX 78712, USA}}

\author{Gaurav Khanna} \affiliation{Department of Physics and Institute for AI \& Computational Research, University of Rhode Island, Kingston, RI 02881}
\affiliation{Department of Physics and Center for Scientific Computing \& Data Science Research, University of Massachusetts, Dartmouth, MA 02747}

%%%%%%%%%%%%%%%%%%%%%%%%%%%%%%%%%%%%%%%%%%%%%%%%%%%%%%%%%%%%%%%%%%%%%%%%%%%

% Because hyperref only gets the *last* author, we need to be explicit.
\hypersetup{pdfauthor={Islam et al.}}

\date{\today}

%==========================================================================
%==========================================================================
%==========================================================================
%==========================================================================
\begin{abstract}
Using approximately $5000$ numerical-relativity (NR) simulations spanning mass ratios up to $q=128$ and $1200$ black-hole-perturbation-theory (BHPT) simulations extending to $q=1000$, we present fully analytic models for the remnant properties of quasi-circular binary black hole mergers. The models \gwModelS{} (for final mass, final spin, peak luminosity, and recoil kick) and \gwModelP{} (for final mass, final spin, and peak luminosity) describe nonprecessing and precessing binaries, respectively. 
Our approach combines analytic insights from post-Newtonian theory and point-particle limits with a data-driven fitting framework which also includes validation-guided AI-agent-assisted optimization.
The resulting fits outperform existing analytic models and achieve accuracies comparable to data-driven models within their domains of validity.
We also construct \gwModel{}, a normalizing-flow model for recoil kicks from precessing binaries, marginalized over the in-plane spin orientations. This model additionally utilizes an enlarged low-fidelity training set, beyond the available NR and BHPT simulations, constructed through a waveform-based data-augmentation procedure.
We also include simple eccentric extensions through leading-order dependence on eccentricity and orbital anomaly.
The models span the equal-mass to extreme-mass-ratio regimes and are publicly available through the \textcolor{linkcolor}{\texttt{gwModels}}\footnote{\href{https://github.com/tousifislam/gwModels}{https://github.com/tousifislam/gwModels}\label{footnote:gwModels}} package for applications in gravitational-wave astronomy, astrophysical population studies, and cosmology.
\end{abstract}
%==========================================================================
%==========================================================================
%==========================================================================
%==========================================================================

\maketitle

%==========================================================================
%==========================================================================
%==========================================================================
\section{Introduction}
\label{sec:introduction}
%==========================================================================
%==========================================================================
%==========================================================================
The merger of two black holes results in a remnant black hole whose mass and spin differ from those of its progenitors~\cite{Maggiore:2007ulw,Maggiore:2018sht}. Conservation of linear momentum can also impart a velocity relative to the initial center-of-mass frame, known as the recoil or kick velocity. In binary black hole (BBH) simulations performed using numerical relativity (NR)~\cite{Baker:2006vn, Baker:2007gi, Baker:2008md,Herrmann:2006cd,Lousto:2007db,Herrmann:2007ac,Herrmann:2007ex,Herrmann:2007cwl,Holley-Bockelmann:2007hmm,Jaramillo:2011re,Koppitz:2007ev,Lousto:2008dn,Lousto:2010xk,Schnittman:2007ij,Sopuerta:2006et,Pollney:2007ss,Rezzolla:2010df,Lousto:2011kp,Lousto:2012gt,Lousto:2012su,Miller:2008en,Tichy:2007hk,Zlochower:2010sn,Healy:2014yta,Lousto:2009mf} or black hole perturbation theory (BHPT)~\cite{Nakano:2010kv,Sundararajan:2010sr,Islam:2023mob,Hughes:2004ck,Price:2013paa,Price:2011fm}, the remnant mass and spin are obtained from horizon or balance-law diagnostics, while the recoil follows from the remnant motion or radiated linear momentum. Such simulations are too computationally expensive for real-time evaluation.
Alternatively, one can use gravitational waveforms generated by models that approximate numerical simulations, either through data-driven surrogates or semi-analytical approaches, and employ flux-based methods to compute the remnant properties~\cite{Gerosa:2018qay,Iozzo:2021vnq,Islam:2023mob}. Although this approach is significantly faster than full numerical simulations, it remains prohibitively expensive for real-time applications. 

\begin{table*}[t]
\centering
\caption{Domains, spin configurations, calibration inputs, and key validation figures of the remnant models presented in this work. Together, the models cover aligned-spin, precessing, eccentric, and point-particle binary configurations.}
\label{tab:domain_validity}

\makebox[\textwidth][c]{%
\begin{tabular}{lccccccc}
\hline
Model & Mass ratio & Spin configuration & Eccentricity & Precession & Calibration & Section & Figures\\
\hline
\gwModelS{} &
$1 \leq q \lesssim 1000$ & Aligned spins & Quasi-circular & No & NR, BHPT &
\ref{sec:nonprec_nonecc} &
\ref{fig:parameter_space_nonprec}, \ref{fig:nonprec_nonecc_error_hist}, \ref{fig:single_spin_mf_chif_convergence}, \ref{fig:smoothness_mf_chif_vs_q}\\

\gwModelP{} &
$1 \leq q \lesssim 1000$ & Generic spins & Quasi-circular & Yes & NR, BHPT &
\ref{sec:prec_nonecc} &
\ref{fig:parameter_space_prec}, \ref{fig:prec_noecc_error_hist_log10}, \ref{fig:jsd}, \ref{fig:prec_kick_validation}\\

\gwModelSE{} &
$1 \leq q \lesssim 1000$ & Aligned spins & Leading-order & No & NR, BHPT &
\ref{sec:ecc} &
\ref{fig:ecc_models}\\

\gwModelPE{} &
$1 \leq q \lesssim 1000$ & Generic spins & Leading-order & Yes & NR, BHPT &
\ref{sec:ecc} &
--\\

\gwModelEMRI{} &
$q \gg 1000$ & Generic Kerr spin & Generic & Yes$^\dagger$ & Analytical &
\ref{sec:emri_model} &
--\\
\hline
\end{tabular}%
}

\vspace{1mm}
\begin{minipage}{0.92\textwidth}
\footnotesize
$^\dagger$ The point-particle model describes a test particle orbiting a Kerr black hole and includes generic orbital inclination rather than comparable-mass spin precession.
\end{minipage}

\end{table*}

Access to remnant properties of BBH mergers is, however, essential for characterizing the remnants of mergers detected by current gravitational-wave (GW) observatories. The ability to compute BBH remnant properties in real time is also necessary for astrophysical dynamical simulations~\cite{Holley-Bockelmann:2007hmm,Berti:2012zp,Gultekin:2004pm,Gerosa:2016vip,Borchers:2025sid,Gerosa:2021hsc,Gerosa:2017kvu,Gerosa:2021hsc,Baibhav:2020xdf,Bouffanais:2019nrw} of star clusters and active galactic nuclei, as well as for large-scale cosmological simulations.
These requirements motivate the development of fast and accurate fitting formulae for remnant properties across different classes of BBH mergers, including quasi-circular, eccentric, precessing, non-precessing, and non-spinning systems~\cite{Hofmann:2016yih,Barausse:2012qz,Barausse:2009uz,Lousto:2008dn,Lousto:2010xk,Lousto:2012gt,Lousto:2012su,Gonzalez:2007hi,Hofmann:2016yih,Planas:2024vnq,Jimenez-Forteza:2016oae,Varma:2019csw,Varma:2018aht,Islam:2021mha,Ravichandran:2026iec,Thomas:2025rje,Boschini:2023ryi,Islam:2023mob,Islam:2025drw}. 
Over the years, considerable effort has been devoted to constructing semi-analytical fits that incorporate analytic point-particle limits and add corrections calibrated to a relatively small set of NR simulations. Many such semi-analytical models also inherit functional forms inspired by post-Newtonian (PN) expressions for the radiated energy, angular momentum, and linear momentum~\cite{Blanchet:2005rj,Sopuerta:2006wj,Sopuerta:2006et,Favata:2004wz,Fitchett:1983qzq,Fitchett:1984qn,Wiseman:1992dv,Kidder:1995zr}. 

Among the most widely used remnant models are the semi-analytic fits developed in Refs.~\cite{Hofmann:2016yih,Barausse:2012qz,Barausse:2009uz}. These fits are anchored to the point-particle limit and subsequently calibrated against a set of NR simulations. Owing to their broad applicability and physical interpretability, they are widely used in astrophysics, cosmology, and GW astronomy. Throughout this paper, we collectively refer to these models as the \HBR{} fits, after the authors. 
Another family of semi-analytic fits for the remnant mass, spin, peak luminosity, and recoil velocity was developed by exploiting insights from PN theory and calibrated primarily to dedicated RIT simulations~\cite{Lousto:2008dn,Lousto:2010xk,Lousto:2012gt,Lousto:2012su,Gonzalez:2007hi}. We refer to these models as the \HLZ{} fits, after the authors. 
A third family of fits for the remnant mass, spin, and peak luminosity was developed using a combination of SXS and BAM/Einstein Toolkit simulations together with numerical BHPT results at mass ratio $q=1000$ where $q:=m_1/m_2$ with $m_1$ ($m_2$) is the mass of the larger (smaller) black hole~\cite{Hofmann:2016yih,Planas:2024vnq,Jimenez-Forteza:2016oae}. We refer to these models as the \UIBa{} fits for nonprecessing binaries and the \UIBb{} fits for single-spin precessing binaries.

These semi-analytical fits, however, are often less accurate, particularly for predicting recoil velocities. Data-driven surrogate models have emerged as an alternative approach for modeling remnant properties by interpolating between existing NR simulations~\cite{Varma:2019csw,Varma:2018aht,Islam:2021mha,Ravichandran:2026iec,Thomas:2025rje}. 
Two of the most widely used examples are \textcolor{linkcolor}{\texttt{NRSur3dq8Remnant}}~\cite{Varma:2018aht}, trained on NR simulations of comparable-mass non-precessing BBH mergers with mass ratios $q\leq 8$, and \textcolor{linkcolor}{\texttt{NRSur7dq4Remnant}}~\cite{Varma:2019csw}, trained on NR simulations of comparable-mass precessing BBH mergers with mass ratios $q\leq 4$.
More recently, surrogate models such as \textcolor{linkcolor}{\texttt{NRSur3dq8EmriRemnant}}~\cite{Boschini:2023ryi} have begun incorporating point-particle-limit information to improve extrapolation beyond the comparable-mass regime. 

In parallel, thousands of new NR simulations have been performed using different codes and made publicly available through the SXS~\cite{Mroue:2013xna,Boyle:2019kee,Scheel:2025jct}, RIT~\cite{Healy:2020vre,Healy:2022wdn,Ficarra:2026cej}, MAYA~\cite{Ferguson:2023vta,Jani:2016wkt}, BAM~\cite{Hamilton:2023qkv,Mahapatra:2026wsp}, and ICCUB~\cite{Trenado:2025ccf} catalogs. Because most existing remnant models were developed several years ago, updated fits can now leverage a substantially larger set of simulations.
Furthermore, the validity of BHPT waveforms has been extended further into the comparable-mass regime through the inclusion of empirical corrections~\cite{Islam:2022laz,Islam:2025tjj,Rink:2024swg,vandeMeent:2020xgc,Islam:2023qyt,Islam:2023jak} and, more recently, second-order perturbative effects~\cite{Wardell:2021fyy,Mathews:2021rod,Mathews:2025nyb}. This progress makes it possible to use perturbative waveforms, supplemented with appropriate corrections, to compute remnant properties in the intermediate mass-ratio regime, where NR simulations remain relatively sparse. 
In fact, Ref.~\cite{Islam:2023mob} used modified BHPT waveforms to develop a model (\textcolor{linkcolor}{\texttt{BHPTNRSurRemnant}}) for remnant properties for BBHs spanning mass ratios from $q=3$ to $q=1000$. However, the accuracy of that model deteriorates for $q \lesssim 10$ when predicting the final mass and spin, and for $q \lesssim 20$ when predicting the recoil velocity.

Recently, Ref.~\cite{Islam:2025drw} developed recoil-velocity models for both nonprecessing quasi-circular binaries (\textcolor{linkcolor}{\texttt{gwModel\_kick\_q200}}) and generic precessing binaries (\textcolor{linkcolor}{\texttt{gwModel\_kick\_prec\_flow}}) by combining NR datasets from multiple catalogs with approximately $100$ BHPT simulations extending to $q=128$. These models combine physically motivated PN-inspired structures with data-driven fitting and have since been incorporated into dynamical star-cluster simulations, where they yield astrophysical predictions that differ from those obtained using previous remnant prescriptions~\cite{Islam:2026iyn,Islam:2026yxx,Islam:2026sjl}.
Furthermore, the presented models are orders of magnitude faster than purely data-driven surrogate models~\cite{Islam:2025drw}, making them well suited for large-scale astrophysical simulations.

Motivated by these developments, we construct a new suite of remnant models for quasi-circular nonprecessing and precessing BBH mergers using a broad collection of numerical and analytic information. Our framework incorporates NR simulations from the SXS, RIT, MAYA, and BAM catalogs ($1 \leq q \leq 128$), supplemented by BHPT simulations spanning the comparable- to large-mass-ratio regimes ($15 \leq q \leq 1000$ for the non-precessing BBHs and $30 \leq q \leq 100$ for the precessing BBHs). We also impose known point-particle limits and draw inspiration from established PN expressions for the radiated energy, angular momentum, and linear momentum.

Based on established remnant-modeling experience and PN intuition, we construct a set of analytic ans\"atze. We then employed an AI-agent-assisted symbolic model-selection workflow based on the propose--evaluate--refine loop of Ref.~\cite{Islam:2026zob}. We implemented this workflow using the \textcolor{linkcolor}{\texttt{Claude Opus 4.6}} model through the \textcolor{linkcolor}{\texttt{Claude Code}} orchestrator. Conceptually, this approach is related to recent large-language-model (LLM)-assisted methods for symbolic regression and scientific equation discovery~\cite{Shojaee2024LLMSRSE,Yang2026ThinkLA,Grayeli2024SymbolicRW,su2026strideselfreflectiveagentframework,xia2025sr,wang2025drsr,guo2025sr,pang2026deliberateevolutionagenticreasoning}, as well as evolutionary coding-agent systems such as AlphaEvolve~\cite{Novikov2025AlphaEvolve}.
The physical limits, symmetry requirements, and analytic ans\"atze specified by us defined the constrained model space explored by the coding agent. Under human supervision, the agent performed a systematic search within this space, proposing modifications to the initial ans\"atze such as alternative decompositions, limiting-behavior prefactors, low-order polynomial terms, and cross terms between physically motivated variables. The agent had access to \textcolor{linkcolor}{\texttt{gwModeller}}\footnote{\href{https://github.com/tousifislam/gwModeller}{https://github.com/tousifislam/gwModeller}}, a private software repository developed over several years that contains code, error metrics, optimization routines, and modeling utilities frequently used in NR-, BHPT-, and PN-based waveform development. Modules and utilities from \textcolor{linkcolor}{\texttt{gwModeller}} have previously been used in several research projects, including Refs.~\cite{Islam:2022laz,Islam:2021mha,Islam:2024rhm,Islam:2024tcs,Islam:2023qyt,Islam:2024zqo,Rink:2024swg,Islam:2025drw,Islam:2023mob,Islam:2026zob}. During model construction, the agent reused this existing functionality for numerical optimization, error evaluation, and physical consistency checks rather than re-implementing established modeling and analysis tools. Furthermore, the final analytic expressions were inspected and selected by the authors.
The AI agent therefore assisted in optimizing, validating, and systematically refining the authors' physics-informed ans\"atze; it did not autonomously derive the underlying physics or determine the final model.

It is also important to note that the recoil kick is a highly nonlinear quantity to model, but it can be computed by integrating the linear-momentum flux obtained from accurate waveform models~\cite{Gerosa:2018qay,Islam:2023mob,Varma:2019csw}. This enables targeted sampling of regions of parameter space that are sparsely covered by direct NR simulations. In this work, we use this strategy to augment the training data for a general-purpose recoil model with kicks computed from surrogate waveforms~\cite{Varma:2019csw}. These surrogate-derived kicks are treated as lower-fidelity information (with smaller weights) than direct NR and BHPT results and are not used as independent validation data. To the best of our knowledge, this is the first use of waveform-model-derived recoil kicks for data augmentation in remnant modeling.

The resulting models, \gwModelS{} and \gwModelP{}, describe nonprecessing and precessing BBH merger remnants, respectively. While \gwModelS{} provides analytic fits for the remnant mass, spin, peak luminosity, and recoil velocity, \gwModelP{} includes fits for the remnant mass, spin, and peak luminosity only. We additionally develop a probabilistic model, \gwModel{}, for recoil kicks in precessing binaries that marginalizes over in-plane spin orientations.
We then develop simple eccentric extensions of these models, \gwModelSE{} and \gwModelPE{}, by incorporating leading-order corrections associated with the eccentricity and orbital anomaly of the binary.
Finally, we provide \gwModelEMRI{}, an analytical point-particle backbone for modeling the remnant properties of eccentric, inclined black hole mergers in the extreme-mass-ratio limit.
Our models span the equal-mass through extreme-mass-ratio regimes. Table~\ref{tab:domain_validity} summarizes their domains of validity.
These models are publicly available via the \textcolor{linkcolor}{\texttt{gwModels}}\textsuperscript{\ref{footnote:gwModels}} package and can be easily integrated into GW inference pipelines, binary evolution studies, and astrophysical simulations.

\begin{figure*}[t]
    \centering
    \subfloat[Nonprecessing data in the $(\eta,\chihat)$ plane.\label{fig:parameter_space_nonprec}]{
        \includegraphics[width=0.47\textwidth]{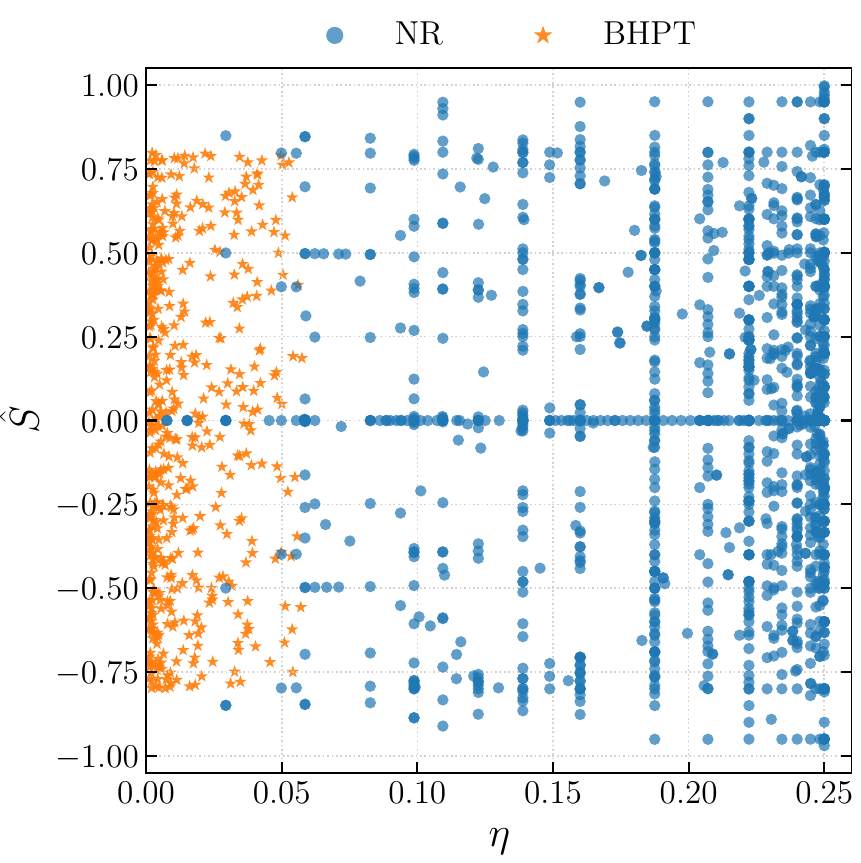}
    }
    \hfill
    \subfloat[Precessing data in the $(\eta,\Sperp)$ plane.\label{fig:parameter_space_prec}]{
        \includegraphics[width=0.47\textwidth]{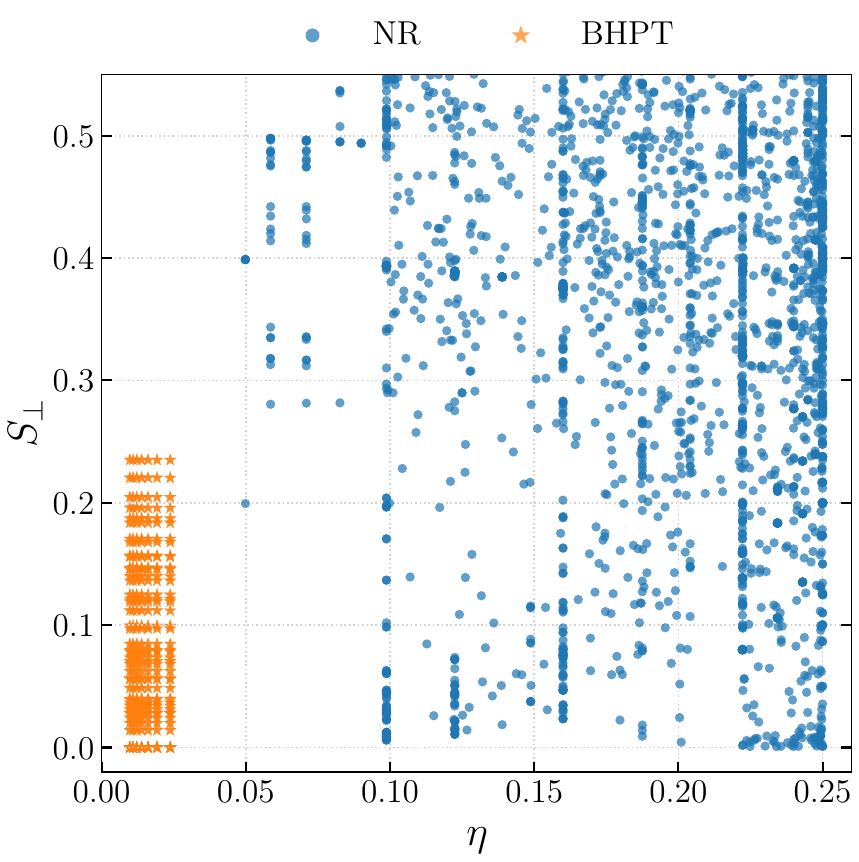}
    }
    \caption{Calibration across the comparable- to extreme-mass-ratio regimes requires complementary numerical methods. The panels show the NR and BHPT simulations used for the nonprecessing \gwModelS{} (left) and precessing \gwModelP{} (right) models. NR simulations densely sample comparable masses, whereas BHPT simulations extend the coverage to large mass ratios; see Secs.~\ref{sec:nonprec_data} and~\ref{sec:prec_nonecc_data}. Together, the datasets provide continuous coverage between the regimes accessible to full numerical simulations and perturbative calculations.}
    \label{fig:parameter_space}
\end{figure*}

The remainder of this paper presents analytic models for quasi-circular nonprecessing binaries (Sec.~\ref{sec:nonprec_nonecc}), extends the framework to generic precessing binaries (Sec.~\ref{sec:prec_nonecc}), and develops perturbative eccentric extensions (Sec.~\ref{sec:ecc}). Section~\ref{sec:emri_model} provides an analytic remnant model for the extreme-mass-ratio regime, Section~\ref{sec:timing} discusses the computational efficiency of our models, and Section~\ref{sec:astro} demonstrates representative astrophysical applications. The appendices provide implementation details, feature-engineering rationale, and further tests of deterministic recoil models for generic precessing BBHs.

%==========================================================================
%==========================================================================
%==========================================================================
\section{Non-precessing quasi-circular models}
\label{sec:nonprec_nonecc}
%==========================================================================
%==========================================================================
%==========================================================================
The remnant properties of BBH mergers are characterized by seven intrinsic parameters: the mass ratio $q$ and the two spin vectors $\vec{\chi}_1$ and $\vec{\chi}_2$. Each spin vector can be written as
\[
\vec{\chi}_1 = (\chi_{1x}, \chi_{1y}, \chi_{1z}), \qquad
\vec{\chi}_2 = (\chi_{2x}, \chi_{2y}, \chi_{2z}).
\]
Alternatively, the spins may be specified by spin angles $\theta_{1,2}$ (measured with respect to the orbital angular momentum axis) and azimuthal angles $\phi_{1,2}$ (describing the in-plane spin orientation). For nonprecessing quasi-circular BBHs, the in-plane spin components vanish, i.e.,
\[
\chi_{1x}=\chi_{1y}=\chi_{2x}=\chi_{2y}=0,
\]
such that the spins are fully characterized by $\chi_{1z}, \chi_{2z} \in [-1,1]$.

%==========================================================================
\subsection{Training dataset}
\label{sec:nonprec_data}
%==========================================================================
To construct our remnant models for nonprecessing BBH mergers, we curate a heterogeneous training dataset comprising NR simulations from the SXS, RIT, MAYA, and BAM catalogs, together with BHPT simulations. 
All BHPT waveforms are generated using the time-domain inspiral-merger-ringdown Teukolsky solver developed in Refs.~\cite{Khanna:2004,Burko:2007,Sundararajan:2008zm,Sundararajan:2010sr,Ori:2000zn,Hughes:2019zmt,Apte:2019txp}, with radiative energy and angular momentum losses computed using the open-source code \textit{GremlinEq}~\cite{gremlin,OSullivan:2014ywd,Drasco:2005kz}.
Specifically, the dataset includes $592$ SXS simulations spanning $1 \leq q \leq 20$ and $-0.97 \leq \chi_{1z,2z} \leq 0.99$, $667$ RIT simulations spanning $1 \leq q \leq 128$ and $-0.95 \leq \chi_{1z,2z} \leq 0.95$, and $141$ MAYA simulations spanning $1 \leq q \leq 15$ and $-0.81 \leq \chi_{1z,2z} \leq 0.81$. In total, the dataset contains approximately $1400$ NR simulations. We further supplement these data with approximately $500$ BHPT simulations spanning $15 \leq q \leq 1000$ and $-0.8 \leq \chi_{1z} \leq 0.8$, with $\chi_{2z}=0$. These BHPT waveforms are modified using mode-dependent empirical scalings designed to capture nonlinear effects~\cite{Islam:2022laz,Islam:2025tjj,Rink:2024swg}, yielding agreement with NR results in the comparable-mass regime. We show the parameter space covered by these simulations in Figure~\ref{fig:parameter_space_nonprec}. For the nonprecessing models, the NR simulations densely sample the comparable-mass regime and span nearly the full range of effective aligned spins, while the BHPT simulations extend the calibration toward small symmetric mass ratios and provide coverage of the extreme-mass-ratio limit. The resulting dataset spans the entire interval $0<\eta\le1/4$ and effectively bridges the gap between comparable-mass NR simulations and perturbative calculations. 

For the SXS and RIT catalogs, we use the remnant properties reported directly in the respective catalogs. The SXS catalog, however, does not provide peak luminosities. For the MAYA catalog, we use the catalog-reported remnant masses and spins whenever available. Because the MAYA catalog does not provide recoil velocities by default, we use the \textcolor{linkcolor}{\texttt{gw\_remnant}}\footnote{\href{https://github.com/tousifislam/gw\_remnant}{https://github.com/tousifislam/gw\_remnant}} package~\cite{Islam:2023mob} to compute missing remnant quantities from the radiated waveform fluxes. We follow the same procedure for the BAM simulations and BHPT data.
When available, \textcolor{linkcolor}{\texttt{gw\_remnant}} uses the initial energy and angular momentum reported in the simulation metadata to compute the remnant properties. In their absence, the package employs state-of-the-art PN expressions, including terms up to 3PN order for spinning eccentric binaries~\cite{Gamboa:2024hli,Khalil:2023kep} and 5PN order for nonspinning binaries~\cite{Blanchet:2023bwj, LeTiec:2011ab}.

For the final-mass and final-spin models, we use $1922$ simulations after quality-control cuts; the recoil model uses $1902$ simulations with reliable kick measurements, and the peak-luminosity model uses $1828$ simulations. We also exploit the mass symmetry at $q=1$ by augmenting the dataset with equal-mass simulations obtained by swapping $\chi_{1z}$ and $\chi_{2z}$.

%==========================================================================
%==========================================================================
\subsection{\gwModelS{} model construction}
\label{sec:nonprec_nonecc_models}
%==========================================================================
%==========================================================================
To facilitate the modeling, we define a set of derived mass and spin parameters motivated by previous modeling efforts~\cite{Healy:2014yta,Hofmann:2016yih,Jimenez-Forteza:2016oae,Varma:2018aht,Lousto:2012gt,Lousto:2012su,Bachhar:2023pir,Price:2023ldu}. We first introduce the symmetric mass ratio $\eta = \frac{q}{(1+q)^2}$
and the mass-difference parameter $\delta_m = \frac{q-1}{q+1}$.
While $\eta$ compresses the wide range of mass ratios considered in this work, $\delta_m$ captures the degree of mass asymmetry in the binary.
We further define the effective spin combinations
\begin{equation}
\hat\chi = \frac{q^2\chi_{1z}+\chi_{2z}}{q^2+1},
\qquad
\chi_a = \frac{\chi_{1z}-\chi_{2z}}{2},
\qquad
\tilde\Delta = \frac{\chi_{1z}-q\,\chi_{2z}}{1+q},
\end{equation}
which naturally interpolate between the equal-mass and point-particle limits. For example, $\hat \chi \rightarrow \chi_{1z}$ as $\eta \rightarrow 0$, while at $q=1$ it reduces to
$\hat \chi = \frac{\chi_{1z}+\chi_{2z}}{2}$.
Here, $\chihat$ denotes an effective spin parameter and $\chia$ characterizes the spin asymmetry. 

Our models are built using a common decomposition framework,
\begin{equation}
Q = Q_{\rm PP} + Q_{\rm EM} + Q_{\rm departure} + Q_{\rm asymmetry},
\end{equation}
where $Q$ denotes the remnant quantity being modeled. In this work, $Q$ corresponds to the remnant mass $M_f$, remnant spin $\chi_{fz}$, peak luminosity $L_{\rm peak}$, and recoil velocity $v_{\rm kick}$. Here, $Q_{\rm PP}$ encodes the point-particle (extreme-mass-ratio) limit and $Q_{\rm EM}$ describes the equal-mass contribution. The term $Q_{\rm departure}$ provides an exchange-symmetric correction to the interpolation between these limiting regimes, whereas $Q_{\rm asymmetry}$ captures mixed mass--spin effects through combinations such as $\dm\chia$ and therefore vanishes when either the component masses or their aligned spins are equal. This separation is a symmetry-constrained organization of the fitting basis rather than a unique physical decomposition. It provides a unified framework for constructing remnant models that reproduce the appropriate analytical limits while remaining accurate across the intermediate parameter space through calibration to NR and BHPT data. We collectively refer to this suite of remnant fits as the \gwModelS{} model.

For the nonprecessing models, the point-particle, equal-mass, departure, and asymmetry sectors provide the fixed physical scaffold for the search. The remaining structural choices include the functions assigned to each sector, the powers retained in the effective-spin and mass-ratio expansions, and the cross terms used to describe spin asymmetry away from the limiting configurations. To inform these choices, in Appendix~\ref{app:feature_importance} we perform a feature-importance analysis to quantify the relative importance of different mass- and spin-related quantities in determining the remnant properties. Guided by these results, we explore candidate structures while requiring the point-particle and equal-mass limits and body-exchange symmetry to remain exact. The final expressions use low-order integer powers, with their organization into separately constrained sectors forming part of the selected ansatz. Candidate coefficients were determined using the numerical procedures described below, while validation residuals and physical consistency checks guided the selection of the final structures. The agentic propose--evaluate--refine workflow was used to assist this exploration.

%==========================================================================
%==========================================================================
\subsubsection{\textcolor{linkcolor}{\texttt{gwModelRemS\_mf}} : model for the remnant mass}
\label{sec:nonprec_nonecc_mf}
%==========================================================================
%==========================================================================
The point-particle contribution is anchored to the specific energy of a test particle at the innermost stable circular orbit (ISCO) of a Kerr black hole~\cite{1972ApJ...178..347B},
\begin{equation}
E_{\rm ISCO}(\chi) = \sqrt{1-\frac{2}{3\,r_{\rm ISCO}(\chi)}},
\end{equation}
where $\chi$ denotes the signed dimensionless Kerr spin parameter of the central black hole and $r_{\rm ISCO}$ is the corresponding ISCO radius, expressed in units of the black hole mass. The ISCO radius is given by
\begin{equation}
r_{\rm ISCO}(\chi) = 3 + Z_2 - d\sqrt{(3-Z_1)(3+Z_1+2Z_2)},
\end{equation}
where $d=\mathrm{sign}(\chi)$ and
\begin{equation}
Z_1 = 1+(1-\chi^2)^{1/3}\Big[(1+\chi)^{1/3}+(1-\chi)^{1/3}\Big],
\qquad
Z_2 = \sqrt{3\chi^2+Z_1^2}.
\end{equation}
Throughout this work, the Kerr quantities are evaluated at the effective spin parameter $\chi=\chihat$.

The remnant mass is expressed in terms of the dimensionless radiated energy,
\begin{equation}
\frac{M_f^{\rm noprec}}{M}=1-\frac{E_{\rm rad}}{M},
\end{equation}
and we further define a rescaled radiated energy,
\begin{equation}
\frac{E_{\rm rad}}{M}=\eta\,\hat E_{\rm rad},
\end{equation}
which factors out the leading-order mass-ratio dependence. Following our general framework, we write
\begin{equation}
\hat E_{\rm rad}
=
Q_{\rm PP}^{E}
+
Q_{\rm EM}^{E}
+
Q_{\rm departure}^{E}
+
Q_{\rm asymmetry}^{E},
\end{equation}
with
\begin{align}
Q_{\rm PP}^{E}
&=(1-4\eta)\bigl[1-E_{\rm ISCO}(\chihat)\bigr],\\
Q_{\rm EM}^{E}
&=4\eta\,\hat E_{\rm EM}(\chihat,\chia),\\
Q_{\rm departure}^{E}
&=\eta(1-4\eta)\,R_E(\eta,\chihat),\\
Q_{\rm asymmetry}^{E}
&=4\eta\,\dm\,\chia\,(u_0+u_1\chihat+u_2\eta).
\end{align}
Combining these terms yields
\begin{align}
\hat E_{\rm rad}(\eta,\chihat,\dm,\chia)
&=(1-4\eta)\bigl[1-E_{\rm ISCO}(\chihat)\bigr]
+4\eta\,\hat E_{\rm EM}(\chihat,\chia)\nonumber\\
&\quad+\eta(1-4\eta)\,R_E(\eta,\chihat)\nonumber\\
&\quad+4\eta\,\dm\,\chia\,(u_0+u_1\chihat+u_2\eta).
\end{align}
The weighting functions $(1-4\eta)=\dm^2$ and $4\eta=1-\dm^2$ form a partition of unity, ensuring a smooth interpolation between the point-particle and equal-mass limits. 
The equal-mass and departure functions are modeled as
\begin{align}
\hat E_{\rm EM}(\chihat,\chia)
&=m_0+m_1\chihat+m_2\chihat^2+m_3\chihat^3+m_4\chihat^4+m_a\chia^2,\\
R_E(\eta,\chihat)
&=r_0+r_1\chihat+r_2\chihat^2+r_3\chihat^3+(1-4\eta)(g_0+g_1\chihat).
\end{align}

By construction, the model satisfies several physically motivated limits. In the point-particle limit, $\eta\rightarrow0$ and $\chihat\rightarrow\chi_{1z}$, yielding
\begin{equation}
\hat E_{\rm rad}\rightarrow1-E_{\rm ISCO}(\chi_{1z}).
\end{equation}
In the equal-mass limit, $q=1$ implies $\dm=0$ and $1-4\eta=0$, giving
\begin{equation}
\hat E_{\rm rad}\rightarrow\hat E_{\rm EM},
\qquad
\frac{M_f}{M}\rightarrow1-\frac{1}{4}\hat E_{\rm EM}.
\end{equation}
Body-exchange symmetry is enforced by requiring every term odd in $\chia$ to be accompanied by $\dm$, while all remaining spin-difference contributions enter through $\chia^2$. The resulting model contains 15 free parameters.

\begin{figure*}
    \centering
    \includegraphics[width=\textwidth]{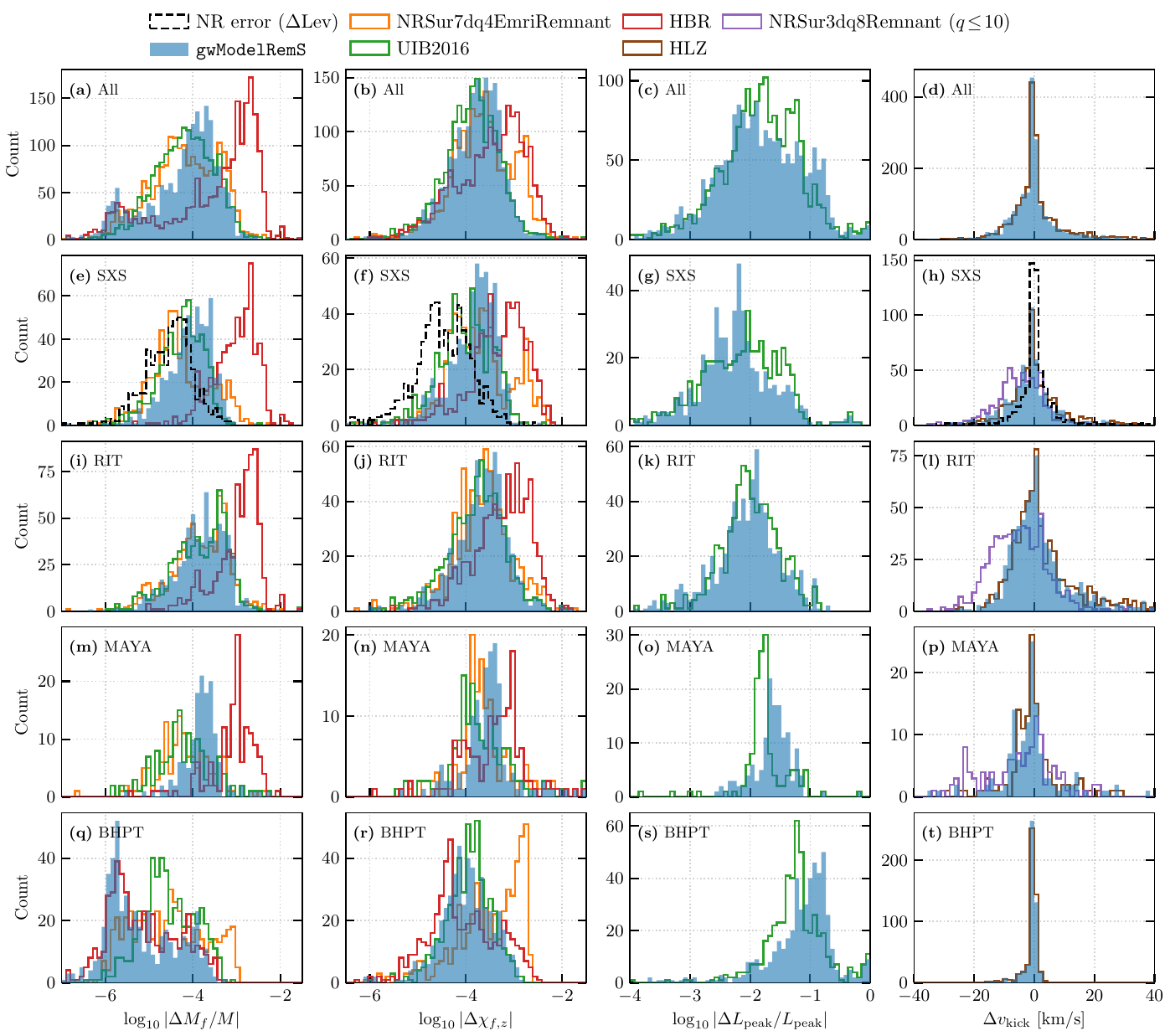}
    \caption{Held-out validation tests the accuracy of the nonprecessing \gwModelS{} across the NR and BHPT regimes. Blue histograms show its fractional or absolute errors; where available, the same data are evaluated with \HBR{}~\cite{Hofmann:2016yih,Barausse:2012qz,Barausse:2009uz}, \UIBa{}~\cite{Jimenez-Forteza:2016oae}, \textcolor{linkcolor}{\texttt{NRSur3dq8Remnant}}~\cite{Varma:2018aht}, and \NRSurEmri{}~\cite{Boschini:2023ryi}. SXS resolution differences provide a numerical-error benchmark. Further details are in Sec.~\ref{sec:nonprec_nonecc_accuracy}, and all models are available through \textcolor{linkcolor}{\texttt{gwModels}}. The comparison shows that \gwModelS{} attains state-of-the-art accuracy for remnant mass, spin, and recoil while remaining competitive for peak luminosity.}
    \label{fig:nonprec_nonecc_error_hist}
\end{figure*}

To determine the model coefficients, we perform nonlinear least-squares optimization using the \textcolor{linkcolor}{\texttt{scipy.curve\_fit}} routine~\cite{Virtanen:2020scipy}. To improve accuracy in the well-resolved NR regime while preserving the correct point-particle behavior, we adopt catalog-dependent weights during the fitting procedure. Simulations from the SXS catalog and BHPT simulations with $q\ge50$ are assigned unit weight, while all remaining simulations are assigned a weight of $0.9$. This weighting scheme balances the influence of the NR and BHPT datasets and ensures accurate performance across the full range $1\le q\lesssim1000$. 

Following Ref.~\cite{Islam:2025drw}, model performance is assessed using 5-fold cross-validation. 
The dataset is divided into five subsets; in each iteration, four subsets are used for training and the remaining subset is reserved for validation. The coefficients reported below correspond to the average values obtained across the five folds, while the quoted uncertainties denote the corresponding standard errors. The best-fit coefficients are $m_0 = +0.194140 \pm 0.000252$, $m_1 = +0.103396 \pm 0.000673$, $m_2 = +0.051366 \pm 0.001761$, $m_3 = +0.059260 \pm 0.001069$, $m_4 = +0.045106 \pm 0.002030$, $m_a = +0.003036 \pm 0.000746$, $r_0 = -0.241133 \pm 0.012030$, $r_1 = -0.178814 \pm 0.034137$, $r_2 = +0.584205 \pm 0.023048$, $r_3 = +0.685399 \pm 0.049230$, $g_0 = +0.270520 \pm 0.031798$, $g_1 = +0.469230 \pm 0.075765$, $u_0 = -0.030179 \pm 0.011551$, $u_1 = -0.002073 \pm 0.004002$, and $u_2 = +0.083935 \pm 0.054835$. The resulting model contains 15 free parameters and achieves an overall root-mean-square (RMS) error of $1.08\times10^{-3}$ in $M_f/M$.

%==========================================================================
%==========================================================================
\subsubsection{\textcolor{linkcolor}{\texttt{gwModelRemS\_chif}}: model for the remnant spin}
\label{sec:nonprec_nonecc_af}
%==========================================================================
%==========================================================================
We first define the inherited spin contribution
\begin{equation}
\tilde S = \frac{m_1^2\chi_{1z}+m_2^2\chi_{2z}}{M^2}
= \frac{q^2\chi_{1z}+\chi_{2z}}{(1+q)^2}
= (1-2\eta)\chihat.
\end{equation}
The point-particle contribution is anchored to the Kerr ISCO angular momentum~\cite{1972ApJ...178..347B},
\begin{equation}
L_{\rm ISCO}(\chi)=\frac{r^2 - 2d a\sqrt{r}+a^2}{r^{3/4}\sqrt{r^{3/2}-3r^{1/2}+2d a}},
\end{equation}
where $a=|\chi|$ is the dimensionless Kerr spin magnitude, $d=\mathrm{sign}(\chi)$, and $r=r_{\rm ISCO}(\chi)$ is the ISCO radius. Throughout this work, the Kerr quantities are evaluated at $\chi=\chihat$. Following Ref.~\cite{Hofmann:2016yih}, we define
\begin{equation}
\ell_{\rm Kerr}(\chi)=L_{\rm ISCO}(\chi)-2\chi\bigl[E_{\rm ISCO}(\chi)-1\bigr].
\end{equation}

\begin{figure}
    \centering
    \includegraphics[width=\columnwidth]{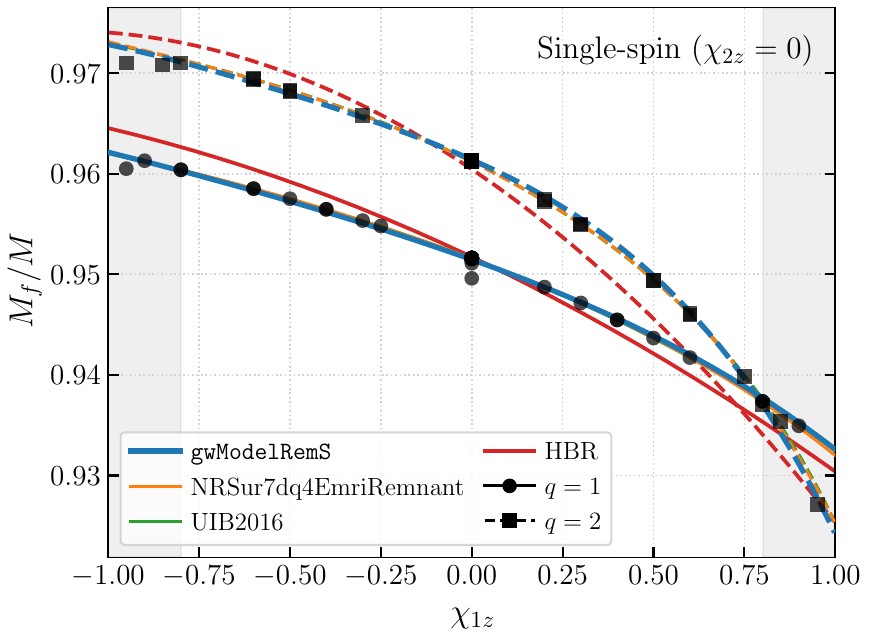}
    \includegraphics[width=\columnwidth]{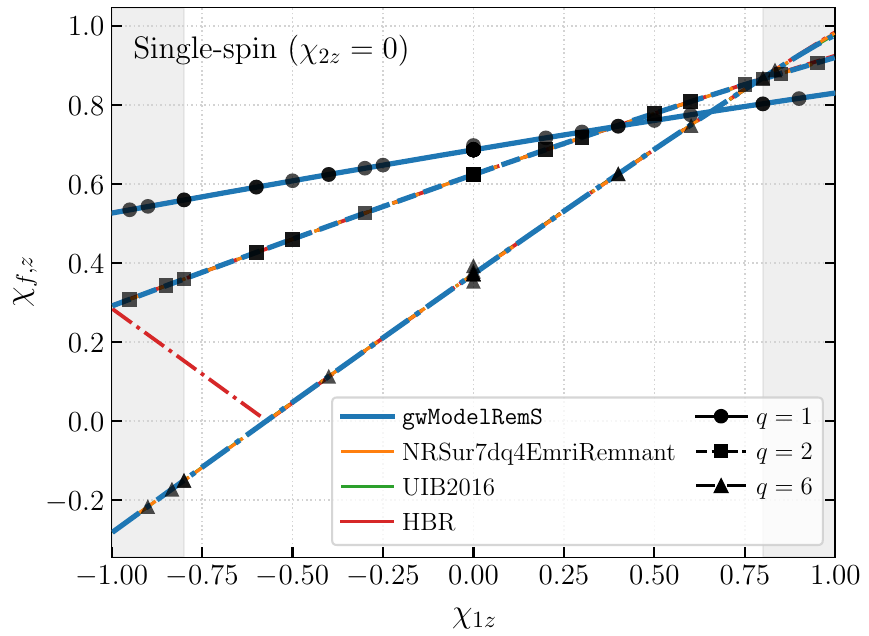}
    \caption{Spin-dependent slices test the models in the high-spin regime, where extrapolation is most demanding. For nonprecessing binaries with $\chi_{2z}=0$ and $q=1,2$, the upper and lower panels show $M_f/M$ and $\chi_{fz}$ as functions of $\chi_{1z}$; shading marks the high-spin region. Predictions from \gwModelS{}, \HBR{}~\cite{Hofmann:2016yih,Barausse:2012qz,Barausse:2009uz}, \UIBa{}~\cite{Jimenez-Forteza:2016oae}, and \textcolor{linkcolor}{\texttt{NRSur3dq8EmriRemnant}}~\cite{Boschini:2023ryi} are compared with NR simulations; see Sec.~\ref{sec:nonprec_nonecc_behavior}. For the lower panel, we also show $q=6$ predictions. The \gwModelS{} predictions improve on \HBR{} for the remnant mass and remain comparable to \UIBa{} and \textcolor{linkcolor}{\texttt{NRSur3dq8EmriRemnant}}.}
    \label{fig:single_spin_mf_chif_convergence}
\end{figure}

The remnant-spin model is written as
\begin{equation}
\chi_{fz}^{\rm noprec}=\tilde S+\eta\,\ell_{\rm orb},
\end{equation}
where $\ell_{\rm orb}$ represents the dimensionless orbital angular momentum contribution retained by the remnant black hole. Following our general decomposition framework and logic presented in Section~\ref{sec:nonprec_nonecc_mf}, we write
\begin{equation}
\ell_{\rm orb}=Q_{\rm PP}^{J}+Q_{\rm EM}^{J}+Q_{\rm departure}^{J}+Q_{\rm asymmetry}^{J},
\end{equation}
with
\begin{align}
Q_{\rm PP}^{J}&=(1-4\eta)\ell_{\rm Kerr}(\chihat),\\
Q_{\rm EM}^{J}&=4\eta\,\ell_{\rm EM}(\chihat,\chia),\\
Q_{\rm departure}^{J}&=\eta(1-4\eta)\mathcal{R}_{\chi}(\eta,\chihat),\\
Q_{\rm asymmetry}^{J}&=4\eta\,\dm\,\chia\,(\upsilon_0+\upsilon_1\chihat+\upsilon_2\eta).
\end{align}
Combining these contributions yields
\begin{align}
\ell_{\rm orb}(\eta,\chihat,\dm,\chia)
&=(1-4\eta)\ell_{\rm Kerr}(\chihat)+4\eta\,\ell_{\rm EM}(\chihat,\chia)\nonumber\\
&\quad+\eta(1-4\eta)\mathcal{R}_{\chi}(\eta,\chihat)\nonumber\\
&\quad+4\eta\,\dm\,\chia\,(\upsilon_0+\upsilon_1\chihat+\upsilon_2\eta).
\end{align}
The equal-mass and departure functions are modeled as
\begin{align}
\ell_{\rm EM}(\chihat,\chia)
&=
\mu_0+\mu_1\chihat+\mu_2\chihat^2+\mu_3\chihat^3+\mu_4\chihat^4+\mu_a\chia^2,\\
\mathcal{R}_{\chi}(\eta,\chihat)
&=
\rho_0+\rho_1\chihat+\rho_2\chihat^2+\rho_3\chihat^3
+(1-4\eta)(\gamma_0+\gamma_1\chihat).
\end{align}
In the point-particle limit, $\eta\rightarrow0$ and $\chihat,\tilde S\rightarrow\chi_{1z}$, yielding
\begin{equation}
\chi_{fz}^{\rm noprec}\rightarrow\chi_{1z}+\eta\,\ell_{\rm Kerr}(\chi_{1z}).
\end{equation}
In the equal-mass limit, $q=1$ implies $\dm=0$ and $1-4\eta=0$, giving
\begin{equation}
\ell_{\rm orb}\rightarrow\ell_{\rm EM}, \qquad \chi_{fz}\rightarrow \tilde S+\frac{1}{4}\ell_{\rm EM}.
\end{equation}

The model contains 15 free parameters and achieves an RMS error of $8.3\times10^{-4}$ in $\chi_{fz}$. To reduce the impact of a small number of outliers, the fit is performed using a robust nonlinear least-squares optimization with a \textcolor{linkcolor}{\texttt{soft\_l1}} loss function implemented in \textcolor{linkcolor}{\texttt{scipy.optimize.least\_squares}}~\cite{Virtanen:2020scipy}. Note that this choice differs slightly from that adopted for the remnant mass fit. The mass residuals are well behaved and contain no significant outliers, making the standard least-squares loss sufficient. In contrast, the spin residuals exhibit a small number of catalog-dependent outliers, for which the \textcolor{linkcolor}{\texttt{soft\_l1}} loss is more robust and prevents them from disproportionately influencing the fit. As before, model performance is assessed using 5-fold cross-validation, and the reported coefficients correspond to the mean values across the five folds, with uncertainties denoting the corresponding standard errors. The best-fit coefficients are $\mu_0 = +2.745811 \pm 0.000475$, $\mu_1 = -0.775098 \pm 0.001274$, $\mu_2 = -0.113350 \pm 0.003331$, $\mu_3 = -0.037128 \pm 0.002017$, $\mu_4 = -0.014675 \pm 0.003829$, $\mu_a = -0.014635 \pm 0.001400$, $\rho_0 = -0.664337 \pm 0.022680$, $\rho_1 = -0.562673 \pm 0.065611$, $\rho_2 = +0.021602 \pm 0.044262$, $\rho_3 = +0.257495 \pm 0.095053$, $\gamma_0 = -1.003577 \pm 0.059931$, $\gamma_1 = -0.340650 \pm 0.144431$, $\upsilon_0 = +0.208862 \pm 0.022024$, $\upsilon_1 = +0.075628 \pm 0.007632$, and $\upsilon_2 = +1.288995 \pm 0.104486$.

%==========================================================================
%==========================================================================
\subsubsection{\textcolor{linkcolor}{\texttt{gwModelRemS\_Lpeak}}: model for the peak luminosity}
\label{sec:nonprec_nonecc_lpeak}
%==========================================================================
%==========================================================================
We next construct a model for the peak GW luminosity $L_{\rm peak}$. To account for the nearly five orders of magnitude dynamic range in $L_{\rm peak}$ across the interval $1\le q\lesssim1000$, the fit is performed in logarithmic space. We assign weights $W_{\rm NR}=5$ to all NR simulations, $W_{\rm BHPT}=1$ for BHPT simulations with $q\ge50$, and $W=0.9$ for all remaining data.
Unlike the remnant mass and spin, no robust analytic point-particle expression is available for the peak luminosity. We therefore adopt a simpler phenomenological ansatz motivated by the leading-order BHPT scaling $L_{\rm peak}\propto\eta^2$ in the extreme-mass-ratio limit,
\begin{equation}
L_{\rm peak}^{\rm noprec}=\eta^2\,\mathcal{L}_{\rm peak},
\end{equation}
where the dimensionless luminosity function is decomposed as
\begin{equation}
\mathcal{L}_{\rm peak}=Q_{\rm PP}^{L}+Q_{\rm EM}^{L}+Q_{\rm departure}^{L}+Q_{\rm asymmetry}^{L}.
\end{equation}
Since the overall factor of $\eta^2$ already enforces the correct point-particle scaling, we set
\begin{align}
Q_{\rm PP}^{L}&=0,\\
Q_{\rm EM}^{L}&=\mathcal{L}_{\rm EM}(\chihat,\chia),\\
Q_{\rm departure}^{L}&=(1-4\eta)\,\mathcal{P}_{L}(\eta,\chihat),\\
Q_{\rm asymmetry}^{L}&=\dm\,\chia\,(\nu_0+\nu_1\chihat+\nu_2\eta).
\end{align}
Combining these contributions yields
\begin{align}
\mathcal{L}_{\rm peak}(\eta,\chihat,\dm,\chia)&=\mathcal{L}_{\rm EM}(\chihat,\chia)
+(1-4\eta)\,\mathcal{P}_{L}(\eta,\chihat)
\nonumber\\
&\quad
+\dm\,\chia\,(\nu_0+\nu_1\chihat+\nu_2\eta).
\end{align}
The equal-mass and departure functions are modeled as
\begin{align}
\mathcal{L}_{\rm EM}(\chihat,\chia)
&=
\pi_0+\pi_1\chihat+\pi_2\chihat^2+\pi_3\chihat^3+\pi_4\chihat^4+\pi_a\chia^2,\\
\mathcal{P}_{L}(\eta,\chihat)
&=
\sigma_0+\sigma_1\chihat+\sigma_2\chihat^2+\sigma_3\chihat^3
+(1-4\eta)(\tau_0+\tau_1\chihat).
\end{align}
In the equal-mass limit, $q=1$ implies $\dm=0$ and $1-4\eta=0$, yielding
\begin{equation}
\mathcal{L}_{\rm peak}\rightarrow\mathcal{L}_{\rm EM},
\qquad
L_{\rm peak}^{\rm noprec}\rightarrow\frac{1}{16}\mathcal{L}_{\rm EM}.
\end{equation}

The fit minimizes
\begin{equation}\sum_i W_i^2\left[\log L_{{\rm peak},i}^{\rm model}-\log L_{{\rm peak},i}^{\rm data}\right]^2,
\end{equation}
and is performed non-hierarchically with all 15 parameters optimized simultaneously. As before, model performance is assessed using 5-fold cross-validation, and the reported coefficients correspond to the mean values across the five folds, with uncertainties denoting the corresponding standard errors. The best-fit coefficients are $\pi_0 = +0.016483 \pm 0.000049$, $\pi_1 = +0.007128 \pm 0.000139$, $\pi_2 = +0.003234 \pm 0.000287$, $\pi_3 = +0.003559 \pm 0.000270$, $\pi_4 = +0.002653 \pm 0.000377$, $\pi_a = +0.000277 \pm 0.000141$, $\sigma_0 = -0.008103 \pm 0.000269$, $\sigma_1 = +0.005559 \pm 0.000680$, $\sigma_2 = +0.007990 \pm 0.000468$, $\sigma_3 = +0.002963 \pm 0.000797$, $\tau_0 = +0.005356 \pm 0.000324$, $\tau_1 = -0.003466 \pm 0.000821$, $\nu_0 = -0.000755 \pm 0.000589$, $\nu_1 = +0.000071 \pm 0.000461$, and $\nu_2 = +0.006371 \pm 0.003373$. The model achieves an overall fractional RMS error of $0.1083$, while the 5-fold cross-validation test error is $0.1078$. Restricting to NR simulations only, the fractional RMS errors are $0.027$ for RIT, $0.030$ for SXS, and $0.035$ for MAYA. For BHPT simulations with $q\ge50$, the fractional RMS error is $0.197$.

%==========================================================================
%==========================================================================
\subsubsection{\textcolor{linkcolor}{\texttt{gwModelRemS\_kick}}: model for the recoil velocity}
\label{sec:nonprec_nonecc_kick}
%==========================================================================
%==========================================================================
Finally, we model the recoil velocity $v_{\rm kick}$ by refitting the aligned-spin recoil model presented in Ref.~\cite{Islam:2025drw} (which draws inspiration from Refs.~\cite{Lousto:2008dn,Lousto:2010xk,Lousto:2012gt,Lousto:2012su,Healy:2014yta}) to the expanded NR and BHPT dataset used in this work, extending the calibration range to mass ratios as large as $q=1000$. As in the remnant-mass, remnant-spin, and peak-luminosity fits, we assign unit weight to SXS simulations and BHPT simulations with $q\ge50$, and weight $0.9$ to all remaining simulations.

For the recoil velocity, the decomposition is most naturally expressed in terms of vector contributions rather than a single scalar quantity. We therefore write
\begin{equation}
v_{\rm kick}^{\rm noprec}=\left|\vec{V}_{\rm PP}+\vec{V}_{\rm EM}+\vec{V}_{\rm mass}+\vec{V}_{\rm spin}\right|,
\end{equation}
where $\vec{V}_{\rm PP}$, $\vec{V}_{\rm EM}$, $\vec{V}_{\rm mass}$, and $\vec{V}_{\rm spin}$ denote the point-particle, equal-mass equal-spin, mass-asymmetry, and spin-asymmetry contributions, respectively. We use the explicit labels ``mass'' and ``spin'' here because the two recoil components have a more direct physical interpretation than the corresponding scalar-model sectors. The point-particle contribution vanishes, $\vec{V}_{\rm PP}=0$, because the recoil scales as $v_{\rm kick}^{\rm noprec}\propto\eta^2$ and therefore approaches zero in the exact point-particle limit. Similarly, the equal-mass equal-spin contribution vanishes, $\vec{V}_{\rm EM}=0$, because such binaries are symmetric and therefore do not experience any net recoil.
Our model therefore becomes
\begin{equation}
v_{\rm kick}^{\rm noprec}=\sqrt{V_{\rm mass}^2+V_{\rm spin}^2+2V_{\rm mass}V_{\rm spin}\cos\xi_{\rm kick}},
\end{equation}
where $\xi_{\rm kick}$ is the angle between the two recoil components $\vec{V}_{\rm mass}$ and $\vec{V}_{\rm spin}$.
The recoil amplitudes are modeled as
\begin{equation}
V_{\rm mass}=A\eta^2\dm(1+B\eta+C\eta^2),
\qquad
V_{\rm spin}=H\eta^2\mathcal{R}_v,
\end{equation}
with
\begin{equation}
\xi_{\rm kick}=\frac{\pi}{180}\left(a_{\rm deg}+b_{\rm deg}\tilde S+c_{\rm deg}\dm\tilde\Delta
\right).
\end{equation}
The spin-induced recoil function is given by
\begin{align}
\mathcal{R}_v
&=
\tilde\Delta
+H_{2a}\tilde S\dm
+H_{2b}\tilde\Delta\tilde S
+H_{3a}\tilde\Delta^2\dm
+H_{3b}\tilde S^2\dm
\nonumber\\
&\quad
+H_{3c}\tilde\Delta\tilde S^2
+H_{3d}\tilde\Delta^3
+H_{3e}\tilde\Delta\dm^2
+H_{4a}\tilde S\tilde\Delta^2\dm
+H_{4b}\tilde S^3\dm
\nonumber\\
&\quad
+H_{4c}\tilde S\dm^3
+H_{4d}\tilde\Delta\tilde S\dm^2
+H_{4e}\tilde\Delta\tilde S^3
+H_{4f}\tilde S\tilde\Delta^3.
\end{align}
Here $A$ and $H$ are measured in $\mathrm{km\,s^{-1}}$, $a_{\rm deg}$, $b_{\rm deg}$, and $c_{\rm deg}$ are measured in degrees, and all remaining coefficients are dimensionless. By construction, the overall $\eta^2$ scaling enforces the correct point-particle behavior, $v_{\rm kick}\propto\eta^2\rightarrow0$ as $\eta\rightarrow0$. In the equal-mass limit, $\dm=0$ and therefore $V_{\rm mass}=0$, implying that the recoil is driven entirely by the spin-difference contribution through $\tilde\Delta$. The model also satisfies the expected symmetry requirement that the recoil vanish for equal-mass, equal-spin binaries.

As before, model performance is assessed using 5-fold cross-validation. The best-fit coefficients are $A = 12929 \pm 464~\mathrm{km\,s^{-1}}$, $B = -2.2280 \pm 0.338$, $C = +4.3961 \pm 1.038$, $H = 7275.1 \pm 54.1~\mathrm{km\,s^{-1}}$, $H_{2a} = +5.8284 \pm 0.072$, $H_{2b} = -0.7398 \pm 0.030$, $H_{3a} = -0.7716 \pm 0.066$, $H_{3b} = -1.6378 \pm 0.166$, $H_{3c} = -1.1596 \pm 0.137$, $H_{3d} = +0.0116 \pm 0.013$, $H_{3e} = +6.7073 \pm 0.139$, $H_{4a} = -0.7910 \pm 0.185$, $H_{4b} = -1.7800 \pm 0.249$, $H_{4c} = +3.5296 \pm 0.186$, $H_{4d} = -2.2385 \pm 0.261$, $H_{4e} = +0.5582 \pm 0.280$, $H_{4f} = +0.1273 \pm 0.081$, $a_{\rm deg} = +147.54 \pm 0.62$, $b_{\rm deg} = +114.08 \pm 2.99$, and $c_{\rm deg} = +144.61 \pm 4.84$.
The refit achieves an RMS error of $10.97~\mathrm{km\,s^{-1}}$, compared to $11.43~\mathrm{km\,s^{-1}}$ obtained using the previously published coefficients in Ref.~\cite{Islam:2025drw}.

For non-precessing BBH mergers, the recoil kick is always directed perpendicular to the orbital angular momentum (and hence to the remnant spin, which is aligned with the orbital angular momentum at merger).

\begin{figure}
    \centering
    \includegraphics[width=\columnwidth]{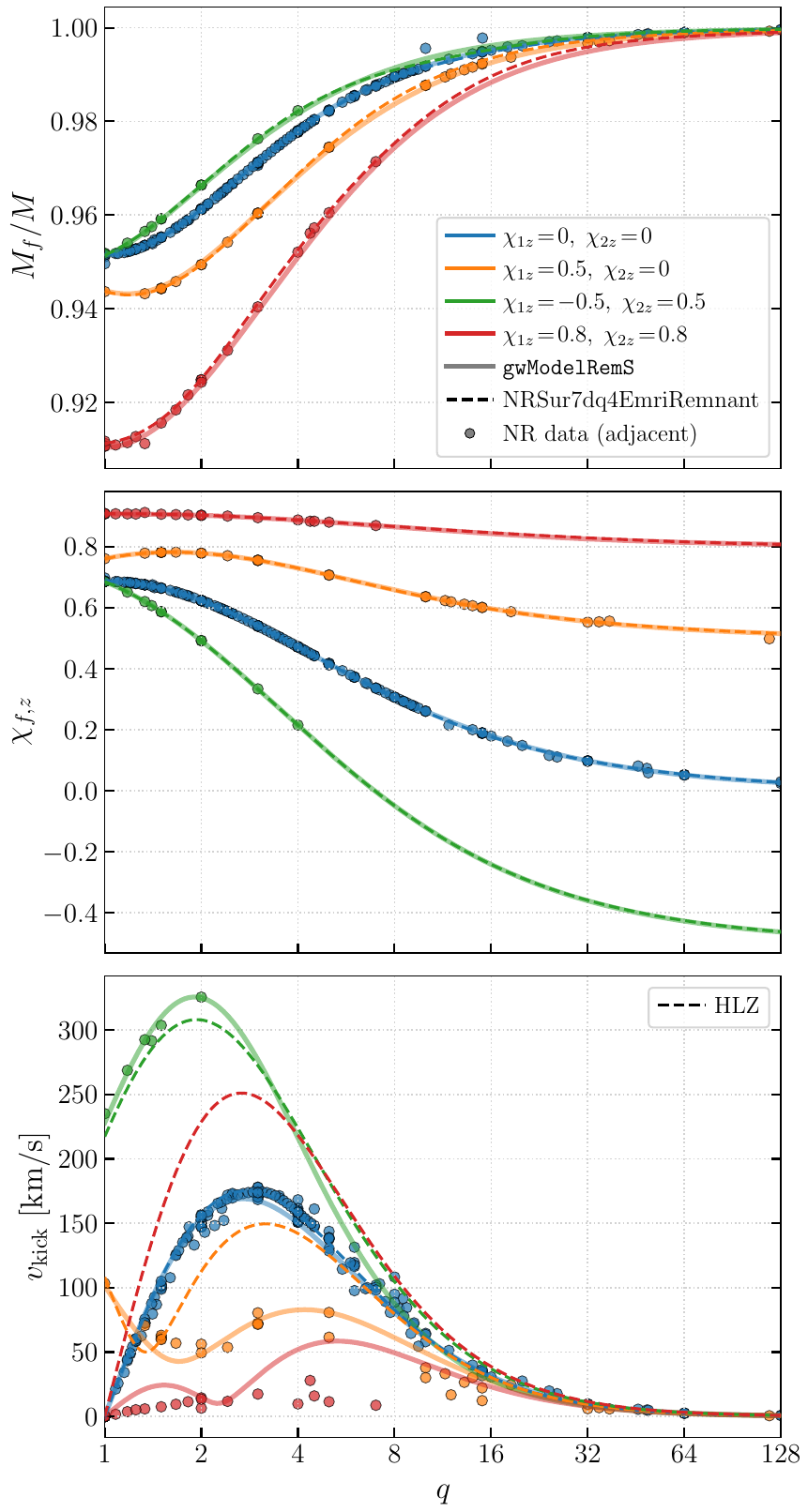}
    \caption{Reliable remnant models must interpolate smoothly from comparable to large mass ratios. The panels show the remnant mass (top), remnant spin (middle), and recoil velocity (bottom) across $1\le q\le128$ for $(\chi_{1z},\chi_{2z})=(0,0)$, $(0.5,0)$, $(-0.5,0.5)$, and $(0.8,0.8)$. Solid \gwModelS{} curves are compared with nearby NR simulations and with \textcolor{linkcolor}{\texttt{NRSur7dq4EmriRemnant}}~\cite{Boschini:2023ryi} for mass and spin or \HLZ{}~\cite{Lousto:2008dn,Lousto:2010xk,Lousto:2012gt,Lousto:2012su,Gonzalez:2007hi} for recoil; see Sec.~\ref{sec:nonprec_nonecc_behavior}. All three \gwModelS{} predictions remain smooth through the transition between the NR and BHPT regimes.}
    \label{fig:smoothness_mf_chif_vs_q}
\end{figure}

%==========================================================================
%==========================================================================
\subsection{\gwModelS{} model accuracy}
\label{sec:nonprec_nonecc_accuracy}
%==========================================================================
%==========================================================================
Figure~\ref{fig:nonprec_nonecc_error_hist} summarizes the validation errors of \gwModelS{} against the full NR and BHPT dataset and compares them with several widely used remnant models. For the remnant mass, \gwModelS{} achieves a median absolute error of $9.1\times10^{-5}$ and a 90th-percentile error of $4.4\times10^{-4}$ over the full validation set, comparable to \UIBa{} and \textcolor{linkcolor}{\texttt{NRSur7dq4EmriRemnant}}, while significantly outperforming \HBR{}. In the BHPT regime, the median and 90th-percentile errors decrease to $4.2\times10^{-6}$ and $1.1\times10^{-4}$, respectively. For the remnant spin, \gwModelS{} achieves median and 90th-percentile errors of $1.8\times10^{-4}$ and $6.5\times10^{-4}$ over the full dataset, with corresponding values of $1.1\times10^{-4}$ and $5.2\times10^{-4}$ in the BHPT regime. Overall, the remnant-mass and remnant-spin models remain competitive with the best existing approaches while maintaining accurate behavior across both the comparable-mass and extreme-mass-ratio regimes.

For the peak luminosity, \gwModelS{} achieves a median fractional error of $1.38\%$ and a 90th-percentile error of $12.79\%$ over the full validation set. On the SXS subset, these values improve to $0.59\%$ and $3.55\%$, respectively, while the BHPT regime remains more challenging, with median and 90th-percentile errors of $9.28\%$ and $21.21\%$. The overall performance is comparable to that of \UIBa{}. For the recoil velocity, \gwModelS{} achieves a median absolute error of $2.78~\mathrm{km\,s^{-1}}$ and a 90th-percentile error of $14.73~\mathrm{km\,s^{-1}}$, essentially matching the performance of the \HLZ{} recoil model and substantially outperforming \textcolor{linkcolor}{\texttt{NRSur3dq8Remnant}}. In the BHPT regime, the median recoil error decreases to only $0.11~\mathrm{km\,s^{-1}}$. These results are consistent with those reported in Ref.~\cite{Islam:2025drw} and demonstrate that \gwModelS{} achieves state-of-the-art accuracy for the remnant mass, spin, and recoil velocity while maintaining competitive performance for peak luminosity.

An important feature of \gwModelS{} is that this accuracy is achieved with a relatively small number of free parameters. The remnant-mass, remnant-spin, and peak-luminosity models each contain 15 parameters, compared to 22, 23, and 22 parameters in the corresponding \UIBa{} models. The recoil model contains 20 parameters, identical to the \HLZ{} recoil model from which it is derived. In total, \gwModelS{} contains 65 parameters across all remnant quantities, comparable to the combined 67 parameters of the \UIBa{} mass, spin, and luminosity models. More importantly, the fits are explicitly constructed through a common decomposition into point-particle, equal-mass, departure, and asymmetry contributions, improving interpretability and extrapolation behavior while providing a unified framework for future extensions.

%==========================================================================
%==========================================================================
\subsection{\gwModelS{} model behavior}
\label{sec:nonprec_nonecc_behavior}
%==========================================================================
%==========================================================================
We further examine the \gwModelS{} predictions for the remnant mass and remnant spin in the comparable-mass regime over the full range of primary spin values, $\chi_{1z}\in[0,1]$, with $\chi_{2z}=0$. As representative examples, we consider mass ratios $q=1$ and $q=2$ (Fig.~\ref{fig:single_spin_mf_chif_convergence}). An interesting feature visible in both the remnant-mass and remnant-spin predictions is the presence of curve crossings. For the remnant mass, the $q=1$ and $q=2$ curves intersect at approximately $\chi_{1z}\simeq0.8$, below which equal-mass binaries produce smaller remnant masses than $q=2$ systems, with the ordering reversing at higher spins. A similar crossing occurs for the remnant spin at approximately $\chi_{1z}\simeq0.4$, where equal-mass binaries initially produce larger remnant spins before the trend reverses. 

For the remnant spin, the physical origin of this crossing (Fig.~\ref{fig:single_spin_mf_chif_convergence}) is relatively straightforward. At low spins, equal-mass binaries retain more orbital angular momentum and therefore tend to produce larger remnant spins than unequal-mass systems. However, as the primary spin increases, the $q=2$ binary acquires a larger effective spin contribution from the heavier black hole. Beyond $\chi_{\rm eff}\simeq0.4$, this additional spin contribution dominates over the orbital-angular-momentum advantage of the equal-mass system, causing the $q=2$ sequence to overtake the $q=1$ sequence. A similar competition underlies the crossing observed in the remnant mass. At low and moderate spins, equal-mass binaries radiate energy more efficiently because they possess the largest symmetric mass ratio, $\eta=1/4$. Consequently, they produce smaller remnant masses than their unequal-mass counterparts. As the spin increases, however, the plunge and merger dynamics become increasingly sensitive to spin effects through the ISCO energy and the associated spin-enhanced radiation. Since the $q=2$ binary possesses a larger effective aligned spin for the same primary spin, the spin-dependent contribution to the radiated energy grows more rapidly than in the equal-mass case. This compensates for the smaller value of $\eta$ and eventually reverses the ordering of the two curves. We find that \gwModelS{}, together with \UIBa{} and \NRSurEmri{}, accurately reproduces these nontrivial trends observed in the NR simulations, while \HBR{} exhibits noticeably larger deviations, particularly for the remnant mass.

Figure~\ref{fig:single_spin_mf_chif_convergence} (lower panel) also highlights a known limitation of the \HBR{} remnant-spin model in the large-mass-ratio regime ($q\gtrsim4$). For the representative configuration $(q,\chi_{2z})=(6,0)$, the \HBR{} prediction exhibits an unphysical change in behavior for sufficiently negative values of $\chi_{1z}$, leading to substantial deviations from the NR data. In contrast, both \gwModelS{} and \textcolor{linkcolor}{\texttt{NRSur7dq4EmriRemnant}} remain in excellent agreement with the available simulations across the entire spin range. This discrepancy likely originates from the limited calibration dataset and phenomenological structure of the \HBR{} model, whereas the incorporation of both NR and BHPT information allows \gwModelS{} to maintain physically consistent behavior across a much broader region of parameter space.

Figure~\ref{fig:smoothness_mf_chif_vs_q} illustrates the behavior of \gwModelS{} across the full mass-ratio range for several representative spin configurations. The model remains smooth from the equal-mass regime to $q=128$ and closely tracks the available NR data throughout the parameter space. The figure also demonstrates the expected convergence of the remnant mass, spin, and recoil velocity toward their point-particle limits at large mass ratios. For the recoil velocity, \gwModelS{} and \HLZ{} often produce similar predictions; however, for several high-spin configurations \HLZ{} exhibits noticeable deviations from the adjacent NR data, while \gwModelS{} remains in closer agreement with the numerical results. 
Further discussion of model behavior is provided in Appendix~\ref{app:behaviour}.

%==========================================================================
%==========================================================================
%==========================================================================
\section{Precessing quasi-circular models}
\label{sec:prec_nonecc}
%==========================================================================
%==========================================================================
%==========================================================================
We now extend the nonprecessing quasi-circular remnant models developed in Sec.~\ref{sec:nonprec_nonecc} to generic precessing BBH mergers.
To characterize precession effects, we define the in-plane spin magnitudes of the individual black holes,
\begin{equation}
\chi_{i\perp}=\sqrt{\chi_{ix}^2+\chi_{iy}^2}, \qquad i\in\{1,2\},
\end{equation}
evaluated at a binary separation of $r=8M$. We then construct the mass-weighted symmetric and antisymmetric combinations,
\begin{align}
\Sperp &=
\frac{\sqrt{(m_1^2\chi_{1\perp})^2+(m_2^2\chi_{2\perp})^2}}
{m_1^2+m_2^2},\\
\Delta_{\perp} &=
\frac{q\,\chi_{1\perp}-\chi_{2\perp}}
{1+q}.
\end{align}
The quantity $\Sperp$ characterizes the overall in-plane spin magnitude, while $\Delta_{\perp}$ captures the antisymmetric in-plane spin contribution.
The spin components $\chi_{1z}^{r=8M}$, $\chi_{2z}^{r=8M}$, $\chi_{1\perp}^{r=8M}$, and $\chi_{2\perp}^{r=8M}$ are obtained by evolving the reference spins from the initial configuration to $r=8M$ using precession-averaged PN inspiral evolution implemented in the \textcolor{linkcolor}{\texttt{precession}} package\footnote{\href{https://github.com/dgerosa/precession/}{https://github.com/dgerosa/precession/}}~\cite{Gerosa:2016sys}. This evolution code includes spin-precession effects through 2PN order and radiation-reaction corrections through 3.5PN (2PN) order for nonspinning (spinning) terms~\cite{Kidder:1995zr}.
The baseline nonprecessing models are evaluated using the aligned spin components at $r=8M$,
\begin{align}
\chi_{fz}^{\rm noprec}&=\textcolor{linkcolor}{\texttt{gwModelRemS\_chif}}
\!\left(q,\chi_{1z}^{r=8M},\chi_{2z}^{r=8M}\right),\\
m_f^{\rm noprec}&=\textcolor{linkcolor}{\texttt{gwModelRemS\_mf}}
\!\left(q,\chi_{1z}^{r=8M},\chi_{2z}^{r=8M}\right),\\
L_{\rm peak}^{\rm noprec}&=\textcolor{linkcolor}{\texttt{gwModelRemS\_Lpeak}}
\!\left(q,\chi_{1z}^{r=8M},\chi_{2z}^{r=8M}\right).
\end{align}
Precession effects are then incorporated through augmentation terms that depend on $\Sperp$ and $\Delta_{\perp}$. In this way, the precessing models preserve the physically motivated structure of the nonprecessing fits while extending their validity to generic spin configurations. 

\begin{figure*}
    \centering
    \includegraphics[width=\textwidth]{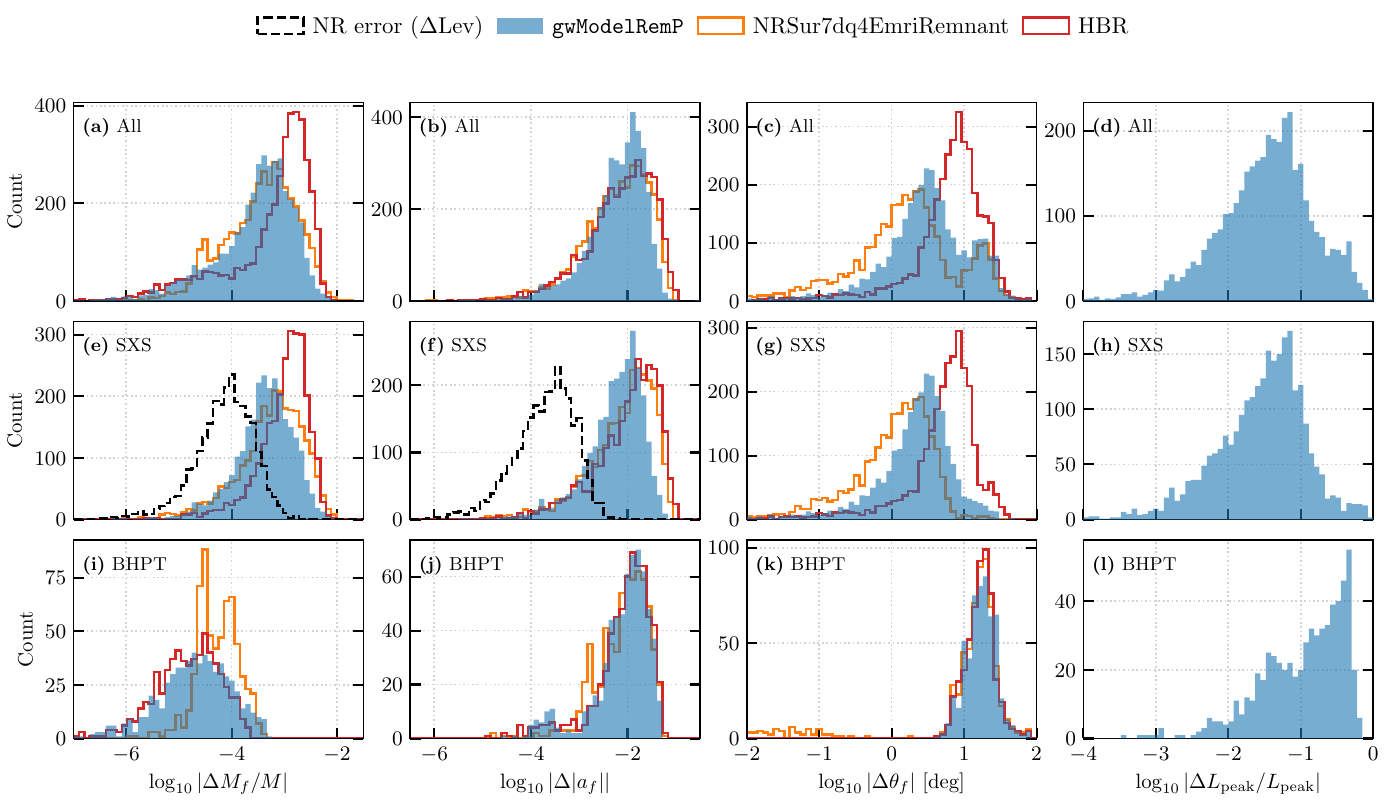}
    \caption{Similar to Fig.~\ref{fig:nonprec_nonecc_error_hist} but showing the held-out validation tests for the precessing models instead of the aligned-spin models. The panels show fractional or absolute errors in remnant mass, spin magnitude, spin tilt angle, and peak luminosity for \gwModelP{} evaluated on precessing NR and BHPT simulations, with SXS resolution differences and predictions from \HBR{} and \NRSurEmri{} included where available. Further details are in Sec.~\ref{sec:prec_nonecc_accuracy}. The \gwModelP{} model is best or joint-best for mass and spin magnitude and remains competitive for spin direction.}
    \label{fig:prec_noecc_error_hist_log10}
\end{figure*}

%==========================================================================
%==========================================================================
\subsection{Training data}
\label{sec:prec_nonecc_data}
%==========================================================================
%==========================================================================
The precessing models are calibrated using all nonprecessing simulations described in Sec.~\ref{sec:nonprec_data}, together with an additional $4125$ precessing NR and BHPT simulations. The precessing dataset comprises $2724$ SXS, $630$ BHPT, $476$ RIT, $186$ MAYA, and $109$ BAM/Einstein Toolkit simulations. Collectively, the calibration data span mass ratios from $q=1$ to $q=1000$, dimensionless spin magnitudes up to $|\chi_i|\simeq0.99$, and mass-weighted in-plane spins up to $\Sperp\simeq0.93$. The SXS catalog provides the broadest coverage of generic precessing configurations, while the BHPT simulations extend the calibration into the large-mass-ratio regime ($40\le q\le100$).
We show the parameter space covered by these simulations in Figure~\ref{fig:parameter_space_prec}. NR simulations cover in-plane spin magnitudes up to $\Sperp\sim0.93$, whereas the BHPT simulations populate the low-$\eta$ regime with more modest in-plane spins. As before, we duplicate the $q=1$ simulations by exploiting the mass symmetry.
During training, SXS and BHPT simulations are assigned unit weight, while RIT, MAYA, and BAM/Einstein Toolkit simulations are assigned weight $0.5$. The final models are selected using 5-fold cross-validation.

%==========================================================================
%==========================================================================
\subsection{\gwModelP{} model construction}
\label{sec:prec_nonecc_models}
%==========================================================================
%==========================================================================
Below, we present the best-fit models for the remnant mass, spin, and peak luminosity of precessing BBH mergers. Collectively, these fits define the \gwModelP{} model.

For the precessing models, the nonprecessing predictions are held fixed as physically constrained baselines. The structural search is therefore restricted to augmentation terms that vanish in the nonprecessing limit. As in the nonprecessing case, we inform these choices through a feature-importance analysis presented in Appendix~\ref{app:feature_importance_prec}. Guided by these results, the final ans\"atze describe the leading in-plane-spin dependence through low-order integer powers of $\Sperp$ and, where supported by the data, $\Delta_\perp$, with coefficient functions built from $\eta$ and the baseline remnant predictions. This construction preserves continuity with \gwModelS{} while allowing validation residuals and physical consistency checks to determine which precession-dependent sectors require additional structure.

%==========================================================================
\subsubsection{\textcolor{linkcolor}{\texttt{gwModelRemP\_mf}} : model for the remnant mass}
\label{sec:prec_nonecc_mf}
%==========================================================================
For the remnant mass, we augment the nonprecessing prediction according to
\begin{equation}
m_f^{\rm prec}=m_f^{\rm noprec}+Q_{\rm prec}^{M},
\end{equation}
where

\begin{equation}
Q_{\rm prec}^{M}=C_m(\eta,\chi_{fz}^{\rm noprec})\,\Sperp^2
+C_{\Delta}^{M}(\eta,\chi_{fz}^{\rm noprec})\,\Delta_\perp^2.
\end{equation}
We then write
\begin{align}
C_m(\eta,\chi_{fz}^{\rm noprec})
&=a_0^{M}+a_1^{M}\eta+a_2^{M}\eta^2
\nonumber\\
&\quad+a_3^{M}(\chi_{fz}^{\rm noprec})^2
+a_4^{M}\eta(\chi_{fz}^{\rm noprec})^2,\\
C_{\Delta}^{M}(\eta,\chi_{fz}^{\rm noprec})
&=b_0^{M}+b_1^{M}\eta
+b_2^{M}\chi_{fz}^{\rm noprec}.
\end{align}
The best-fit coefficients are
$a_0^{M}=+0.014219\pm0.002253$,
$a_1^{M}=-0.157484\pm0.020035$,
$a_2^{M}=+0.481226\pm0.050415$,
$a_3^{M}=+0.011819\pm0.003045$,
$a_4^{M}=-0.114662\pm0.014462$,
$b_0^{M}=-0.011178\pm0.001966$,
$b_1^{M}=+0.047514\pm0.009889$, and
$b_2^{M}=-0.007593\pm0.001586$.
This model achieves an RMS error of $1.23\times10^{-3}$ on the full dataset and $9.1\times10^{-5}$ on the BHPT subset.

%==========================================================================
\subsubsection{\textcolor{linkcolor}{\texttt{gwModelRemP\_chif}} : model for the remnant spin}
\label{sec:prec_nonecc_af}
%==========================================================================
For the remnant spin magnitude, we augment the nonprecessing prediction according to
\begin{equation}
|a_f|_{\rm prec}
=
\sqrt{(\chi_{fz}^{\rm noprec})^2
+C_a(\eta,\chi_{fz}^{\rm noprec})\,\Sperp^2
+C_{\Delta}^{a}(\eta,\chi_{fz}^{\rm noprec})\,\Delta_\perp^2},
\end{equation}
where
\begin{align}
C_a(\eta,\chi_{fz}^{\rm noprec})
&=
a_0^{a}+a_1^{a}\eta+a_2^{a}\eta^2
+a_3^{a}(\chi_{fz}^{\rm noprec})^2,\\
C_{\Delta}^{a}(\eta,\chi_{fz}^{\rm noprec})
&=
b_0^{a}+b_1^{a}\eta+b_2^{a}\chi_{fz}^{\rm noprec}.
\end{align}
The best-fit coefficients are
$a_0^{a}=+0.727520\pm0.033851$,
$a_1^{a}=-2.834119\pm0.300329$,
$a_2^{a}=+0.950143\pm0.693734$,
$a_3^{a}=-0.104717\pm0.013330$,
$b_0^{a}=+0.250740\pm0.029394$,
$b_1^{a}=-0.833394\pm0.151444$, and
$b_2^{a}=+0.077093\pm0.016856$.
To ensure physically admissible predictions, we cap the model at the Kerr limit by setting
\begin{equation}
|a_f|_{\rm prec}\leftarrow
\min\!\left(|a_f|_{\rm prec},\,1\right).
\end{equation}
The resulting model achieves an RMS error of $1.47\times10^{-2}$ over the full calibration set.

To predict the remnant spin direction, we augment the nonprecessing prediction for the final spin component parallel to the orbital angular momentum according to
\begin{equation}
\chi_{f,z}=\chi_{fz}^{\rm noprec}+C_{\theta}(\eta,\chi_{fz}^{\rm noprec})\,\Sperp^2,
\end{equation}
where
\begin{equation}
C_{\theta}(\eta,\chi_{fz}^{\rm noprec})=a_0^{\theta}+a_1^{\theta}\eta+a_2^{\theta}\eta^2+a_3^{\theta}\chi_{fz}^{\rm noprec}.
\end{equation}
The best-fit coefficients are $a_0^{\theta}=+0.154097\pm0.014260$, $a_1^{\theta}=+1.311602\pm0.184080$,
$a_2^{\theta}=-3.814038\pm0.529204$, and $a_3^{\theta}=-0.398386\pm0.009448$.
The remnant spin tilt angle is then obtained from
\begin{equation}
\theta_f=\arccos\!\left(\frac{\chi_{f,z}}{|a_f|_{\rm prec}}\right).
\end{equation}
This model yields an RMS error of $0.0202$ in $\chi_{f,z}$ and approximately $10^\circ$ in the remnant spin tilt angle. Note that the remnant spin direction model does not include terms proportional to $\Delta_\perp^2$, as their inclusion improves the RMS error by only $\sim0.2\%$.

%==========================================================================
\subsubsection{\textcolor{linkcolor}{\texttt{gwModelRemP\_Lpeak}} : model for the peak luminosity}
\label{sec:prec_nonecc_luminosity}
%==========================================================================
Finally, the precessing peak-luminosity model is constructed by augmenting the nonprecessing prediction multiplicatively,
\begin{equation}
L_{\rm peak}^{\rm prec}=L_{\rm peak}^{\rm noprec}\exp\!\left(Q_{\rm prec}^{L}\right),
\end{equation}
where
\begin{equation}
Q_{\rm prec}^{L}=\Sperp^2\left(b_0+b_1\chihat+b_2(1-4\eta)+b_3\chihat^2\right)+b_4\Sperp^4.
\end{equation}
The best-fit coefficients are $b_0=-2.344\times10^{-4}$, $b_1=+1.983\times10^{-3}$, $b_2=-9.223\times10^{-2}$, $b_3=+8.290\times10^{-1}$, and $b_4=+8.825\times10^{-1}$.
The exponential form guarantees $L_{\rm peak}^{\rm prec}>0$ and reduces to a linear correction for small values of $\Sperp$. The resulting model achieves a fractional RMS error of $0.0852$, compared to $0.1354$ for the corresponding nonprecessing model, representing a $37\%$ improvement.

By construction, all augmentation terms vanish in the nonprecessing limit $\Sperp\rightarrow0$. Consequently,
\begin{equation}
m_f^{\rm prec}\rightarrow m_f^{\rm noprec},
\qquad
|a_f|_{\rm prec}\rightarrow\chi_{fz}^{\rm noprec},
\qquad
L_{\rm peak}^{\rm prec}\rightarrow L_{\rm peak}^{\rm noprec},
\end{equation}
ensuring a smooth transition between the nonprecessing and precessing models. The remnant-mass and spin-magnitude fits retain $\Delta_\perp$-dependent terms where supported by the data, whereas the spin-direction and peak-luminosity fits omit them because they do not improve the validation errors.

\begin{figure*}
    \centering
    \includegraphics[width=\columnwidth]{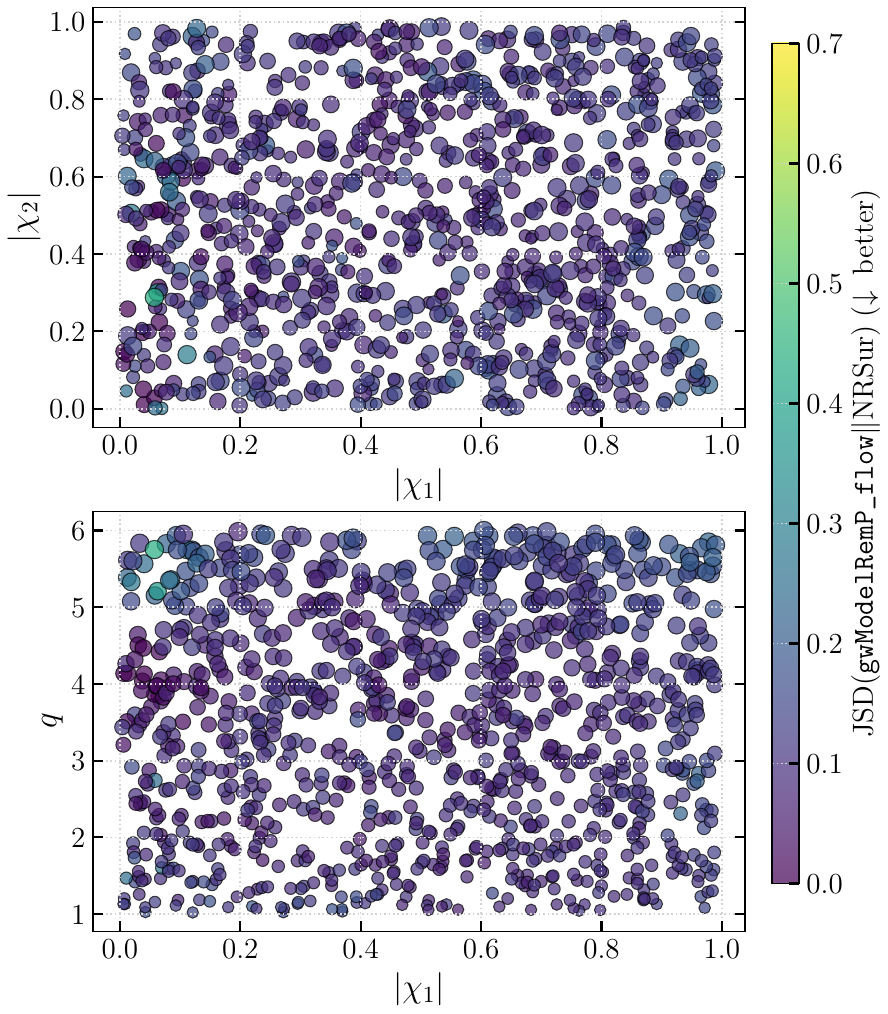}
    \includegraphics[width=\columnwidth]{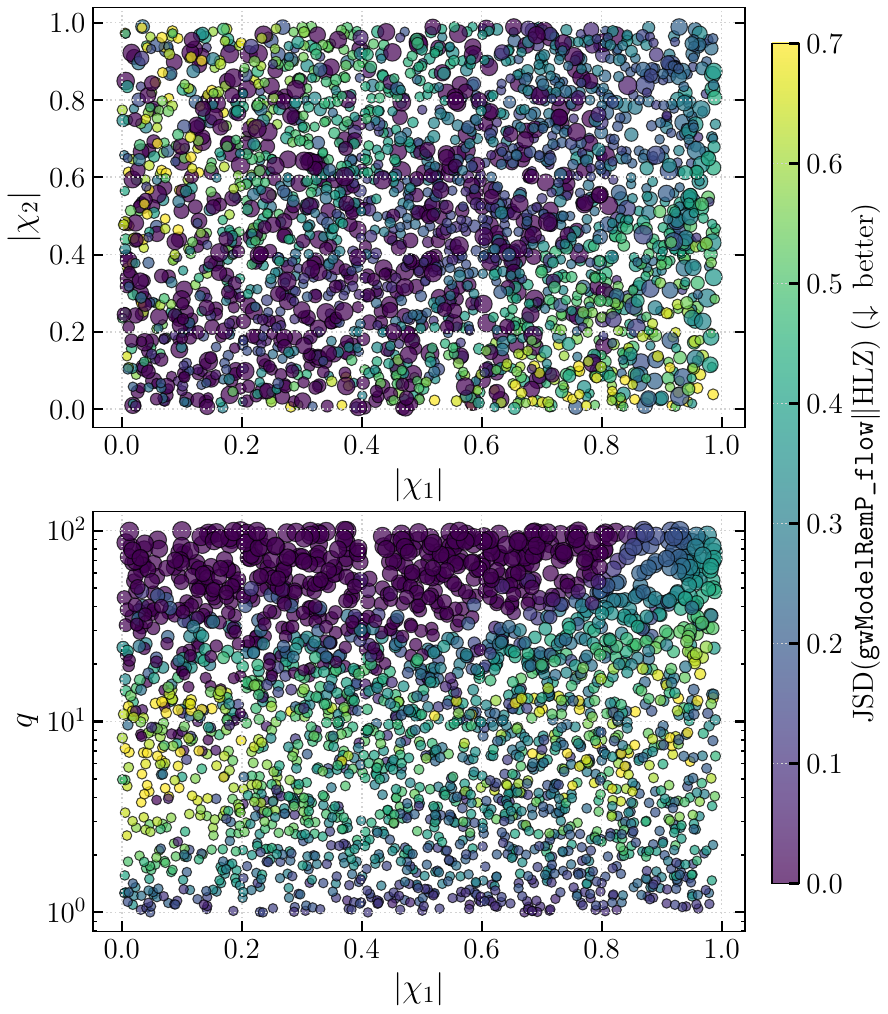}
    \caption{Existing precessing-recoil models trade accuracy against parameter-space coverage. The Gaussian-process-regression-based \NRSur{} model~\cite{Varma:2019csw} is available over $q\in[1,6]$, whereas the fully analytic \HLZ{} prescription~\cite{Lousto:2008dn,Lousto:2010xk,Lousto:2012gt,Lousto:2012su,Gonzalez:2007hi} extends to all mass ratios but is comparatively less accurate. We test our \gwModelP{} normalizing flow using the Jensen--Shannon distance (JSD): the left column compares it with \NRSur{} for $1000$ precessing binaries spanning $q\in[1,6]$, and the right column compares it with \HLZ{} for $2500$ binaries spanning $q\in[1,100]$; both samples cover $|\chi_1|,|\chi_2|\in[0,1]$. The upper panels show the spin-magnitude dependence, and the lower panels show mass ratio versus primary-spin magnitude; see Sec.~\ref{sec:prec_nonecc_accuracy}. Agreement is stronger with \NRSur{} (median JSD $0.100$) than with \HLZ{} (median JSD $0.258$), with the largest discrepancies concentrated in asymmetric-spin configurations. Thus, the flow combines \NRSur{}-like behavior in the low-mass-ratio regime with applicability to substantially larger mass ratios.}
    \label{fig:jsd}
\end{figure*}

%==========================================================================
\subsubsection{\textcolor{linkcolor}{\texttt{gwModelRemP\_flow}} : probabilistic model for the recoil velocity}
\label{sec:prec_flow}
%==========================================================================
We do not construct a corresponding deterministic recoil model for precessing BBH mergers. By deterministic, we mean a single-valued map
\begin{equation}
v_{\rm kick}=f\!\left(q,\vec{\chi}_1,\vec{\chi}_2\right),
\end{equation}
that assigns a unique recoil prediction to each binary configuration. Writing the two spin vectors in spherical coordinates makes the dimensionality explicit: the full map depends on the seven inputs $(q,|\chi_1|,|\chi_2|,\theta_1,\theta_2,\phi_1,\phi_2)$. A deterministic fit must therefore resolve the detailed dependence on all four spin-orientation angles. Although several phenomenological prescriptions for precessing recoils exist in the literature, we were unable to identify a formulation that simultaneously reproduces the available NR data and achieves an accuracy comparable to that of our remnant-mass, remnant-spin, and peak-luminosity models. We therefore defer the construction of a deterministic precessing recoil model to future work.

Instead, following Ref.~\cite{Islam:2025drw}, we develop a probabilistic recoil model that marginalizes over unresolved spin-orientation information. For populations with isotropic spin directions, as commonly assumed for dynamically assembled binaries in dense stellar environments~\cite{Rodriguez:2016vmx}, marginalizing over all four spin angles would reduce the conditioning variables to $(q,|\chi_1|,|\chi_2|)$. This replaces a seven-dimensional point-estimation problem with the lower-dimensional task of learning a conditional probability distribution. The present model follows the same principle but retains the richer five-dimensional context defined below, including remnant-property and in-plane-spin summaries, while marginalizing over the degrees of freedom not retained in that context. It extends Ref.~\cite{Islam:2025drw} by using the mass and spin models developed in Secs.~\ref{sec:nonprec_nonecc} and~\ref{sec:prec_nonecc} to exploit correlations between the recoil velocity and other remnant properties.
Like the other remnant models, the probabilistic recoil model is trained using a combination of precessing NR simulations from the SXS, MAYA, and RIT catalogs together with precessing BHPT simulations. We use two distinct forms of data augmentation. First, to ensure that the learned distribution smoothly approaches the nonprecessing limit, we augment the training set with $2000$ synthetic nonprecessing recoil samples generated using \gwModelS{}. For these synthetic binaries, the mass ratio is sampled uniformly over $1 \leq q \leq 1000$, while the aligned spin components, $\chi_{1z}$ and $\chi_{2z}$, are drawn independently from a uniform distribution on $[-1,1]$. These analytic samples encode the nonprecessing boundary condition rather than additional independent simulations.

For each binary, we first evaluate the \gwModelP{} remnant models to obtain predictions for the remnant mass and spin magnitude. These quantities are then combined with the binary parameters to define a low-dimensional context vector,
\begin{equation}
\mathbf{c}=\left(M_f^{\rm model},|\chi_f|^{\rm model},\eta,\Sperp,\Delta_\perp\right).
\end{equation}
We then train a conditional normalizing flow to model the probability distribution
\begin{equation}
P\!\left(\log_{10} \frac{1}{v_{\rm kick}}\,\big|\,\mathbf{c}\right),
\end{equation}
We model $\log_{10}(1/v_{\rm kick})$ rather than $v_{\rm kick}$ directly. The logarithm compresses the several orders of magnitude spanned by recoil velocities in generic precessing BBH mergers, while the reciprocal fixes the orientation of the distribution tails. Any flexible density model assigns small but nonzero probability beyond the range spanned by the training data, and the parametrization determines where that probability goes. Recoil magnitudes have an upper edge set by the maximum superkick and a soft lower edge because $v_{\rm kick}\to0$ for symmetric configurations. The reciprocal parametrization directs the excess tail toward $v_{\rm kick}\to0$, which is physically harmless, whereas modeling $\log_{10}v_{\rm kick}$ directs it toward unphysically large recoils.
We also impose a floor of $v_{\rm kick}\ge1~{\rm km\,s^{-1}}$ before taking the logarithm, so that configurations with vanishing recoil map to $\log_{10}(1/v_{\rm kick})=0$, the upper endpoint of the target range. This regularization affects only near-symmetric configurations for which the physical kick is negligibly small. It is worth noting that normalizing flows have been widely used across GW data analysis, including rapid parameter estimation~\cite{Green2020ANF,Dax2021RTGW,Wildberger2022NoiseShift,Wong2023FastPE,Polanska2024AccelPE,Shen2019StatDL,Lanchares2025GP15}, population and hierarchical inference~\cite{Ruhe2022HierPop,Leyde2023PopCosmo,Mould2025RapidPop,Wong2020PopFlow}, glitch-robust inference~\cite{Xiong2024RobustGlitch,Zhang2023TempSpecFlow}, and space-based/pulsar-timing-array applications~\cite{Du2023TaijiFlow,Liang2024CNFMBHB,Vallisneri2024PTAFlow}.

Our second form of data augmentation addresses the sparse coverage of low-spin systems, because existing NR simulations sample the high-spin region more densely~\cite{Islam:2025drw}. Although \NRSur{} remains the most accurate deterministic recoil model currently available, its kick predictions are less accurate than its remnant-mass or remnant-spin predictions; Appendix~\ref{app:kick} quantifies these errors. Recoil is intrinsically more difficult to model than the waveform because it depends on nonlinear couplings among waveform modes through the linear-momentum flux~\cite{Ruiz:2007yx,RevModPhys.52.299}. We generate precessing waveforms with \textcolor{linkcolor}{\texttt{NRSur7dq4}} and evaluate their recoils with \textcolor{linkcolor}{\texttt{gw\_remnant}}. From a candidate set of surrogate evaluations, we select $5000$ samples using a farthest-point criterion in the five-dimensional context space to improve coverage of regions sparsely populated by direct simulations. Because these labels inherit modeling systematics from the waveform surrogate and flux integration, they receive a reduced training weight of $0.5$, compared with unit weight for SXS and BHPT simulations. They supplement the fit within the surrogate-informed region but are not treated as independent ground truth or used to establish the model's nominal domain of validity. In Appendix~\ref{app:gwremannt}, we quantify the accuracy of waveform-based recoil calculations.

The probabilistic recoil model is implemented using a \textit{Rational-Quadratic Neural Spline Flow (RQ-NSF)}~\cite{2019arXiv190604032D} with eight masked autoregressive spline transformations as implemented in \textcolor{linkcolor}{\texttt{nflows}}~\cite{nflows}. Each transformation employs two residual blocks with 64 hidden features, eight spline bins, linear tails with a bound of $6.0$, and a dropout rate of $0.05$. Reverse permutations are applied between successive flow layers, resulting in a model with approximately $1.6\times10^5$ trainable parameters. Both the target ($\log_{10}[1/\max(v_{\rm kick},1~{\rm km\,s^{-1}})]$) and conditioning variables, $(M_f^{\rm model}, |\chi_f|^{\rm model}, \eta, \Sperp, \Delta_\perp)$, are standardized to zero mean and unit variance before training.
The model is optimized using \textit{Adam}~\cite{kingma2014adam} and achieves a test negative-log-likelihood (NLL) of $-1.431$. . Further details are provided in Appendix~\ref{app:flow_details} and Ref.~\cite{Islam:2025drw}.

Note that the resulting recoil model, \textcolor{linkcolor}{\texttt{gwModelRemP\_flow}}, is used in conjunction with the final mass and spin models developed in the preceding sections, enabling direct sampling of recoil kick velocities for arbitrary binary configurations.

\begin{figure*}[t]
    \centering
    \subfloat[Representative recoil-velocity distributions.\label{fig:prec_kick_posteriors}]{
        \includegraphics[width=0.47\textwidth]{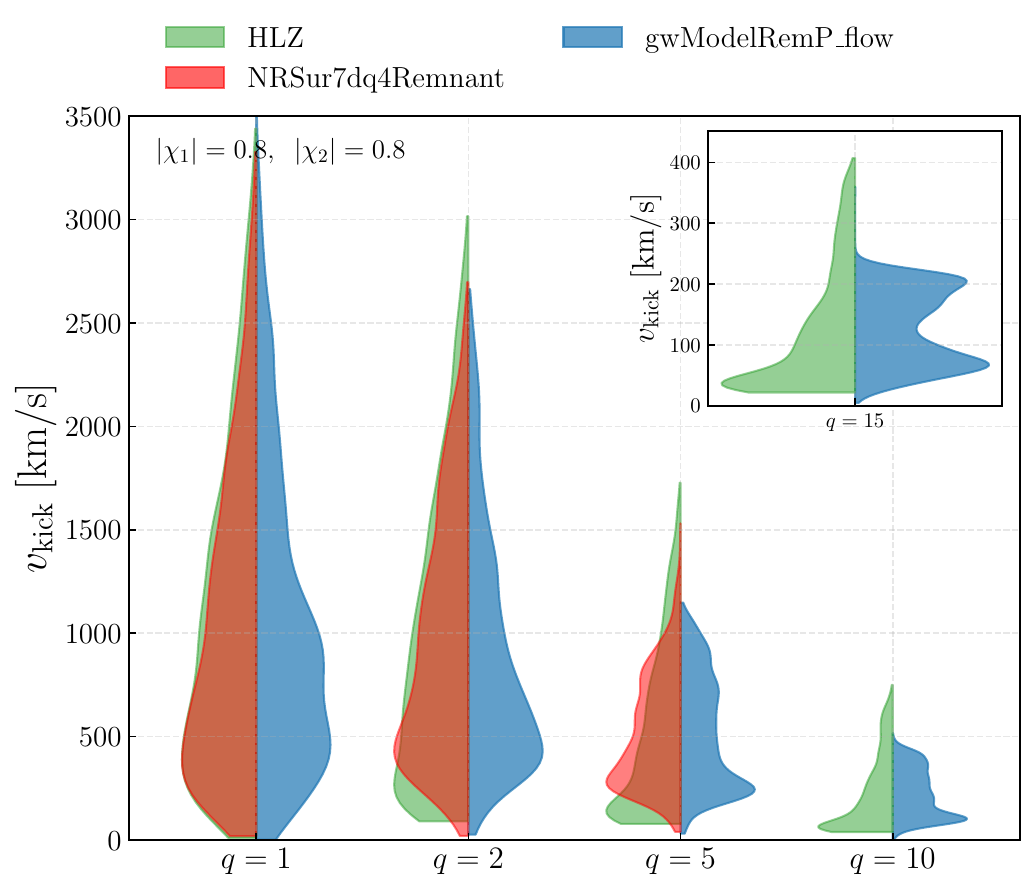}
    }
    \hfill
    \subfloat[Calibration of the predicted recoil distributions.\label{fig:pp_plot}]{
        \includegraphics[width=0.43\textwidth]{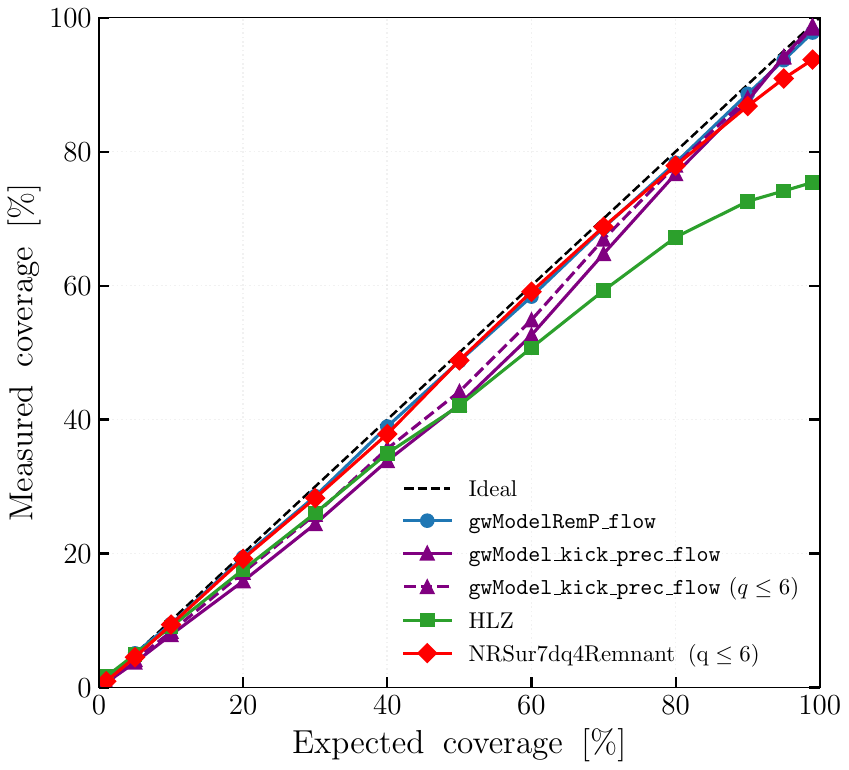}
    }
    \caption{Building on the parameter-space comparison in Fig.~\ref{fig:jsd}, we test representative recoil distributions and uncertainty calibration. The left panel compares distributions from \gwModelP{} (blue), \HLZ{} (green), and \NRSur{} (red) for binaries with $|\chi_1|=|\chi_2|=0.8$ and $5000$ isotropic spin orientations at each mass ratio; \NRSur{} is shown only within its supported range. The right panel shows a probability--probability test, where the dashed diagonal denotes ideal coverage, and includes our previous \textcolor{linkcolor}{\texttt{gwModel\_kick\_prec\_flow}} model~\cite{Islam:2025drw}. The \gwModelP{} flow and \NRSur{} within $q\leq6$ track the diagonal, whereas deterministic \HLZ{} predictions are biased toward under-coverage; see Sec.~\ref{sec:prec_nonecc_accuracy}. Thus, \gwModelP{} retains \NRSur{}-like behavior at low mass ratios while providing calibrated recoil distributions over a substantially broader domain.}
    \label{fig:prec_kick_validation}
\end{figure*}

%==========================================================================
%==========================================================================
\subsection{\gwModelP{} model accuracy}
\label{sec:prec_nonecc_accuracy}
%==========================================================================
%==========================================================================
In Fig.~\ref{fig:prec_noecc_error_hist_log10}, we show the fractional/absolute validation errors of \gwModelP{} for the remnant mass, remnant spin magnitude, remnant spin tilt angle, and peak luminosity when evaluated against the NR and BHPT datasets used in this work. When available, we also show the corresponding errors for existing models (\HBR{} and \NRSurEmri{}) computed on the same validation data. For the remnant mass, \gwModelP{} achieves a median absolute error of $3.9\times10^{-4}$, with 90th- and 95th-percentile errors of $1.8\times10^{-3}$ and $2.3\times10^{-3}$, respectively, outperforming \HBR{} and remaining competitive with \NRSurEmri{}. For the remnant spin magnitude, \gwModelP{} attains a median error of $7.9\times10^{-3}$ and exhibits substantially tighter tail errors than both comparison models, with 90th- and 95th-percentile errors of $2.6\times10^{-2}$ and $3.2\times10^{-2}$, respectively. Histograms of the estimated SXS resolution errors are also shown as a benchmark.
For the remnant spin direction, \NRSurEmri{} achieves the smallest median tilt-angle error of $1.6^\circ$, while \gwModelP{} follows closely with a median error of $2.9^\circ$, substantially outperforming \HBR{} ($7.5^\circ$). 
For the peak luminosity, \gwModelP{} achieves a median fractional error of $3.5\%$, with 90th- and 95th-percentile errors of $20\%$ and $37\%$, respectively. Overall, \gwModelP{} provides the best or joint-best performance for the remnant mass and spin magnitude across the full parameter space while remaining competitive with state-of-the-art surrogate models for the remnant spin direction.

We then compare the kick-velocity distributions for a set of $1000$ binaries spanning $q\in[1,6]$, $|\chi_1|\in[0,1]$, and $|\chi_2|\in[0,1]$. For each binary configuration $(q,|\chi_1|,|\chi_2|)$, we generate $5000$ isotropically distributed spin orientations and compute the corresponding kick distributions. To quantify their differences, we compute the Jensen--Shannon distances (JSDs) between the recoil distributions predicted by \gwModelP{} and \NRSur{}, as shown in the left panel of Fig.~\ref{fig:jsd}. We find good overall agreement across the calibration region, with the majority of binaries exhibiting JSD values below $\sim0.2$. The distribution has a median JSD of $0.10$ and a standard deviation of $0.041$. The largest discrepancies are confined to a small number of configurations with $q\gtrsim5$ and asymmetric low-spin configurations, where one of the component spins is close to zero. The two largest outliers (JSD $>0.37$) both occur at $q\simeq5.5$--$5.8$ with $|\chi_1|\simeq0.06$, close to the upper edge of the \NRSur{} calibration domain where the surrogate itself is expected to become less reliable. Consequently, the larger JSD values in this region likely reflect limitations of both the surrogate and the flow model rather than deficiencies of \gwModelP{} alone. Outside this small region, the JSD values remain uniformly low, indicating that the recoil distributions predicted by \gwModelP{} closely track those of \NRSur{}.

We also compare the recoil-velocity distributions predicted by \gwModelP{} with those implied by the analytic \HLZ{} recoil model over a wider range of mass ratios. The right panel of Figure~\ref{fig:jsd} shows the JSD between the two recoil distributions for a set of $2500$ binaries spanning $q\in[1,100]$, $|\chi_1|\in[0,1]$, and $|\chi_2|\in[0,1]$. Compared to the \NRSur{} validation, the agreement with \HLZ{} is substantially weaker, with a median JSD of $0.258$, compared to $0.100$ for \NRSur{}. The JSD distribution is also considerably broader, with a standard deviation of $0.204$ versus $0.041$, indicating that the discrepancies between the probabilistic flow model and the deterministic \HLZ{} prescription vary significantly across parameter space. The disagreement is driven primarily by spin rather than mass ratio. The largest JSD values occur for highly asymmetric spin configurations, particularly those with one nearly extremal spin and the other comparatively small, where recoil distributions are most sensitive to superkick physics. In contrast, the dependence on mass ratio is relatively weak over the range considered, with the agreement improving modestly at large $q$, where the recoil magnitude is intrinsically suppressed. These results suggest that while the \HLZ{} model captures the overall recoil scale, it does not reproduce the full probability distribution learned from NR data in the strongly spinning regime.

In Fig.~\ref{fig:prec_kick_posteriors}, we provide examples of recoil-velocity distributions predicted by \gwModelP{}, the analytic \HLZ{} model, and \textcolor{linkcolor}{\texttt{NRSur7dq4Remnant}} for representative precessing binaries with spin magnitudes $(|\chi_1|, |\chi_2|)=(0.8,0.8)$.
For each mass ratio, we generate $5000$ isotropically distributed spin orientations and compute the resulting kick distribution. The main panel shows $q=1$, $2$, $5$, and $10$, while the inset shows $q=15$. \textcolor{linkcolor}{\texttt{NRSur7dq4Remnant}} distributions are shown only for $q\in\{1,2,5\}$. The flow model closely matches the \textcolor{linkcolor}{\texttt{NRSur7dq4Remnant}} kick distributions at low mass ratios, with JSD of $\mathrm{JSD}=0.009$, $0.012$, and $0.017$ for $q=1$, $2$, and $5$, respectively, well below the $\mathrm{JSD}\lesssim0.1$ threshold for statistical similarity ($0\leq\mathrm{JSD}\leq\log 2\approx0.693$). The JSD between \gwModelP{} and \HLZ{} is also small at $q=1$ ($\mathrm{JSD}=0.006$) but grows to $\sim0.1$ at higher mass ratios. At large $q$, the \gwModelP{} and \HLZ{} distributions show qualitative agreement but differ in detail.

While the JSD measures distributional similarity configuration by configuration, it does not test whether the flow's predictive uncertainties are statistically calibrated. We therefore construct a probability--probability (PP) plot. For each training and test configuration in an 80/20 split, we draw $N=500$ samples from $P(v_{\rm kick}\mid\mathbf{c})$ and compute the central $p\%$ credible interval (CI), bounded by the $(100-p)/2$-th and $(100+p)/2$-th percentiles, for $p\in\{1,5,10,20,\ldots,90,95,99\}$. The empirical coverage is the fraction of configurations whose true recoil velocity lies within the corresponding CI. A calibrated model follows the diagonal in the PP plot, while deviations below (above) indicate under-coverage (over-coverage).

As shown in Fig.~\ref{fig:pp_plot}, the \gwModelP{} flow closely follows the diagonal, achieving $87.9\%$ coverage at the nominal $90\%$ level and $97.5\%$ at $99\%$, demonstrating well-calibrated predictive uncertainties. For comparison, we construct recoil distributions from the \HLZ{} formula by marginalizing over $\Delta\phi$ and $\Theta$, and from \NRSur{} by marginalizing over $\Delta\phi$ for the $1675$ test configurations with $q\le6$. Both exhibit significant under-coverage: \HLZ{} reaches only $73.5\%$ coverage at the nominal $99\%$ level, while the corresponding \NRSur{} construction reaches $70.7\%$. Both \gwModelP{} and our previous \textcolor{linkcolor}{\texttt{gwModel\_kick\_prec\_flow}} model substantially outperform \HLZ{}, which saturates near $75\%$. Although conditioned only on $(q,a_1,a_2)$, the earlier flow already avoids this saturation and performs comparably to \NRSur{} within its domain of validity. By additionally conditioning on tilt-sensitive in-plane-spin summaries, the five-dimensional \gwModelP{} closely follows the ideal diagonal over the full mass-ratio range, indicating that the remaining miscalibration of the earlier model primarily arose from its reduced conditioning rather than limitations of the flow architecture.
Further discussion of model behavior is provided in Appendix~\ref{sec:prec_nonecc_behavior}.

%==========================================================================
%==========================================================================
%==========================================================================
\section{Eccentric models}
\label{sec:ecc}
\begin{figure*}[t]
    \centering
    \subfloat[Validation of quasi-circular remnant models.\label{fig:ecc_residuals}]{
        \begin{minipage}[b]{0.35\textwidth}
            \centering
            \includegraphics[width=\linewidth]{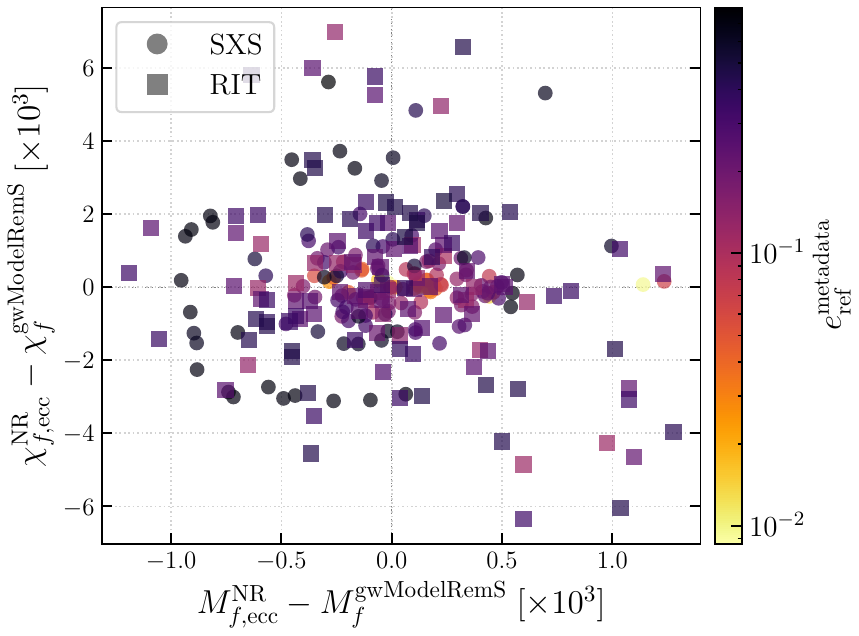}\\[0.6em]
            \includegraphics[width=\linewidth]{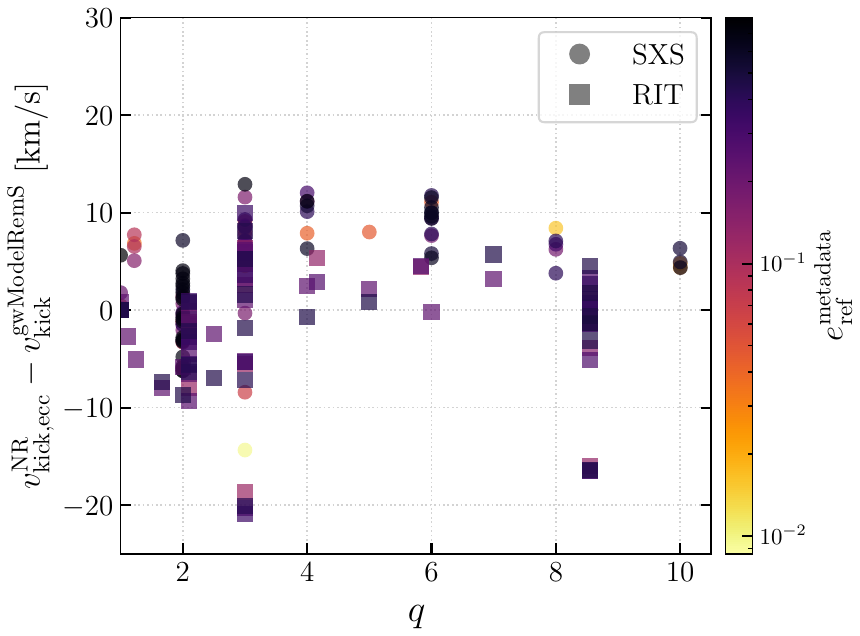}
        \end{minipage}
    }
    \hfill
    \subfloat[Predictions of the eccentric remnant models.\label{fig:ecc_remnant_2d}]{
        \begin{minipage}[b]{0.63\textwidth}
            \centering
            \includegraphics[width=\linewidth]{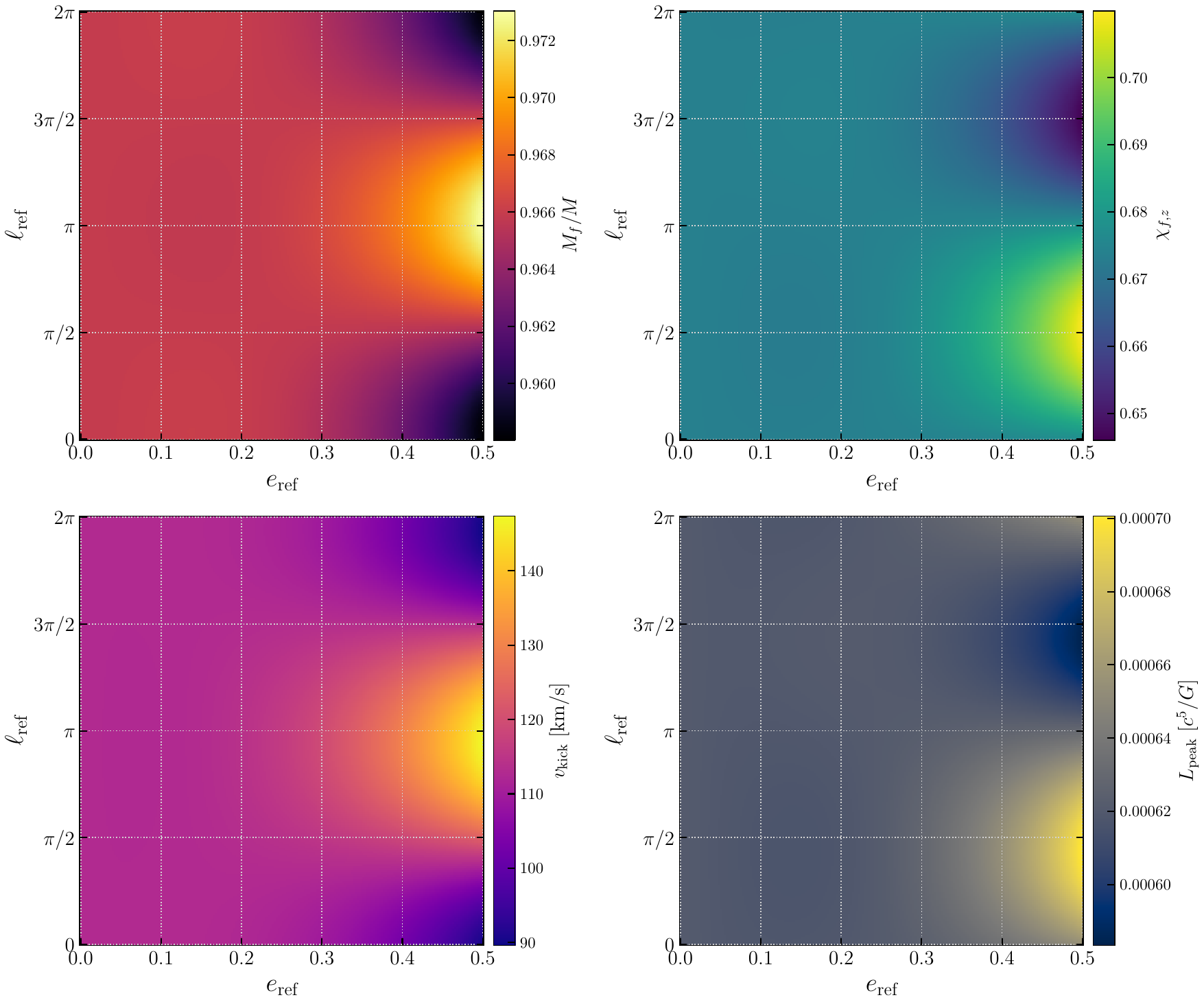}
        \end{minipage}
    }
    \caption{We construct eccentric remnant models as perturbative extensions of the quasi-circular baseline. The upper and lower panels in the left column show, respectively, mass--spin and recoil residuals obtained by applying the quasi-circular \gwModelS{} to eccentric SXS and RIT simulations; color denotes $e_{\rm ref}$ obtained from NR metadata. The corrections vary smoothly, vanish toward the circular limit, and remain modest relative to the leading quasi-circular behavior. The four-panel block on the right shows \gwModelSE{} predictions versus $(e_{\rm ref},\ell_{\rm ref})$ for $(q,\chi_{1z},\chi_{2z})=(3,0.3,0.1)$: remnant mass, remnant spin, recoil velocity, and peak luminosity. Further details are in Sec.~\ref{sec:ecc}.}
    \label{fig:ecc_models}
\end{figure*}

%==========================================================================
%==========================================================================
%==========================================================================
Before constructing eccentric corrections, it is useful to assess how well the quasi-circular models perform when applied directly to eccentric nonprecessing NR simulations from the SXS and RIT catalogs. Figure~\ref{fig:ecc_residuals} shows the residuals between the eccentric NR remnant properties and the corresponding \gwModelS{} predictions. Most eccentric NR simulations differ from the \gwModelS{} predictions by less than $\sim5\times10^{-3}$ in remnant mass and $\sim1.5\times10^{-3}$ in remnant spin, demonstrating that eccentricity introduces only a subdominant correction relative to the leading quasi-circular behavior. A similar trend is observed for the recoil velocity, where most residuals remain within $\sim10~{\rm km\,s^{-1}}$.
For this study, we use the eccentricities reported in the NR metadata, which are typically measured during the inspiral and are expected to decrease significantly by the time of merger. Consequently, the observed eccentricity corrections are generally small and are unlikely to affect most current astrophysical applications. 

We find that the residuals induced by eccentricity are often comparable to, or smaller than, the intrinsic modeling errors of the quasi-circular remnant fits themselves. Nevertheless, eccentric BBH mergers are expected to become increasingly relevant as GW detector sensitivity improves, and several events have already been reported to exhibit evidence for residual eccentricity. Motivated by these considerations, we develop simple eccentric extensions of our remnant models that treat eccentricity as a perturbative correction to the quasi-circular predictions. Our primary goal is not to construct the most accurate eccentric remnant model currently possible, but rather to develop a physically motivated framework that captures the leading qualitative effects of eccentricity on the remnant properties.

%==========================================================================
%==========================================================================
\subsection{Model construction}
\label{sec:ecc_models}
%==========================================================================
%==========================================================================
We now construct leading-order eccentric extensions of the quasi-circular remnant models \gwModelS{} and \gwModelP{}, denoted \gwModelSE{} and \gwModelPE{}, respectively.
The model takes as input $(q,\chi_{1z},\chi_{2z},e_{\rm ref},\ell_{\rm ref})$, where $e_{\rm ref}$ and $\ell_{\rm ref}$ denote the eccentricity and mean anomaly at a chosen reference time. 
In this work, we use \textcolor{linkcolor}{\texttt{gwModels}} to estimate the eccentricity and mean anomaly at a reference time of $t=-2500M$ following the prescription in Ref.~\cite{Shaikh:2023ypz}.
Our model construction is motivated by leading-order PN expectations for eccentric corrections to the radiated energy, angular momentum, and linear momentum, together with the point-particle separatrix limit.
Here the quasi-circular prediction again supplies the fixed baseline, and the ansatz search is confined to corrections that vanish as $e_{\rm ref}\rightarrow0$. The retained structures combine a low-order polynomial envelope in $(e_{\rm ref},\eta)$ with a periodic dependence on $\ell_{\rm ref}$ and an $\eta$-dependent phase. The polynomial order controls the secular eccentric correction, while the harmonic term captures the oscillatory dependence on the orbital state at the reference time. The comparison of alternative anomaly harmonics is described in Sec.~\ref{sec:necc_harmonic}.

One possible modeling choice is to replace the circular point-particle backbone with an eccentric separatrix backbone. In the Schwarzschild limit, the separatrix is given by $p_{\rm sep}=6+2e_{\rm s}$, with
\begin{equation}
E_{\rm sep}^{\rm Schw}(e_{\rm s})=\sqrt{\frac{8}{9-e_{\rm s}^2}},
\qquad
L_{\rm sep}^{\rm Schw}(e_{\rm s})=\frac{6+2e_{\rm s}}{\sqrt{3+2e_{\rm s}-e_{\rm s}^2}}.
\end{equation}
Here, $e_{\rm s}$ is the eccentricity at the separatrix. We use a phenomenological Kerr extension that interpolates between the Schwarzschild separatrix and the Kerr ISCO,
\begin{align}
E_{\rm sep}(e_{\rm s},\chihat)
&=E_{\rm sep}^{\rm Schw}(e_{\rm s})
\nonumber\\
&\quad+
\left[E_{\rm ISCO}(\chihat)-E_{\rm ISCO}(0)\right]
\frac{9}{9-e_{\rm s}^2},
\\
L_{\rm sep}(e_{\rm s},\chihat)
&=L_{\rm sep}^{\rm Schw}(e_{\rm s})
\nonumber\\
&\quad+
\left[L_{\rm ISCO}(\chihat)-L_{\rm ISCO}(0)\right]
\frac{\sqrt{3}}{\sqrt{(3-e_{\rm s})(1+e_{\rm s})}}.
\end{align}
The corresponding angular-momentum contribution is
\begin{equation}
\ell_{\rm sep}(e_{\rm s},\chihat)=L_{\rm sep}(e_{\rm s},\chihat)-2\chihat\left[E_{\rm sep}(e_{\rm s},\chihat)-1\right].
\end{equation}
By construction, $E_{\rm sep}\rightarrow E_{\rm ISCO}$ and $\ell_{\rm sep}\rightarrow\ell_{\rm Kerr}$ in the circular limit $e_{\rm s}\rightarrow0$.
Although we briefly explored this approach, we do not adopt it in the present work because it would require introducing two distinct eccentricity scales: one characterizing the inspiral at the reference point ($t=-2500M$) and another describing the binary near merger. For this first eccentric remnant model, we instead favor a simpler formulation based on a single reference eccentricity. Nevertheless, \textcolor{linkcolor}{\texttt{gwModels}} provides the eccentric separatrix backbone as an optional component for users who wish to experiment with such models, and we plan to investigate this approach in more detail in future work.

%==========================================================================
\subsubsection{Model for the remnant mass}
\label{sec:necc_mf}
%==========================================================================
We note that PN calculations predict eccentricity-dependent corrections to the radiated energy~\cite{Arun:2009mc,Gamboa:2024hli}, while Refs.~\cite{Nee:2025zdy,Wang:2023vka,Ravichandran:2026iec} show that the remnant properties exhibit an approximately sinusoidal dependence on the mean anomaly. Motivated by these findings, we adopt the following phenomenological model for the eccentric remnant mass that captures the leading-order trend:
\begin{equation}
\frac{M_f^{\rm ecc}}{M}=\frac{M_f^{\rm circ}}{M}\left[1+\delta_M(e_{\rm ref},\ell_{\rm ref},\eta)\right],
\end{equation}
where
\begin{equation}
\delta_M=\mathcal{P}_M(e_{\rm ref},\eta)\left[1+\alpha_M e_{\rm ref}\sin\!\left(\ell_{\rm ref}+\varphi_M(\eta)\right)\right],
\end{equation}
with
\begin{equation}
\varphi_M(\eta)=\varphi_0^M+\varphi_1^M\eta,
\end{equation}
and
\begin{equation}
\mathcal{P}_M(e_{\rm ref},\eta)=(a_1^M+b_1^M\eta)e_{\rm ref}+(a_2^M+b_2^M\eta)e_{\rm ref}^2.
\end{equation}
This form preserves the circular limit by construction, since $\delta_M\to0$ as $e_{\rm ref}\to0$. We calibrate the model against the combined aligned-spin eccentric NR datasets from the SXS and RIT catalogs. The best-fit eccentric parameters are $a_1^M=-7.572773\times10^{-4}$, $a_2^M=4.301594\times10^{-3}$, $b_1^M=6.935577\times10^{-3}$, $b_2^M=-3.661705\times10^{-2}$,
$\alpha_M=4.210183\times10^{1}$, $\varphi_0^M=-1.294636\times10^{1}$, and $\varphi_1^M=4.336959\times10^{1}$.

%==========================================================================
\subsubsection{Model for the remnant spin}
\label{sec:necc_chif}
%==========================================================================
Similarly, the eccentric remnant-spin model is written as a multiplicative correction to the circular prediction,
\begin{equation}
\chi_f^{\rm ecc}=\chi_f^{\rm circ}\left[1+\delta_{\chi}(e_{\rm ref},\ell_{\rm ref},\eta)\right],
\end{equation}
where
\begin{equation}
\delta_{\chi}=\mathcal{P}_{\chi}(e_{\rm ref},\eta)
\left[1+\alpha_{\chi}e_{\rm ref}\cos\!\left(\ell_{\rm ref}+\varphi_{\chi}(\eta)\right)\right],
\end{equation}
with
\begin{equation}
\varphi_{\chi}(\eta)=\varphi_0^{\chi}+\varphi_1^{\chi}\eta,
\end{equation}
and
\begin{equation}
\mathcal{P}_{\chi}(e_{\rm ref},\eta)=(a_1^{\chi}+b_1^{\chi}\eta)e_{\rm ref}+(a_2^{\chi}+b_2^{\chi}\eta)e_{\rm ref}^2.
\end{equation}
This form preserves the circular limit by construction, since
$\delta_{\chi}\rightarrow0$ as $e_{\rm ref}\rightarrow0$.
The best-fit eccentric parameters are $a_1^{\chi}=-2.225555\times10^{-2}$, $a_2^{\chi}=1.165720\times10^{-1}$, $b_1^{\chi}=6.385577\times10^{-2}$,
$b_2^{\chi}=-3.756145\times10^{-1}$, $\alpha_{\chi}=1.484316\times10^{1}$,
$\varphi_0^{\chi}=1.630530\times10^{1}$, and $\varphi_1^{\chi}=-9.549006\times10^{1}$.

%==========================================================================
\subsubsection{Model for the recoil kick}
\label{sec:necc_kick}
%==========================================================================
For the recoil velocity, we model the eccentric recoil as a multiplicative correction to the circular prediction,
\begin{equation}
v_{\rm kick}^{\rm ecc}=v_{\rm kick}^{\rm circ}\left[1+\delta_k(e_{\rm ref},\ell_{\rm ref},\eta)\right],
\end{equation}
where
\begin{equation}
\delta_k=\mathcal{P}_k(e_{\rm ref},\eta)\left[1+\alpha_k e_{\rm ref}\cos\!\left(\ell_{\rm ref}+\varphi_k(\eta)\right)\right],
\end{equation}
with
\begin{equation}
\varphi_k(\eta)=\varphi_0^k+\varphi_1^k\eta,
\end{equation}
and
\begin{equation}
\mathcal{P}_k(e_{\rm ref},\eta)=(a_1^k+b_1^k\eta)e_{\rm ref}+(a_2^k+b_2^k\eta)e_{\rm ref}^2.
\end{equation}
This form preserves both the circular limit and the equal-mass limit by construction, since $v_{\rm kick}^{\rm circ}=0$ for $q=1$ and $\delta_k\rightarrow0$ as $e_{\rm ref}\rightarrow0$. The best-fit eccentric parameters are
$a_1^k=-1.561135\times10^{-1}$, $a_2^k=1.017796$, $b_1^k=6.682131\times10^{-1}$, $b_2^k=-4.005937$,
$\alpha_k=1.000054\times10^{1}$, $\varphi_0^k=-6.802120$, and $\varphi_1^k=2.049863\times10^{1}$.

%==========================================================================
\subsubsection{Model for the peak luminosity}
\label{sec:necc_lpeak}
%==========================================================================
Similarly, the eccentric peak luminosity is modeled as a multiplicative correction to the circular prediction,
\begin{equation}
L_{\rm peak}^{\rm ecc}=L_{\rm peak}^{\rm circ}\left[1+\delta_L(e_{\rm ref},\ell_{\rm ref},\eta)\right],
\end{equation}
where
\begin{equation}
\delta_L=\mathcal{P}_L(e_{\rm ref},\eta)\left[1-\alpha_L e_{\rm ref}\sin\!\left(\ell_{\rm ref}+\varphi_L(\eta)\right)\right],
\end{equation}
with
\begin{equation}
\varphi_L(\eta)=\varphi_0^L+\varphi_1^L\eta,
\end{equation}
and
\begin{equation}
\mathcal{P}_L(e_{\rm ref},\eta)=(a_1^L+b_1^L\eta)e_{\rm ref}+(a_2^L+b_2^L\eta)e_{\rm ref}^2.
\end{equation}
This form preserves the circular limit by construction, since
$\delta_L\rightarrow0$ as $e_{\rm ref}\rightarrow0$.
The best-fit eccentric parameters are
$a_1^L=-1.397943\times10^{-1}$, $a_2^L=6.543722\times10^{-1}$, $b_1^L=4.109220\times10^{-1}$,
$b_2^L=-2.097417$, $\alpha_L=-5.568125$, $\varphi_0^L=5.773265$, and $\varphi_1^L=-2.966536\times10^{1}$.

%==========================================================================
\subsubsection{Choice of anomaly harmonic}
\label{sec:necc_harmonic}
%==========================================================================
At leading Newtonian quadrupole order, the instantaneous GW fluxes for an eccentric orbit of mean anomaly $\ell$ are
\begin{align}
\dot{E} &\propto 1 + 6\,e_{\rm ref}\cos\ell_{\rm ref} + \mathcal{O}(e^2), \\
\dot{J} &\propto 1 + \tfrac{31}{8}\,e_{\rm ref}\cos\ell_{\rm ref} + \mathcal{O}(e^2), \\
|\dot{\vec{P}}| &\sim e_{\rm ref}\cos 2\ell_{\rm ref} + \mathcal{O}(e^2).
\end{align}
Since the energy and angular-momentum fluxes modulate at the orbital frequency, $M_f$ and $\chi_f$ are expected to inherit an $n=1$ harmonic in $\ell_{\rm ref}$. The instantaneous linear-momentum flux instead contains an $n=2$ harmonic, suggesting $\cos 2\ell_{\rm ref}$ for the recoil. However, the accumulated kick is dominated by the final orbits, and its magnitude depends on the merger-phase anomaly, a $2\pi$-periodic function of $\ell_{\rm ref}$, so the instantaneous $n=2$ structure need not survive integration.

We test this against the NR data by fitting each remnant quantity $X(\ell_{\rm ref})$ with a Fourier series and comparing the median fractional amplitudes of each harmonic. We also compare the three-parameter model $c_0[1+A\cos(n\ell_{\rm ref}+\varphi)]$ for $n=1$ and $n=2$. We find that the $n=1$ harmonic is preferred for all four quantities. The preference is strongest for the peak luminosity and recoil, where the $n=2$ model increases the residual RMS by factors of $40$ and $3.3$, respectively, while the final mass and spin retain only a subdominant $n=2$ contribution. We therefore adopt $\cos(\ell_{\rm ref}+\varphi_k)$ rather than $\cos(2\ell_{\rm ref}+\varphi_k)$ for the recoil, and use a single $n=1$ harmonic throughout. The choice of $\sin$ or $\cos$ in Secs.~\ref{sec:necc_mf}--\ref{sec:necc_lpeak} is simply a phase convention absorbed into $\varphi_0$.

%==========================================================================
%=========================================================================
\subsection{Model accuracy and behavior}
\label{sec:ecc_accuracy}
%==========================================================================
%=========================================================================
Relative to the quasi-circular baseline, the eccentric correction reduces the residuals by $9.9\%$ for the remnant mass and $13.7\%$ for the remnant spin over the full eccentric dataset, with comparable improvements for systems with $e_{\rm ref}>0.05$. For the recoil velocity, the improvement is more modest, at approximately $2.2\%$, consistent with the larger intrinsic scatter in kick predictions. In all cases, the correction vanishes smoothly in the circular limit $e_{\rm ref}\rightarrow0$, ensuring that \gwModelSE{} reduces exactly to \gwModelS{} for nonprecessing quasi-circular BBH mergers.

In Figure~\ref{fig:ecc_remnant_2d}, we show the behavior of the eccentric remnant model as a function of the reference eccentricity and mean anomaly while fixing $q=3$, $\chi_{1z}=0.3$ and $\chi_{2z}=0.1$. Consistent with the perturbative nature of the eccentric corrections, the remnant mass and spin exhibit only modest variations over the range $0\le e_{\rm ref}\le0.5$. For the representative configuration shown, the remnant mass varies by approximately $10^{-3}$ and the remnant spin by approximately $10^{-2}$ across the full eccentricity and anomaly range. The recoil velocity exhibits a stronger dependence on both quantities, with variations of several $\mathrm{km\,s^{-1}}$. The figure also highlights the distinct anomaly dependence predicted by the model: the remnant mass is modulated primarily through the $\sin\ell_{\rm ref}$ correction, the remnant spin through the $\cos\ell_{\rm ref}$ correction, and the recoil velocity through the $\cos\ell_{\rm ref}$ correction. As expected, all eccentric corrections vanish smoothly in the limit $e_{\rm ref}\rightarrow0$, recovering the quasi-circular predictions. Similar behavior can be obtained for the \gwModelPE{} model.

\begin{table}[t]
\centering
\caption{Computational cost for evaluating $5000$ BBH configurations, averaged over five repetitions. Fast models use $100$ inner repetitions for timing resolution. The last column reports the speedup relative to \textcolor{linkcolor}{\texttt{NRSur7dq4Remnant}}; the analytic models are up to four orders of magnitude faster. Further details are in Sec.~\ref{sec:timing}.}
\label{tab:timing}
\begin{tabular}{lcc}
\hline
Model & Time (s) & Speedup \\
\hline
\gwModelS{} & $0.002\pm0.000$ & $\sim1.1\times10^4$ \\
\textcolor{linkcolor}{\texttt{UIB2016}} & $0.002\pm0.000$ & $\sim9.8\times10^3$ \\
\gwModelSE{} & $0.002\pm0.000$ & $\sim9.7\times10^3$ \\
\HBR{}/\HLZ{} & $0.002\pm0.000$ & $\sim7.9\times10^3$ \\
\gwModelP{} & $0.027\pm0.002$ & $\sim620$ \\
\gwModelEMRI{} & $0.028\pm0.002$ & $\sim595$ \\
\textcolor{linkcolor}{\texttt{NRSur3dq8Remnant}} & $2.318\pm0.079$ & $\sim7$ \\
\textcolor{linkcolor}{\texttt{NRSur7dq4Remnant}} & $16.878\pm0.401$ & $1$ \\
\textcolor{linkcolor}{\texttt{NRSur7dq4EmriRemnant}} & $20.888\pm0.490$ & $\sim0.8$ \\
\hline
\end{tabular}
\end{table}

%==========================================================================
\section{Point-particle remnant model}
\label{sec:emri_model}
%==========================================================================
Although the primary focus of this work is the construction of remnant models for comparable- and large-mass-ratio BBH mergers, the underlying framework is anchored to the exact point-particle limit wherever possible. Consequently, the models presented in this work (\gwModelS{}, \gwModelP{}, \gwModelPE{}, and \gwModelSE{}) numerically recover the correct leading-order behavior as $\eta\rightarrow0$.
Nevertheless, for applications in the genuine extreme-mass-ratio regime, it is preferable to use an implementation that directly evaluates the exact point-particle limit. We therefore provide a dedicated EMRI remnant model, \gwModelEMRI{}, within the package.

The model takes as input the mass ratio $q$, the dimensionless spin of the primary Kerr black hole $\chi$, the orbital inclination angle $\theta$, and the eccentricity at the separatrix $e_{\rm sep}$. The remnant mass and spin are computed from the conserved energy $E_{\rm sep}(\chi,\theta,e_{\rm sep})$ and angular momentum $L_{\rm sep}(\chi,\theta,e_{\rm sep})$ of the small body at the onset of plunge. At leading order in $\eta$,
\begin{align}
\frac{M_f}{M}&=1-\eta\left[1-E_{\rm sep}\right]+\mathcal{O}(\eta^2),\\
\chi_f&=\chi+\eta\left[L_{\rm sep}-2\chi(E_{\rm sep}-1)\right]+\mathcal{O}(\eta^2).
\end{align}
Unlike the other remnant models presented in this paper, \gwModelEMRI{} contains no phenomenological corrections fitted to NR simulations. It is intended for the genuine extreme-mass-ratio regime ($q\gg1000$), where linear BHPT accurately describes the dynamics. The model also provides a reference implementation of the point-particle backbone used throughout this work and allows users to investigate alternative prescriptions for the separatrix energy and angular momentum independently of the calibrated BBH models.

%==========================================================================
%==========================================================================
\section{Computational cost}
\label{sec:timing}
%==========================================================================
%==========================================================================
Table~\ref{tab:timing} compares the computational cost of the remnant models considered in this work. The benchmark evaluates $5000$ BBH configurations and averages over five repetitions. The deterministic models \gwModelS{} and \gwModelSE{} each require approximately $0.002~{\rm s}$, providing speedups of $\sim1.1\times10^4$ and $\sim9.7\times10^3$, respectively, relative to \textcolor{linkcolor}{\texttt{NRSur7dq4Remnant}}. The probabilistic precessing model \gwModelP{} requires $0.027\pm0.002~{\rm s}$, corresponding to a speedup of approximately $620$. Its additional cost arises because recoil samples are generated with a PyTorch normalizing flow rather than a closed-form expression. Nevertheless, all models introduced here remain substantially faster than the surrogate models while covering a larger parameter space.

For applications requiring repeated single-system evaluations, \gwModelS{}, \gwModelSE{}, and \gwModelP{} require approximately $192~\mu{\rm s}$, $213~\mu{\rm s}$, and $2.4~{\rm ms}$ per binary, respectively. These runtimes are sufficiently small for large-scale population synthesis, hierarchical inference, and Monte Carlo studies involving millions of binary evaluations.

All model evaluations and timing benchmarks reported in this work were performed on CPU on a MacBook Pro with an Apple M3 Pro chip (5 performance and 6 efficiency cores) and 36\,GB of unified memory, running Python 3.10.18 with NumPy 2.2.6 and PyTorch 2.5.1.

\begin{figure}[t]
    \centering
    \includegraphics[width=\columnwidth]{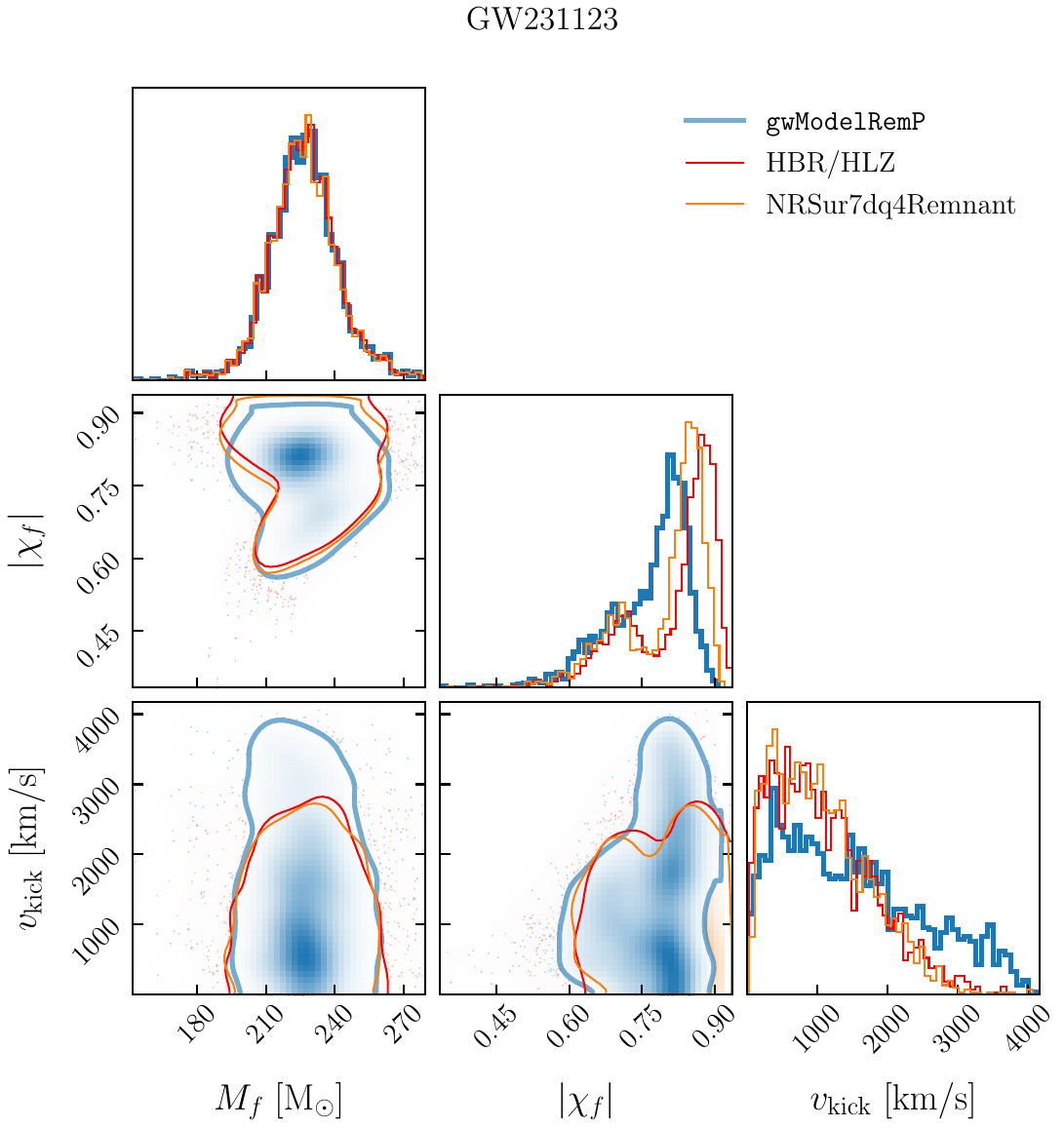}
    \caption{We test our recoil model using data from the recent gravitational-wave event GW231123 by propagating its public \textcolor{linkcolor}{\texttt{NRSur7dq4}} posterior samples through the remnant models. We compare $M_f$, $|\chi_f|$, and $v_{\rm kick}$ inferred with \gwModelP{}, \HBR{}/\HLZ{}, and \textcolor{linkcolor}{\texttt{NRSur7dq4Remnant}}; see Sec.~\ref{sec:astro_gwtc}. The models give nearly identical remnant masses and broadly consistent spins, but \gwModelP{} predicts a larger median recoil and a broader high-kick tail.}
    \label{fig:event_level_remnants}
\end{figure}

%==========================================================================
%==========================================================================
\section{Astrophysical applications}
\label{sec:astro}
%==========================================================================
%==========================================================================
To demonstrate the utility of our models, we now consider several representative astrophysical applications.

%==========================================================================
\subsection{Inference of remnant properties for the GW events}
\label{sec:astro_gwtc}
%==========================================================================
We consider the recently reported highly-spinning intermediate-mass BBH merger event GW231123~\cite{LIGOScientific:2025rsn}, with a total source-frame mass of approximately $241\,M_\odot$, mass ratio $q\in[1.0,3.2]$, and large component spins ($|\chi_1|\sim0.89$, $|\chi_2|\sim0.83$). We propagate the public \textcolor{linkcolor}{\texttt{NRSur7dq4}} posterior samples through \gwModelP{}, \HBR{}/\HLZ{}, and \textcolor{linkcolor}{\texttt{NRSur7dq4Remnant}}, and show the resulting remnant-property posteriors in Fig.~\ref{fig:event_level_remnants}.
The three models predict nearly identical remnant masses, with median values of approximately $226\,M_\odot$, and broadly consistent remnant spin magnitudes of $|\chi_f|\simeq0.8$--$0.9$. The largest differences again arise in the recoil velocity. We find that \gwModelP{} predicts a median recoil of approximately $1322\,{\rm km\,s^{-1}}$, compared to $958\,{\rm km\,s^{-1}}$ from \HBR{}/\HLZ{} and $965\,{\rm km\,s^{-1}}$ from \textcolor{linkcolor}{\texttt{NRSur7dq4Remnant}}. In addition, the probabilistic flow model predicts a substantially broader high-kick tail, with the $90\%$ credible interval extending to approximately $3481\,{\rm km\,s^{-1}}$, compared to roughly $2400\,{\rm km\,s^{-1}}$ for the deterministic models. 

\begin{figure*}[t]
    \centering
    \includegraphics[width=\textwidth]{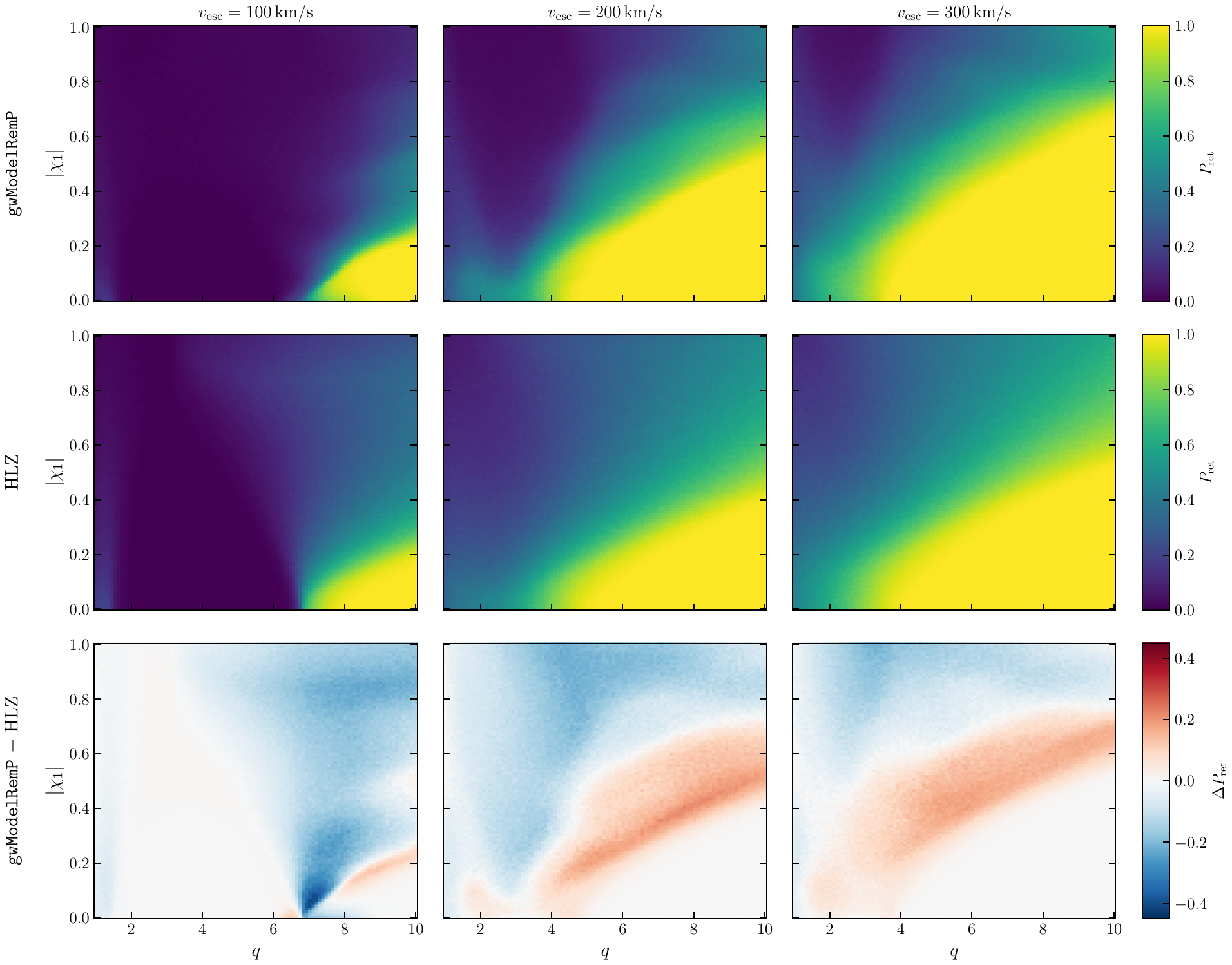}
    \caption{Recoil predictions determine whether hierarchical-merger remnants are retained in their host environments. Maps of $p_{\rm ret}$ from \gwModelP{} (top row) and analytic \HLZ{} (middle row) are shown versus mass ratio and primary-spin magnitude for $v_{\rm esc}=100$, $200$, and $300~{\rm km\,s^{-1}}$. At each point, $|\chi_2|$ is drawn uniformly over $[0,1]$, both spin directions are sampled isotropically, and $5000$ realizations are used. The bottom row shows $\Delta p_{\rm ret}=p_{\rm ret}^{\gwModelP}-p_{\rm ret}^{\HLZ}$; see Sec.~\ref{sec:astro_retention}. \gwModelP{} predicts systematically higher retention, with the largest differences at moderate mass ratios as the escape velocity increases.}
    \label{fig:retention_qchi1}
\end{figure*}

%==========================================================================
\subsection{Remnant-retention probabilities}
\label{sec:astro_retention}
%==========================================================================
One of the primary astrophysical applications of recoil models is determining whether merger remnants are retained within their host environments. Several population-synthesis and cluster-dynamics frameworks have been developed to model compact-object demographics in dynamical formation channels, including Cluster Monte Carlo (\textcolor{linkcolor}{\texttt{CMC}})~\cite{Rodriguez:2019huv,Kremer:2019iul}, \textcolor{linkcolor}{\texttt{McFACTS}}~\cite{McKernan:2024kpr}, \rapster{}~\cite{Kritos:2022non}, \textcolor{linkcolor}{\texttt{fastcluster}}~\cite{Mapelli:2021gyv}, \textcolor{linkcolor}{\texttt{cBHBd}}~\cite{Antonini:2019ulv}, and \textcolor{linkcolor}{\texttt{B-POP}}~\cite{Sedda:2021vjh}. Retention calculations in such studies commonly employ established analytic recoil prescriptions, including the \HLZ{} model. As discussed in Sec.~\ref{sec:prec_nonecc_accuracy} and illustrated in Figs.~\ref{fig:jsd} and~\ref{fig:prec_kick_validation}, \HLZ{} was calibrated using a substantially smaller NR dataset than is available today and differs from our flow model in parts of the precessing parameter space. Such differences can propagate directly into predictions for hierarchical-merger retention and compact-object demographics.
To isolate the effect of the primary spin on retention, we evaluate the retention probability $p_{\rm ret}$ on a $100\times100$ grid in mass ratio $q\in[1,10]$ and primary spin magnitude $|\chi_1|\in[0,1]$. At each grid point, the secondary spin magnitude is drawn from $|\chi_2|\sim\mathcal{U}(0,1)$ and both spin directions are sampled isotropically, with $5000$ realizations per cell. Figure~\ref{fig:retention_qchi1} compares the resulting retention maps from \gwModelP{} and \HLZ{} for three fiducial escape velocities, $v_{\rm esc}=100$, $200$, and $300~{\rm km\,s^{-1}}$.
For context, typical escape velocities span $0$--$100~{\rm km\,s^{-1}}$ for globular clusters, $15$--$600~{\rm km\,s^{-1}}$ for nuclear star clusters, and $200$--$1500~{\rm km\,s^{-1}}$ for elliptical galaxies~\cite{Merritt:2004xa,Antonini:2016gqe}.

At $v_{\rm esc}=100~{\rm km\,s^{-1}}$, typical of globular clusters, both models predict that nearly all mergers are ejected except at high mass ratios with small primary spins. As the escape velocity increases to $200$ and $300~{\rm km\,s^{-1}}$, a clear difference emerges: \gwModelP{} predicts systematically higher retention than \HLZ{} across the parameter space, with the largest differences appearing at moderate mass ratios ($q\sim4$--$8$). At $v_{\rm esc}=300~{\rm km\,s^{-1}}$, relevant for nuclear star clusters and low-mass elliptical galaxies, \gwModelP{} predicts substantial retention extending to moderate primary spins, while \HLZ{} confines high retention to a narrower band at low $|\chi_1|$.

This offset reflects the fact that \HLZ{} overpredicts kick magnitudes in parts of the parameter space relative to the NR-calibrated flow model. Whereas population-averaged retention curves may wash out these differences by marginalizing over all binary parameters, conditioning on $(q,|\chi_1|)$ reveals that the choice of kick model can appreciably change the predicted retention fraction. Our result is directly relevant for hierarchical-merger rate calculations in dense stellar environments, where the retention of individual merger products depends sensitively on both the mass ratio and the spin of the primary.

%==========================================================================
%==========================================================================
%==========================================================================
\section{Final remarks}
\label{sec:final}
%==========================================================================
%==========================================================================
%==========================================================================
In this work, we combined a broad collection of publicly available NR simulations with an extended BHPT dataset and a unified modeling framework to construct analytic remnant models for BBH mergers. Empirical corrections extend the BHPT information into the comparable-mass regime~\cite{Islam:2022laz,Rink:2024swg}. Our models predict the remnant mass, spin, recoil velocity, and peak luminosity for nonprecessing and precessing binaries. Wherever possible, the fits are anchored to PN structure and point-particle limits while remaining accurate against the training and validation data.

The resulting models, \gwModelS{} and \gwModelP{}, are applicable to nonprecessing and precessing BBH mergers, respectively, across the entire range from the equal-mass regime to the extreme-mass-ratio limit. For nonprecessing binaries, we find median validation errors of approximately $2\times10^{-4}$ in the remnant mass, $2\times10^{-4}$ in the remnant spin, $2.8~{\rm km\,s^{-1}}$ in the recoil velocity, and $1.3\%$ in the peak luminosity.
For precessing binaries, the corresponding median errors are approximately $3\times10^{-4}$ in the remnant mass, $8\times10^{-3}$ in the remnant spin magnitude, $3.1^\circ$ in the remnant spin tilt angle, and $4.2\%$ in the peak luminosity. Across their domains of validity, the models achieve accuracies comparable to state-of-the-art surrogate models such as \NRSur{}, while simultaneously remaining valid from the equal-mass regime to the extreme-mass-ratio limit.

For precessing recoil velocities, constructing a deterministic analytic fit with comparable accuracy was not practical. We instead developed the conditional normalizing-flow model \textcolor{linkcolor}{\texttt{gwModelRemP\_flow}}, which exploits correlations between recoil velocity and other remnant properties. It reproduces the recoil distributions predicted by existing surrogate models while remaining computationally efficient and applicable over a broader parameter range.

To explore eccentric mergers, we introduced the quasi-circular extensions \gwModelSE{} and \gwModelPE{}. These intentionally simple models include leading-order anomaly-dependent corrections motivated by PN theory and capture the dominant trends in the available eccentric NR simulations. The package also provides an optional point-particle separatrix backbone for future investigation.

A major advantage of the present framework is computational efficiency. When the probabilistic recoil model is included, \gwModelP{} is nearly three orders of magnitude faster than the corresponding \NRSur{} models. Without the flow-based recoil component, the fully analytic remnant models are nearly four orders of magnitude faster. Because most of the framework consists of closed-form analytic expressions, the models are straightforward to interpret, modify, and incorporate into existing GW, astrophysical, and cosmological simulation pipelines.

More broadly, this work illustrates two new modelling aspects. First, validated waveform surrogates can provide lower-fidelity labels derived from conserved fluxes, allowing targeted augmentation of sparsely sampled regions while reserving direct NR and BHPT results for calibration and validation. Second, the agent-assisted propose--evaluate--refine workflow can systematize the search over analytic decompositions, limiting-behavior prefactors, and polynomial-like correction terms. The resulting expressions remain compact and interpretable, and their coefficients are obtained through conventional optimization. In this role, the AI agent served as a controlled, validation-guided modeling aid rather than an end-to-end scientific decision maker: candidate proposals were accepted only after quantitative evaluation, physical checks, and author review. This separation between generation and acceptance provides a practical safeguard against plausible but scientifically incorrect agent outputs.

To demonstrate potential applications, we analyzed a recent event GW231123. Although different remnant prescriptions can yield similar population-level distributions, appreciable differences emerge for individual configurations, particularly in recoil predictions outside the calibration range of existing surrogate models. These differences propagate into remnant-retention probabilities in stellar clusters and galactic nuclei. We provide additional applications for a synthetic BBH population in Appendix~\ref{sec:astro_pop}.

All models developed in this paper are available through the \textcolor{linkcolor}{\texttt{gwModels}} package (\href{https://github.com/tousifislam/gwModels}{https://github.com/tousifislam/gwModels}). 
In future, we plan to extend this framework to model the remnant properties of binary neutron star and black hole--neutron star mergers.

%%%%%%%%%%%%%%%%%%%%%%%%%%%%%%%%%%%%%%%%%%%%%%%%%%%%%%%%%%%%%%%%%%%%%%%%%%%%%%%%%%%%%%%%%%%%%%%%%%%%%%%%
%%%%%%%%%%%%%%%%%%%%%%%%%%%%%%%%%%%%%%%%%%%%%%%%%%%%%%%%%%%%%%%%%%%%%%%%%%%%%%%%%%%%%%%%%%%%%%%%%%%%%%%%
\begin{acknowledgments}
T.I. is supported in part by the National Science Foundation under Grant No. NSF PHY-2309135 and the Gordon and Betty Moore Foundation Grant No. GBMF7392. G.K. acknowledges support from NSF Grants No. PHY-2307236 and DMS-2309609.
This work used computational resources provided through the ACCESS program under allocation PHY260280.
Use was made of computational facilities purchased with funds from the National Science Foundation (CNS-1725797) and administered by the Center for Scientific Computing (CSC). The CSC is supported by the California NanoSystems Institute and the Materials Research Science and Engineering Center (MRSEC; NSF DMR 2308708) at UC Santa Barbara. 
Some computations were performed on the UMass-URI UNITY HPC/AI cluster at the Massachusetts Green High-Performance Computing Center (MGHPCC).
\end{acknowledgments}
%%%%%%%%%%%%%%%%%%%%%%%%%%%%%%%%%%%%%%%%%%%%%%%%%%%%%%%%%%%%%%%%%%%%%%%%%%%%%%%%%%%%%%%%%%%%%%%%%%%%%%%%
%%%%%%%%%%%%%%%%%%%%%%%%%%%%%%%%%%%%%%%%%%%%%%%%%%%%%%%%%%%%%%%%%%%%%%%%%%%%%%%%%%%%%%%%%%%%%%%%%%%%%%%%

%%%%%%%%%%%%%%%%%%%%%%%%%%%%%%%%%%%%%%%%%%%%%%%%%%%%%%%%%%%%%%%%%%%%%%%%%%%%%%%%%%%%%%%%%%%%%%%%%%%%%%%%
\bibliography{references}
%%%%%%%%%%%%%%%%%%%%%%%%%%%%%%%%%%%%%%%%%%%%%%%%%%%%%%%%%%%%%%%%%%%%%%%%%%%%%%%%%%%%%%%%%%%%%%%%%%%%%%%%

\newpage

\begin{figure*}[t]
    \centering
    \subfloat[Non-precessing BBHs.\label{fig:feature_importance_nonprec}]{
        \includegraphics[width=0.42\textwidth]{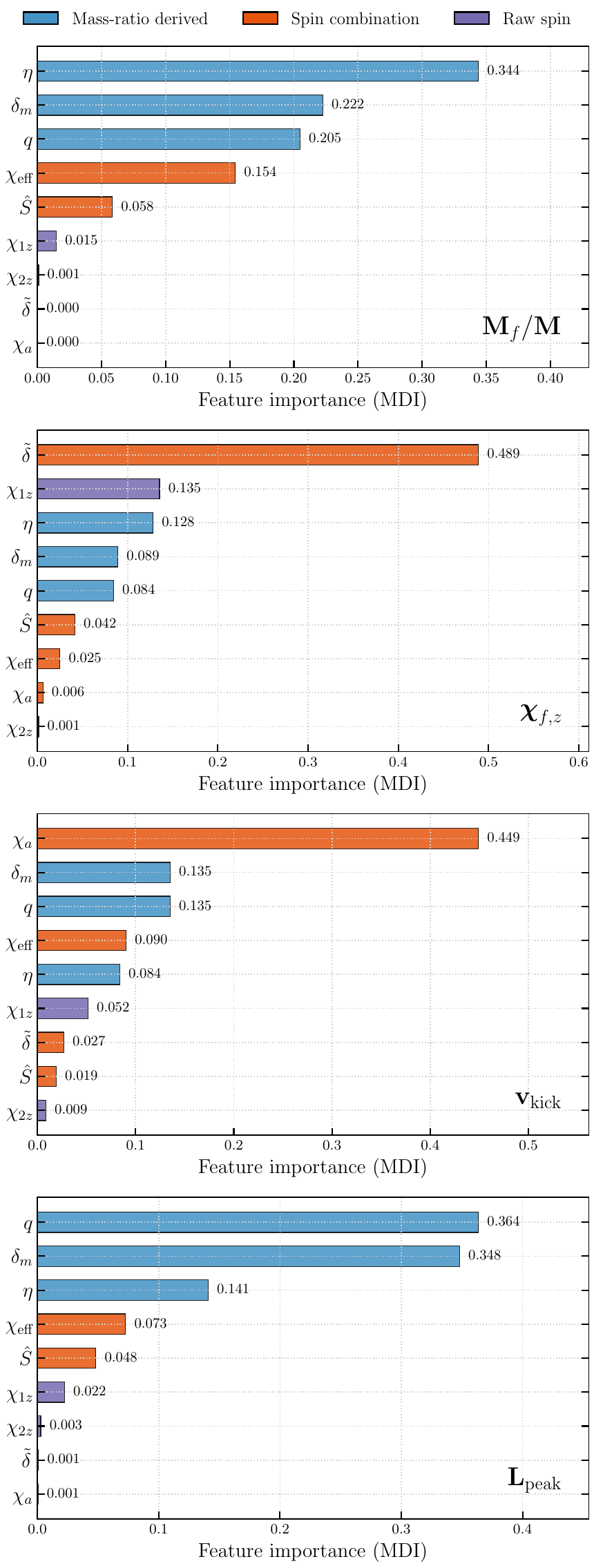}
    }
    \hspace{0.5em} % adjust this
    \subfloat[Precessing BBHs.\label{fig:feature_importance_prec}]{
        \includegraphics[width=0.42\textwidth]{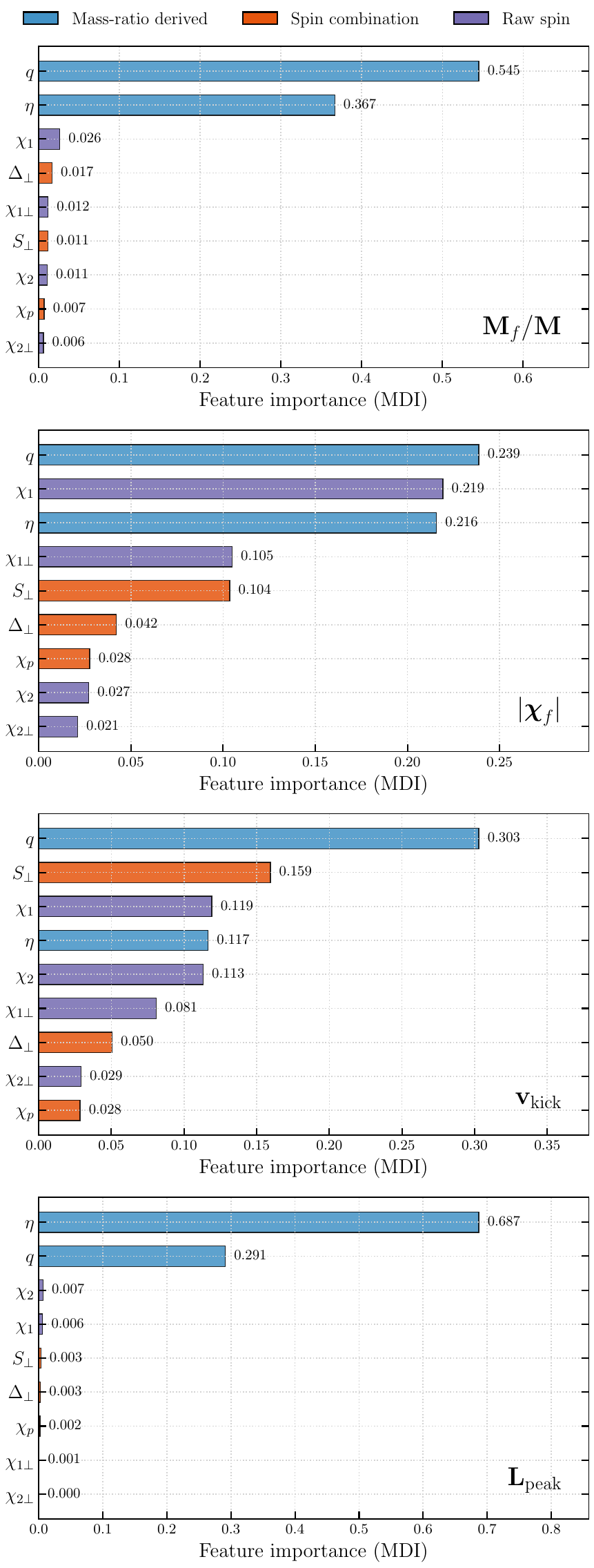}
    }
    \caption{Feature importance identifies the variables most useful for constructing compact analytic ansatzes. For each remnant quantity, we train a separate random-forest regressor on the same NR and BHPT calibration data used to construct the corresponding analytic model, and rank the candidate inputs using the mean decrease in impurity. Results are shown for nonprecessing (a) and precessing (b) binaries, grouping mass-ratio variables (blue), physically motivated spin combinations (orange), and individual spin components (purple). The random forests are used only as an interpretability and feature-selection diagnostic; they are not components of the final analytic models and do not generate their predictions. Further details are in Appendices~\ref{app:feature_importance} and~\ref{app:feature_importance_prec}. Mass ratio dominates the remnant mass and peak luminosity, whereas spin combinations become more important for remnant spin and recoil.}
    \label{fig:feature_importance}
\end{figure*}

\appendix
%==========================================================================
%==========================================================================
\section{Feature importance for non-precessing binaries}
\label{app:feature_importance}
%==========================================================================
%==========================================================================

To gain insight into the relative importance of the physical parameters governing the remnant properties, we perform a feature-importance analysis using random-forest regressors implemented in \textcolor{linkcolor}{\texttt{scikit-learn}} and trained on the same datasets used throughout this work. We quantify the importance of each input using the \emph{mean decrease in impurity} (MDI), which measures the average reduction in prediction error produced when a feature is used to split the decision trees. 

In Figure~\ref{fig:feature_importance_nonprec}, we summarize the feature importance for the nonprecessing remnant quantities. For the remnant mass, the dominant predictors are the mass-ratio-related quantities $(\eta,\delta_m,q)$, which together account for more than $75\%$ of the total importance, while the effective aligned spin $\chi_{\rm eff}$ provides the largest spin contribution. This behavior is consistent with the expectation that the total radiated energy is governed primarily by the mass ratio, with spin producing a secondary correction.
The remnant-spin model exhibits a markedly different behavior. The spin-asymmetry parameter $\tilde{\Delta}$ dominates the prediction, followed by the primary aligned spin $\chi_{1z}$ and the symmetric mass ratio $\eta$. This reflects the fact that the final spin depends on both the inherited spin angular momentum and the orbital angular momentum transferred during the merger.
For the recoil velocity, the spin-difference parameter $\chi_a$ is by far the most important feature, followed by the mass-asymmetry quantities $(\delta_m,q,\eta)$. This result agrees with the well-known physical picture that unequal masses and unequal aligned spins are the primary sources of recoil in nonprecessing binaries. In contrast, the peak luminosity depends almost entirely on the mass ratio, with only modest contributions from the effective spin combinations.

%==========================================================================
%==========================================================================
\section{Feature importance for precessing binaries}
\label{app:feature_importance_prec}
%==========================================================================
%=================
Figure~\ref{fig:feature_importance_prec} presents the corresponding analysis for the precessing models. The remnant mass remains overwhelmingly controlled by the mass ratio, with $q$ and $\eta$ accounting for more than $90\%$ of the total importance. The in-plane spin combinations, including $S_\perp$ and $\Delta_\perp$, contribute only weakly, consistent with the perturbative augmentation adopted in Sec.~\ref{sec:prec_nonecc_models}.
The remnant spin magnitude depends on a more balanced combination of parameters. Besides the mass ratio, the dominant predictors include the primary spin magnitude $\chi_1$, the in-plane spin of the primary black hole $\chi_{1\perp}$, and the mass-weighted in-plane spin $S_\perp$, reflecting the important role of spin precession in determining the final Kerr parameter. By contrast, the peak luminosity remains almost entirely determined by the mass ratio, with the aligned and in-plane spin variables contributing only a few percent of the total importance. The recoil velocity depends much more strongly on the in-plane spin combinations.

We use this feature-importance analysis solely as an interpretability tool to guide the design of our analytic fitting functions. The choice of fitting variables is guided by both physical insight and statistical considerations. By expressing the models in terms of combinations such as $\eta$, $\delta_m$, $\hat{\chi}$, $\chi_a$, $\tilde{\Delta}$, $S_\perp$, and $\Delta_\perp$, which naturally capture the leading mass-ratio and spin dependencies of the merger dynamics, the resulting analytic fits become significantly more compact, interpretable, and accurate than fits constructed directly from $(q,\boldsymbol{\chi}_1,\boldsymbol{\chi}_2)$.

\begin{figure*}[t]
    \centering
    \subfloat[Nonprecessing predictions across the parameter space.\label{fig:nonprec_2d_colormap}]{
        \includegraphics[width=0.65\textwidth]{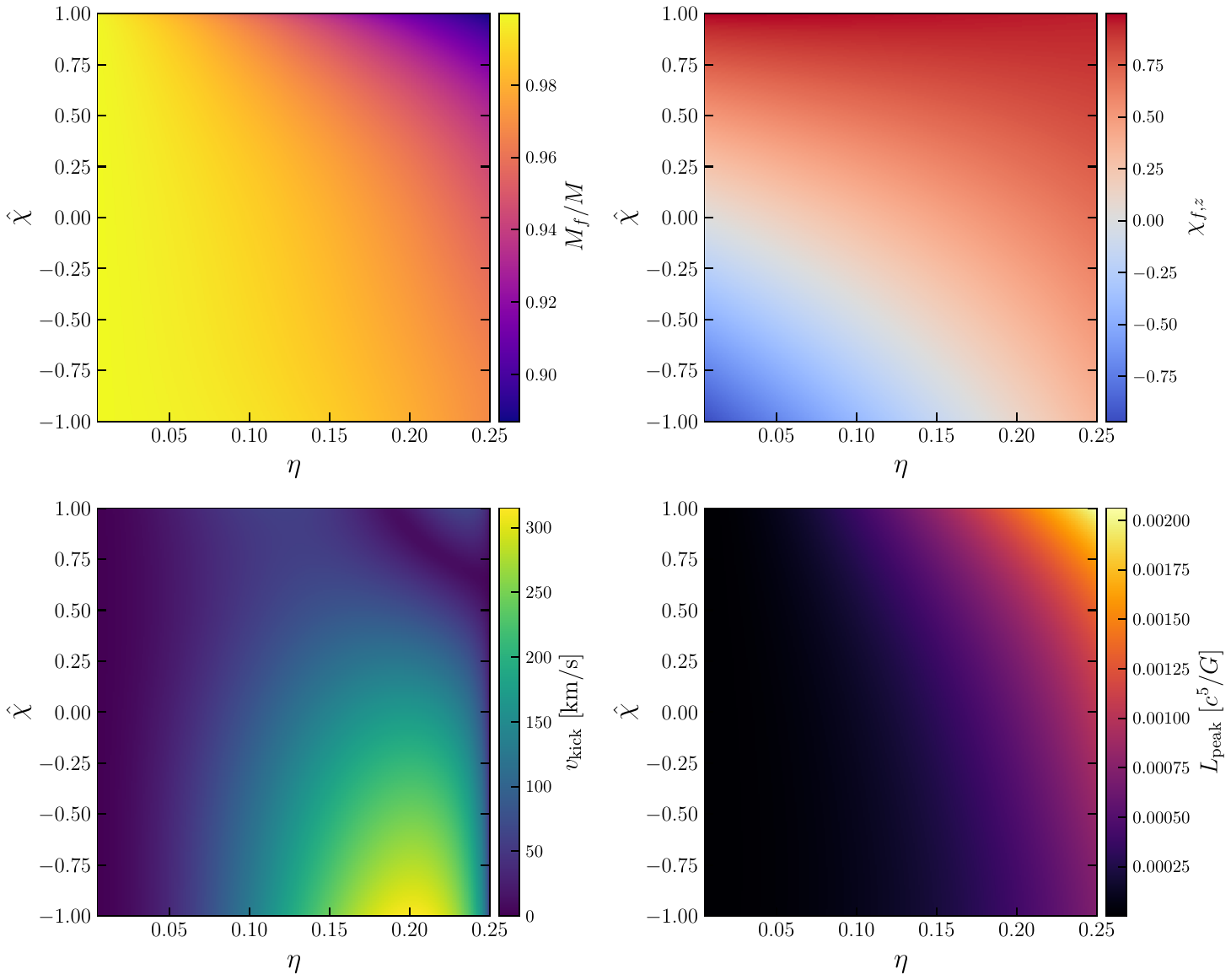}
    }
    \hfill
    \subfloat[Dependence on spin orientation.\label{fig:prec_smoothness_theta1}]{
        \includegraphics[width=0.33\textwidth]{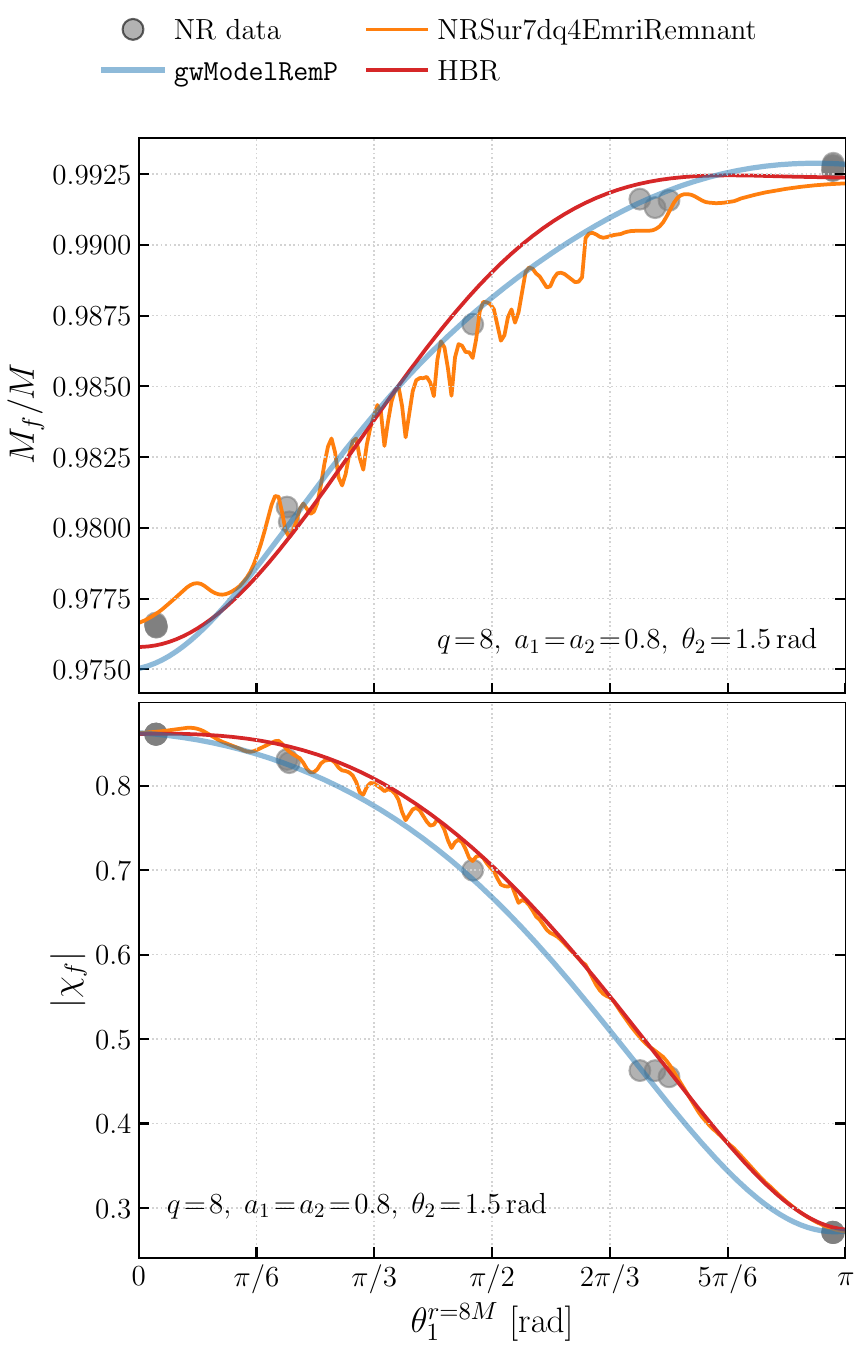}
    }
    \caption{A global parameter-space view and a spin-orientation scan test whether the analytic fits preserve physical trends away from the calibration points. The left panel shows \gwModelS{} predictions for $M_f/M$, $\chi_{f,z}$, $v_{\rm kick}$, and $L_{\rm peak}$ across $(\eta,\chihat)$ for equal-spin nonprecessing binaries with $\chi_{1z}=\chi_{2z}=\chihat$; the predictions vary smoothly and recover the symmetry-required vanishing recoil at equal mass. The right panel shows the remnant mass and spin as the primary-spin tilt varies for the highly spinning configuration $(q,a_1,a_2,\theta_2)=(8,0.8,0.8,1.5~{\rm rad})$, comparing \gwModelP{}, \HBR{}, and \NRSurEmri{} with NR simulations; \gwModelP{} varies smoothly over the full tilt range and accurately reproduces the trends in the NR data. See Appendices~\ref{app:behaviour} and~\ref{sec:prec_nonecc_behavior} for details.}
    \label{fig:model_behaviour}
\end{figure*}

%==========================================================================
%=========================================================================
\section{\gwModelS{} model behavior}
\label{app:behaviour}
%==========================================================================
%=========================================================================

Figure~\ref{fig:nonprec_2d_colormap} provides a global view of the \gwModelS{} predictions across the nonprecessing parameter space for binaries with $\chi_{1z}=\chi_{2z}=\chihat$. The remnant mass increases toward the extreme-mass-ratio limit and decreases for increasingly anti-aligned spins, reflecting the enhanced energy losses associated with mergers of rapidly spinning black holes. The remnant spin exhibits a similarly strong dependence on the effective spin parameter, varying smoothly from negative values for strongly anti-aligned configurations to nearly extremal values for highly aligned binaries. The recoil velocity vanishes along the equal-mass boundary, as required by symmetry, and reaches its largest values for unequal-mass binaries with anti-aligned spins. Finally, the peak luminosity increases with both symmetric mass ratio and aligned spin, attaining its largest values near the equal-mass, highly spinning limit. The absence of discontinuities or sharp features in any of the panels demonstrates the smooth interpolation properties of the model across the full parameter space.

%==========================================================================
\section{\gwModelP{} model behavior}
\label{sec:prec_nonecc_behavior}
%==========================================================================

We performed analogous global checks (not shown) of the \gwModelP{} predictions across the precessing parameter space, including variations with the symmetric mass ratio and in-plane spin magnitude $\Sperp$. We find that all remnant quantities vary smoothly throughout the explored parameter space and exhibit the expected physical trends. In particular, the remnant mass depends primarily on the symmetric mass ratio with a weaker dependence on $\Sperp$, while the remnant spin magnitude and recoil velocity show a stronger dependence on the in-plane spin. The peak luminosity remains dominated by its mass-ratio dependence, with comparatively weaker variations with $\Sperp$. We find no discontinuities, sharp features, or other pathological behavior, demonstrating that the \gwModelP{} models provide a smooth extension of the underlying nonprecessing framework across the precessing parameter space.

As a representative example of these smoothness checks, we show one particularly stringent configuration in Fig.~\ref{fig:prec_smoothness_theta1}. The figure illustrates the dependence of the precessing remnant mass and spin on the primary-spin tilt for a high-spin, unequal-mass configuration with $(q,a_1,a_2)=(8,0.8,0.8)$ and $\theta_2=1.5~{\rm rad}$. The NR results vary smoothly between aligned and anti-aligned configurations, and \gwModelP{} reproduces this trend across the full tilt range. The agreement is particularly notable for the remnant spin, which changes by more than a factor of three as the primary-spin tilt varies from $0$ to $\pi$. In contrast, \NRSurEmri{} exhibits visible oscillations in the remnant-mass prediction, while \HBR{} shows larger systematic deviations from the NR data. Thus, the \gwModelP{} augmentation preserves the expected smooth spin-orientation dependence while remaining consistent with the available simulations.

\begin{figure}[t]
    \centering
    \includegraphics[width=0.4\textwidth]{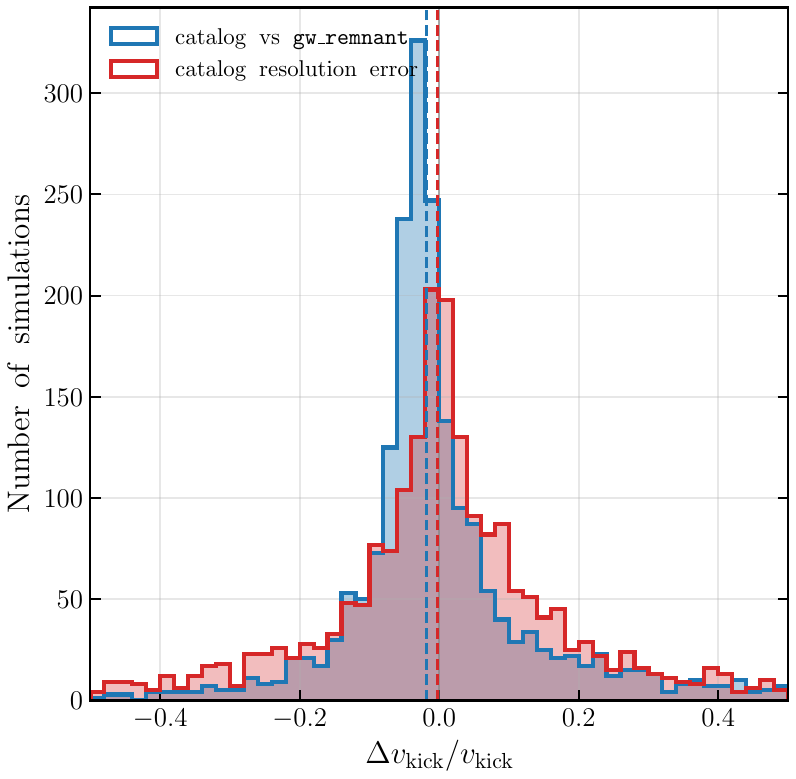}
    \caption{To augment the sparse training data for black-hole kicks, we generate surrogate waveforms and derive additional recoil labels by integrating their radiated linear-momentum flux. This figure validates the same extraction procedure on SXS waveforms, for which independent catalog kicks are available. For $2156$ non-eccentric SXS simulations with two resolutions, blue shows the fractional difference between recoils integrated from waveform modes through $\ell=4$ using \textcolor{linkcolor}{\texttt{gw\_remnant}} and the catalog values, while red shows the difference between the two highest SXS resolutions; dashed lines mark the medians. The waveform-derived recoils have smaller scatter than the SXS resolution differences, supporting the use of surrogate-waveform-derived kicks as down-weighted training labels. Further details are in Appendix~\ref{app:gwremannt}.}
    \label{fig:sxs_gwremannt_waveform_kick_error}
\end{figure}

%==========================================================================
%=========================================================================
\section{Accuracy of kick estimates from waveform modes}
\label{app:gwremannt}
%==========================================================================
%=========================================================================
To examine whether recoil velocities can be recovered directly from the radiated waveform, we use \textcolor{linkcolor}{\texttt{gw\_remnant}} to integrate the linear-momentum flux, retaining SXS strain modes through $\ell=4$, for $2668$ recent non-eccentric simulations with recorded remnant velocities. The recovered kicks closely match the catalog values, with a median ratio
$v_{\rm kick}^{\rm waveform}/v_{\rm kick}^{\rm catalog}=0.984$; $61\%$ agree within $10\%$ and $78\%$ within $25\%$.
Figure~\ref{fig:sxs_gwremannt_waveform_kick_error} compares these differences with the intrinsic SXS resolution uncertainty. For the $2156$ simulations with two available resolutions, the waveform-derived kicks have a $68$th-percentile scatter of $0.111$, smaller than the $0.164$ scatter between the two highest SXS resolutions, and fall within the corresponding resolution uncertainty for $55\%$ of the binaries. The small median offset ($-0.018$) is consistent with momentum lost from truncating the waveform at $\ell=4$.
The waveform-derived recoils have smaller scatter than the SXS resolution differences, supporting the use of surrogate-waveform-derived kicks as down-weighted training labels in Section~\ref{sec:prec_flow}. 

\begin{figure}
    \centering
    \includegraphics[width=0.9\columnwidth]{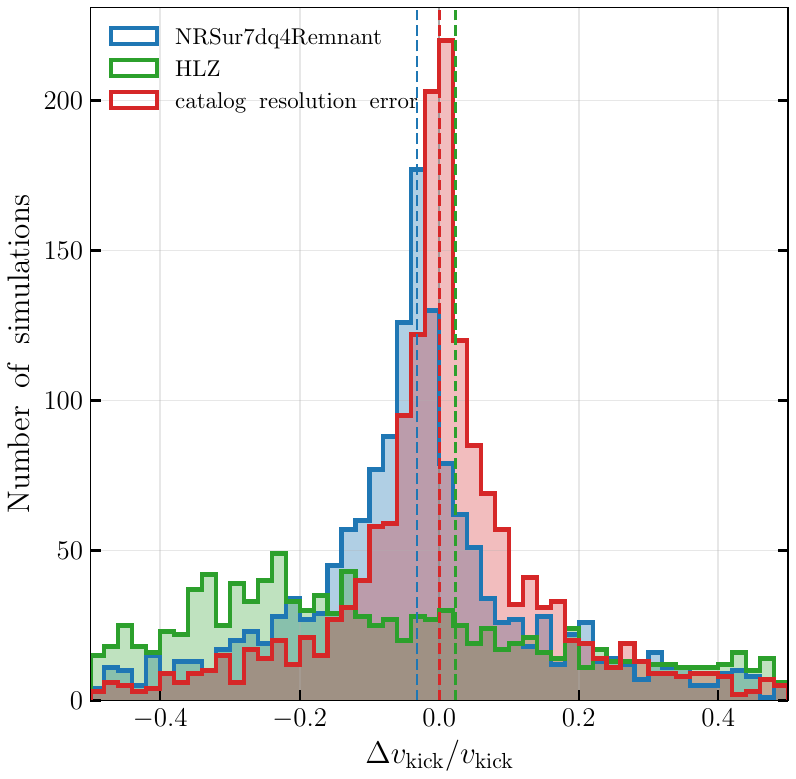}
    \caption{Distribution of the fractional recoil error,
    $\Delta v_{\rm kick}/v_{\rm kick}$, for generic precessing SXS
    simulations. For \NRSur{}, the recoil is evaluated directly for each
    binary configuration. For \HLZ{}, we evaluate 100 values of the
    undetermined phase $\Theta$ for each configuration and use the median
    predicted recoil. The SXS catalog resolution uncertainty, obtained from
    the two highest available numerical resolutions, is shown for comparison.}
    \label{fig:precessing_kick_error_hist}
\end{figure}

%==========================================================================
%=========================================================================
\section{Accuracy of recoil models for precessing BBHs}
\label{app:kick}
%==========================================================================
%=========================================================================

Next, we quantify the accuracy of \NRSur{} and \HLZ{} recoil models for $q \leq 6$.
Figure~\ref{fig:precessing_kick_error_hist} shows the distribution of the fractional recoil error, $\Delta v_{\rm kick}/v_{\rm kick}$, for the \NRSur{} and \HLZ{} recoil models evaluated against the full set of precessing SXS simulations. For \NRSur{}, we directly evaluate the predicted recoil for each NR configuration. The \HLZ{} model requires an additional phase parameter $\Theta$ that is not specified by the intrinsic binary parameters. We therefore evaluate the \HLZ{} recoil for $100$ values of $\Theta$ for each NR configuration, take the median predicted recoil, and use this median to compute the fractional error shown in the histogram. For reference, we also show the SXS catalog resolution uncertainty, obtained from the difference between the recoil velocities at the two highest available numerical resolutions.

Both recoil models exhibit substantially broader error distributions than the NR resolution uncertainty. The \NRSur{} distribution is relatively concentrated around zero but has broad asymmetric tails, while the \HLZ{} distribution is considerably broader and develops a pronounced negative tail. Importantly, the NR resolution uncertainty itself also exhibits a non-negligible distribution with extended tails, highlighting the numerical difficulty of accurately extracting recoil velocities for generic precessing mergers. Nevertheless, the model-error distributions extend substantially beyond the typical NR uncertainty, with fractional errors of tens of percent occurring for both models. Thus, while part of the observed disagreement is associated with the intrinsic numerical uncertainty of the NR recoils, it cannot account for the full modeling error. This behavior reflects the intrinsic difficulty of predicting generic precessing recoils deterministically, since the kick depends sensitively on the in-plane spin geometry and, in the case of \HLZ{}, on the additional phase $\Theta$.

Over the full SXS dataset, \NRSur{} achieves a median absolute error of approximately $30~\mathrm{km\,s^{-1}}$ and a mean absolute error of $311~\mathrm{km\,s^{-1}}$, while \HLZ{} achieves a median absolute error of approximately $15~\mathrm{km\,s^{-1}}$ and a mean absolute error of $370~\mathrm{km\,s^{-1}}$. The large difference between the median and mean errors is consistent with the extended tails visible in Fig.~\ref{fig:precessing_kick_error_hist}: although many configurations are predicted reasonably well, a smaller population of systems produces very large recoil errors. Thus, neither model provides uniformly accurate deterministic kick predictions across the full precessing parameter space.

The behavior of \NRSur{} depends strongly on whether the binary lies within its calibration domain. For systems satisfying $q\le4$, \NRSur{} is essentially unbiased, with a median error of only $-1.3~\mathrm{km\,s^{-1}}$. Outside this regime the model develops a systematic positive bias, with a median error of $84~\mathrm{km\,s^{-1}}$. The effect becomes particularly pronounced for mass ratios $q>4$, where the median error increases to approximately $86~\mathrm{km\,s^{-1}}$. Similar trends appear when the simulations are binned by recoil velocity, with both \NRSur{} and \HLZ{} showing increasingly large systematic deviations for kicks above $\sim1000~\mathrm{km\,s^{-1}}$. Together with the broad fractional-error distributions in Fig.~\ref{fig:precessing_kick_error_hist}, these results motivate our decision to model precessing recoil velocities probabilistically rather than through a single deterministic fitting formula.

%==========================================================================
%=========================================================================
\section{Details about the normalizing flow model}
\label{app:flow_details}
\begin{figure}
    \centering
    \includegraphics[width=0.48\textwidth]{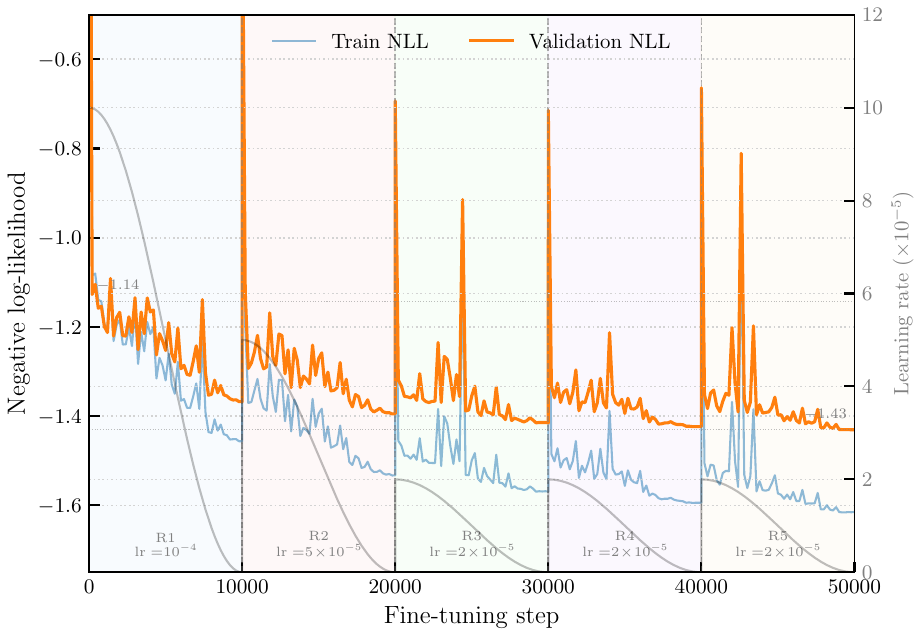}
    \caption{Warm restarts are used to improve optimization of the conditional normalizing-flow recoil model \textcolor{linkcolor}{\texttt{gwModelRemP\_flow}}. Training (blue) and validation (orange) negative log likelihoods are shown across five rounds with progressively reduced learning rates; see Appendix~\ref{app:flow_details}. The restart schedule reaches a test NLL of $-1.431$, improving on the $-1.394$ obtained with continuous fixed-rate training.}
    \label{fig:finetune_curve}
\end{figure}

%==========================================================================
%=========================================================================
We train the normalizing flow using a multi-round warm-restart strategy, dividing the optimization into five successive rounds of $10{,}000$ steps with learning rates $\{10^{-4}, 5 \times 10^{-5}, 2 \times 10^{-5}, 2 \times 10^{-5}, 2 \times 10^{-5}\}$. At the start of each round, the optimizer state (gradient moments and learning-rate scheduler) is reset while the network weights are retained from the previous round's best checkpoint. This procedure periodically refreshes the optimizer's momentum history, while the decreasing learning rates progressively narrow the search around the best solution. Figure~\ref{fig:finetune_curve} shows the test-set NLL as a function of training step. A continuous $40{,}000$-step run with a fixed learning rate converges to a test NLL of $-1.394$, whereas the warm-restart strategy reaches $-1.431$ after $50{,}000$ total steps. Table~\ref{tab:flow_architecture} summarizes the architecture and hyperparameters.

\begin{table}[t]
\centering
\caption{Architecture and hyperparameters of the \gwModelP{} normalizing flow for recoil-velocity prediction.}
\label{tab:flow_architecture}
\begin{tabular}{ll}
\hline\hline
Parameter & Value \\
\hline
Base distribution & Standard normal \\
Transform & Masked piecewise rational-quadratic \\
Number of layers & 8 \\
Hidden features per layer & 64 \\
Residual blocks per layer & 2 \\
Number of spline bins & 8 \\
Tail bound & 6.0 \\
Dropout probability & 0.05 \\
Context dimension & 5 \\
Target dimension & 1 \\
Total trainable parameters & $\sim155{,}000$ \\
\hline\hline
\end{tabular}
\end{table}

%==========================================================================
\section{Inference of remnant properties for a population of BBH mergers}
\label{sec:astro_pop}
\begin{figure}
    \centering
    \includegraphics[width=\columnwidth]{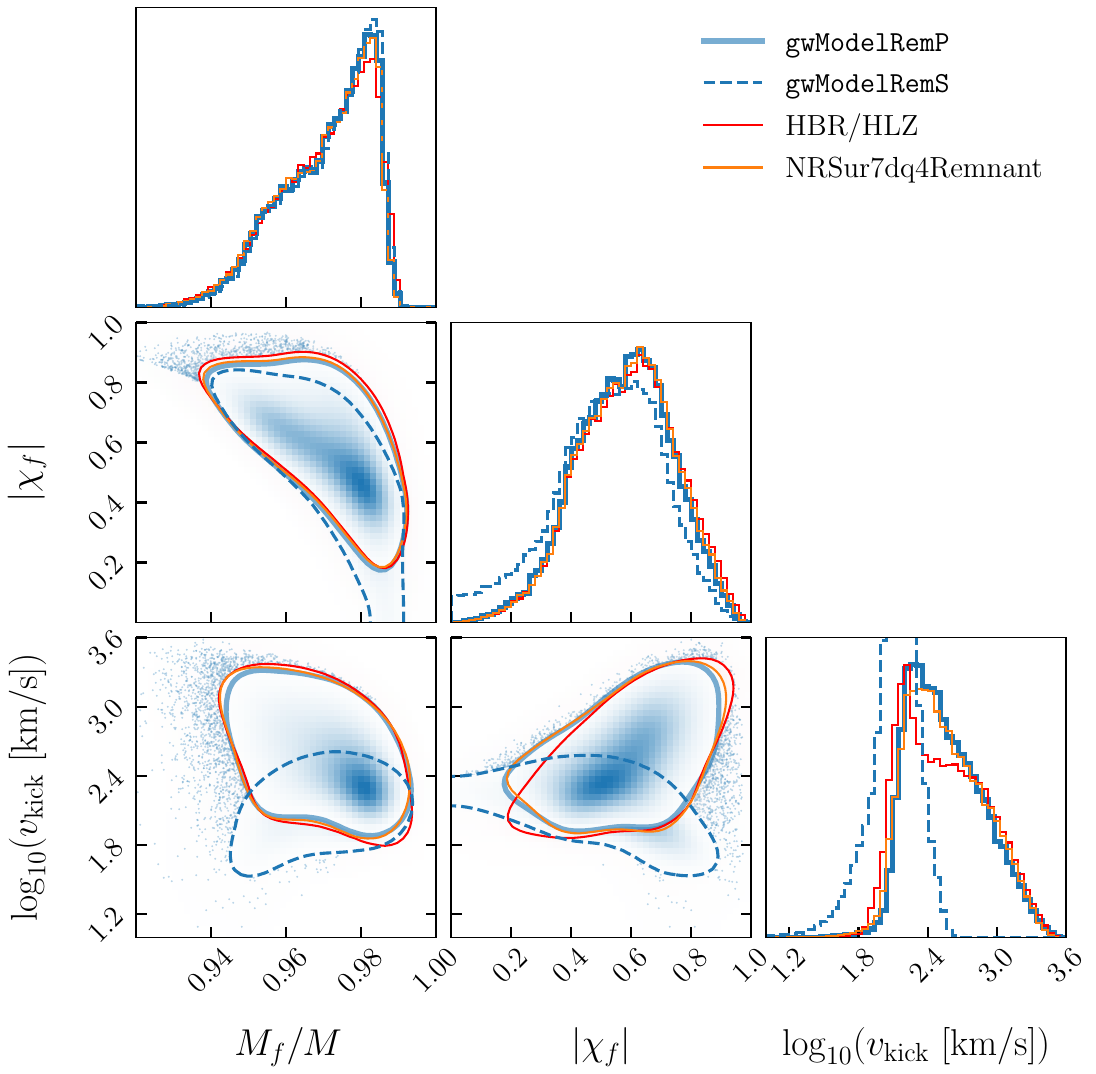}
    \caption{Population-level comparisons test whether configuration-dependent model differences survive marginalization over an astrophysical ensemble. For $5000$ synthetic precessing binaries with $q\in[1,6]$, $|\chi_1|,|\chi_2|\in[0.001,1]$, and isotropic spin orientations, we compare \gwModelP{}, \HBR{}/\HLZ{}, and \textcolor{linkcolor}{\texttt{NRSur7dq4Remnant}}. Diagonal panels show the marginal distributions of $M_f/M$, $|\chi_f|$, and $v_{\rm kick}$, and off-diagonal panels show their correlations; see Appendix~\ref{sec:astro_pop}. The models agree closely for remnant mass and spin at the population level, while recoil retains the largest model dependence.}
    \label{fig:astro_remnant_corner}
\end{figure}

%==========================================================================
To assess the impact of remnant-model choices in population-level applications, we generate a synthetic population of $5000$ precessing BBHs by sampling $q\in[1,6]$, $|\chi_1|,|\chi_2|\in[0.001,1]$, and isotropically distributed spin directions. Figure~\ref{fig:astro_remnant_corner} compares the resulting distributions of remnant mass, remnant spin magnitude, and recoil velocity predicted by \gwModelP{}, \HBR{}/\HLZ{}, and \textcolor{linkcolor}{\texttt{NRSur7dq4Remnant}}. For reference, we also include predictions from the aligned-spin model \gwModelS{}. All three precessing models produce broadly consistent distributions for the remnant mass and spin magnitude, whereas larger differences emerge in the recoil velocity. This behavior is expected: the remnant mass depends primarily on the total radiated energy and is only weakly affected by spin precession, while the recoil velocity is highly sensitive to the in-plane spin configuration and nonlinear mode coupling. Consequently, the aligned-spin model reproduces the remnant-mass distribution well but exhibits noticeable differences in the remnant-spin and recoil distributions. 
These results show that localized model differences in specific regions of parameter space, such as high mass ratios, large spins, or strong precession, as discussed in Secs.~\ref{sec:nonprec_nonecc} and~\ref{sec:prec_nonecc}, do not necessarily translate into significant differences in population-level summary statistics, even when appreciable discrepancies exist for individual configurations.

\end{document}